\documentclass[aps,prd,groupaddress,floatfix,tighten,nofootinbib,showpacs,amsfont,superscriptaddress]{revtex4}
\usepackage{url}
\usepackage{xcolor}
\usepackage{hyperref}

\colorlet{shadecolor}{gray!15}
\definecolor{greenLinks}{rgb}{0, 0.6, 0}
\definecolor{blueLinks}{rgb}{0, 0, 0.6}
\definecolor{redLinks}{rgb}{0.6, 0, 0}
\definecolor{tempText}{rgb}{0.55, 0.10,0.67}
\definecolor{eprintLinks}{rgb}{0.4, 0.4, 0.4}
\definecolor{journalLinks}{rgb}{0.6, 0, 0}
\usepackage{amssymb}
\usepackage{amsmath,color}
\usepackage{epsfig}
\usepackage{graphicx,epsf,epsfig}
\usepackage{footnote}
\usepackage{multirow}
\usepackage{enumitem}
\usepackage{color}
\def\dg{\dagger}     

\newcommand{\mc}{\mathcal}

\def\slc#1{\setbox0=\hbox{$#1$}                  
    \dimen0=\wd0                                 
    \setbox1=\hbox{/} \dimen1=\wd1               
    \ifdim\dimen0>\dimen1                        
       \rlap{\hbox to \dimen0{\hfil/\hfil}}      
       #1                                        
    \else                                        
       \rlap{\hbox to \dimen1{\hfil$#1$\hfil}}   
       /                                         
    \fi}

\def\be{\begin{equation}}
\def\ee{\end{equation}}
\def\gs{\mathrel{
   \rlap{\raise 0.511ex \hbox{$>$}}{\lower 0.511ex \hbox{$\sim$}}}}
\def\ls{\mathrel{
   \rlap{\raise 0.511ex \hbox{$<$}}{\lower 0.511ex \hbox{$\sim$}}}}

\newcommand{\ba}{\begin{array}{c}}
\newcommand{\baz}{\begin{array}{cc}}
\newcommand{\barrr}{\begin{array}{rrr}}
\newcommand{\bad}{\begin{array}{ccc}}
\newcommand{\bav}{\begin{array}{cccc}}
\newcommand{\baf}{\begin{array}{ccccc}}
\newcommand{\bea}{\begin{equation} \begin{array}{c}}
\newcommand{\eea}{ \end{array} \end{equation}}
\newcommand{\ea}{\end{array}}

\usepackage[normalem]{ulem}

\def\21{$\mathrm{SU(2)_L \otimes U(1)_Y}$ }

\newcommand {\ignore}[1]{}
\newcommand{\lra}{\mu \leftrightarrow \tau}
\newcommand{\vt}{\vert}
\newcommand{\nn}{\nonumber}
\allowdisplaybreaks \allowdisplaybreaks[2]
\newcommand{\AddrIFUNAM}{
 Instituto~de F{\'{\i}}sica, 
 Universidad~Nacional Aut\'onoma de M\'exico, \\
 Apdo. Postal 20-364, CDMX 01000, M\'exico.}
  
\newcommand{\AddrFCEBUAP}{
 Fac. de Cs. de la Electr\'onica, 
 Benem\'erita Universidad Aut\'onoma de Puebla, 
 Apdo. Postal 542,\\ Puebla, Pue. 72000, M\'exico.}

\newcommand{\AddrCINVESTAV}{
 Departamento de F\'isica, 
 Centro de Investigaci\'on y de Estudios Avanzados del I. P. N.,\\
 Apdo. Post. 14-740, 07000, Ciudad de M\'exico, M\'exico.}

\newcommand{\AddrCIFFU}{Centro Internacional de F\'{\i}sica Fudamental, 
 Benem\'erita Universidad Aut\'onoma de Puebla.
} 

\newcommand{\AddrCECYT}{Centro de Estudios Cient\'ificos y Tecnol\'ogicos No 16, Instituto Polit\'ecnico Nacional, Pachuca: Ciudad del Conocimiento y la Cultura, Carretera Pachuca Actopan km 1+500, San Agust\'in Tlaxiaca, Hidalgo, M\'exico.\\ 
	}
  
\begin{document}
\title{Flavored Non-Minimal Left-Right Symmetric Model\\
	Fermion Masses and Mixings} 
%
\author{E. A. Garc\'es}
 \email{egarces@fis.cinvestav.mx}
 \affiliation{\AddrIFUNAM}
\author{Juan Carlos G\'omez-Izquierdo}
 \email{jcgizquierdo1979@gmail.com}
 \affiliation{\AddrCECYT}
 \affiliation{\AddrCINVESTAV}
 \affiliation{\AddrIFUNAM}
\author{F. Gonzalez-Canales}
 \email{felixfcoglz@gmail.com}
 \affiliation{\AddrFCEBUAP}
 \affiliation{\AddrCIFFU}

\date{\bf \today} 

\begin{abstract}\vspace{2cm}
 A complete study on the fermion masses and flavor mixing is presented in a non-minimal 
 left-right symmetric model (NMLRMS) where the 
 ${\bf S}_{3}\otimes {\bf Z}_{2}\otimes {\bf Z}^{e}_{2}$ flavor symmetry drives the Yukawa 
 couplings. In the quark sector, the mass matrices possess a kind of the generalized Fritzsch 
 textures that allow us to fit the CKM mixing matrix in good agreement to the latest
 experimental data. In the lepton sector, on the other hand, a soft breaking of the 
 $\mu\leftrightarrow \tau$ symmetry provides a non zero and non maximal reactor and 
 atmospheric angles, respectively. The inverted and degenerate hierarchy are favored in the 
 model where a set of free parameters is found to be consistent with the current neutrino 
 data.
\end{abstract}

\begin{flushright}
CIFFU-18-XX
\end{flushright}
%
\maketitle
\section{Introduction}
%
 In particle physics, flavor 
 symmetries~\cite{Ishimori:2010au,Grimus:2011fk,Ishimori:2012zz,King:2013eh} have played an 
 important role in the understanding of the quark and lepton flavor mixings through the 
 CKM~\cite{Cabibbo:1963yz, Kobayashi:1973fv} and PMNS~\cite{Maki:1962mu, Pontecorvo:1967fh} 
 mixing matrices, respectively. 
 According to the experimental data, the values for the magnitudes of all CKM entries obtained 
 from a global fit are~\cite{Patrignani:2016xqp}: 
\begin{align}\label{ckm}
 {\bf V}_{\mathrm{CKM}}=
 \begin{pmatrix}
  0.97434^{+0.00011}_{-0.00012} & 
   0.22506 \pm 0.00050 & 
   0.00357 \pm 0.00015 \\ 
  0.22492 \pm 0.00050 & 
   0.97351 \pm 0.00013 & 
   0.0411 \pm 0.0013 \\ 
  0.00875^{+0.00032}_{-0.00033} & 
   0.0403 \pm 0.0013 & 
   0.99915 \pm 0.00005
\end{pmatrix}.
\end{align}
The Jarlskog invariant is $J = \left( 3.04_{-0.20}^{+0.21} \right) \times 10^{-5}$. 
In the lepton sector, on the other hand, we know that active neutrinos have a small, but not 
negligible mass which can be understood by type I see-saw 
mechanism~\cite{Minkowski:1977sc,GellMann:1980vs,Mohapatra:1979ia,Schechter:1980gr,Mohapatra:1980yp,Schechter:1981cv}. 
The mixings turn out to be non-trivial, so in the theoretical framework of three active 
neutrinos, the numerical values for the squared neutrino masses and flavor mixing angles 
obtained from a global fit to the current experimental data on neutrino oscillations~\cite{deSalas:2017kay,Esteban:2016qun,Capozzi:2018ubv}, at Best 
Fit Point (BFP) $\pm 1 \sigma$ and $3 \sigma$ ranges, 
are~\cite{deSalas:2017kay,globalfit} 
\begin{equation}\label{ndata}
\begin{array}{ll}\vspace{3mm} 
\Delta m^{2}_{21} \left( 10^{-5} \, \textrm{eV}^{2} \right) = 
7.60_{-0.18}^{+0.19}, \, 7.11 - 8.18, &
\left| \Delta m^{2}_{31} \right| \left( 10^{-3} \, \textrm{eV}^{2} \right) =  
\left\{ \begin{array}{l}\vspace{2mm}
2.48_{-0.07}^{+0.05}, 2.30 - 2.65  \\ 
2.38_{-0.06}^{+0.05}, 2.20 - 2.54
\end{array} \right. , \\ \vspace{3mm}
\sin^{2} \theta_{12} / 10^{-1} = 3.23 \pm 0.16, \, 2.78 - 3.75 , &
\sin^{2} \theta_{23} / 10^{-1} = 
\left\{ \begin{array}{l}\vspace{2mm}
5.67_{-1.24}^{+0.32} , \, 3.93 - 6.43 \\ 
5.73_{-0.39}^{+0.25} , \, 4.03 - 6.40
\end{array} \right.  , \\ \vspace{2mm}
\sin^{2} \theta_{13} / 10^{-2}  = 
\left\{ \begin{array}{l}\vspace{2mm} 
2.26 \pm 0.12 , \, 1.90 - 2.62  \\  
2.29 \pm 0.12 , \, 1.93 - 2.65
\end{array}  \right. . 
\end{array}
\end{equation}
The upper and lower rows are for a normal and inverted hierarchy of 
the neutrino mass spectrum, respectively. At the same time, there is not yet solid evidence on the Dirac CP-violating phase. So, 
from these data  it is obtained (for inverted ordering) that the magnitude of the leptonic mixing matrix elements have 
the following values at $3\sigma$~\cite{Esteban:2016qun}
\begin{equation}\label{Eq:PMNS:fits-2}
	\left( \begin{array}{ccc}
	0.799 - 0.844 & 0.516 - 0.582 & 0.141 - 0.156 \\
	0.242 - 0.494 & 0.467 - 0.678 & 0.639 - 0.774 \\
	0.284 - 0.521 & 0.490 - 0.695 & 0.615 - 0.754 
	\end{array}  \right) .
	\end{equation} 
Understanding the contrasted values between the CKM and PMNS mixing matrices is still a challenge in particle physics. 
In this line of thought, many flavor models such as 
$S_{3}$~\cite{Pakvasa:1977in,Kubo:2003iw,Kubo:2003pd,Kobayashi:2003fh,
Chen:2004rr, Kubo:2005sr, Mondragon:2006hi, Felix:2006pn,Mondragon:2007af, Mondragon:2007nk, Mondragon:2007jx, Meloni:2010aw,Dicus:2010iq,Bhattacharyya:2010hp,Canales:2011ug,
Dong:2011vb,Dias:2012bh,Canales:2012ix,Canales:2012dr,GonzalezCanales:2012za, Canales:2013ura, Canales:2013cga,Ma:2013zca,Kajiyama:2013sza,Hernandez:2013hea,Das:2014fea,Ma:2014qra,Hernandez:2014vta,Hernandez:2014lpa,Gupta:2014nba,Das:2015sca,Hernandez:2015dga,Hernandez:2015zeh,Arbelaez:2016mhg,Hernandez:2015hrt,CarcamoHernandez:2016pdu,Pramanick:2016mdp,Gomez-Izquierdo:2017rxi,Barradas-Guevara:2017iyt, Cruz:2017add,Das:2017zrm,Espinoza:2018itz, Ge:2018ofp,Gomez-Izquierdo:2018jrx}, $A_{4}$~\cite{Ma:2001dn,He:2006dk,Chen:2009um,Ahn:2012tv,Memenga:2013vc,
Felipe:2013vwa,Varzielas:2012ai,Ishimori:2012fg,Hernandez:2013dta,Babu:2002dz,Altarelli:2005yx,Gupta:2011ct,Altarelli:2005yp,Kadosh:2010rm,Kadosh:2013nra,delAguila:2010vg,Campos:2014lla,Vien:2014pta,Karmakar:2014dva,Karmakar:2015jza,Joshipura:2015dsa,Hernandez:2015tna,Bhattacharya:2016lts,Karmakar:2016cvb,Bhattacharya:2016rqj,Chattopadhyay:2017zvs,CarcamoHernandez:2017kra,CentellesChulia:2017koy,Bjorkeroth:2017tsz,CarcamoHernandez:2018aon}, 
$S_{4}$~\cite{Patel:2010hr,Mohapatra:2012tb,BhupalDev:2012nm,Varzielas:2012pa,Ding:2013hpa,Ishimori:2010fs,Ding:2013eca,Hagedorn:2011un,Campos:2014zaa,Dong:2010zu,VanVien:2015xha,deAnda:2017yeb}, 
$D_{4}$~\cite{Frampton:1994rk,Grimus:2003kq,Grimus:2004rj,Frigerio:2004jg,Adulpravitchai:2008yp,Ishimori:2008gp,Hagedorn:2010mq,Vien:2013zra}, 
$Q_{6}$~\cite{Babu:2004tn,Kajiyama:2005rk,Kajiyama:2007pr,Kifune:2007fj,Babu:2009nn,Kawashima:2009jv,Kaburaki:2010xc,Babu:2011mv,Araki:2011zg,Gomez-Izquierdo:2013uaa,Gomez-Izquierdo:2017med}, 
$T_{7}$~\cite{Luhn:2007sy,Hagedorn:2008bc,Cao:2010mp,Luhn:2012bc,Kajiyama:2013lja,Vien:2014gza,Vien:2015koa,Hernandez:2015cra,Arbelaez:2015toa}, 
$T_{13}$~\cite{Ding:2011qt,Hartmann:2011dn,Hartmann:2011pq,Kajiyama:2010sb}, 
$T^{\prime}$~\cite{Sen:2007vx,Chen:2007afa,Frampton:2008bz,Eby:2011ph,Frampton:2013lva,Chen:2013wba}, 
$\Delta(27)$~\cite{Ma:2007wu,Varzielas:2012nn,Bhattacharyya:2012pi,Ma:2013xqa,Nishi:2013jqa,Varzielas:2013sla,Ma:2014eka,Abbas:2014ewa,Abbas:2015zna,Varzielas:2015aua,Bjorkeroth:2015uou,Chen:2015jta,Vien:2016tmh,Hernandez:2016eod,CarcamoHernandez:2017owh,deMedeirosVarzielas:2017sdv,Bernal:2017xat,CarcamoHernandez:2018iel}, 
and $A_{5}$~\cite{Everett:2008et,Feruglio:2011qq,Cooper:2012bd,Varzielas:2013hga,Gehrlein:2014wda,Gehrlein:2015dxa,DiIura:2015kfa,Ballett:2015wia,Gehrlein:2015dza,Turner:2015uta,Li:2015jxa} 
have been proposed to face this open question. 

From a phenomenological point of view, the CKM mixing matrix may be accommodated by the Fritzsch
\cite{Fritzsch:1977vd, Fritzsch:1979zq, Fritzsch:1985eg} and the Nearest Neighbor Interaction textures (NNI)
\cite{Branco:1988iq, Branco:1994jx, Harayama:1996am, Harayama:1996jr}, however, only the latter can
fit with good accuracy the CKM matrix. On the other hand, as can be seen from the PMNS values, the lepton sector seems to obey approximately the $\mu \leftrightarrow \tau$ symmetry ~\cite{Mohapatra:1998ka, Lam:2001fb, Kitabayashi:2002jd, Grimus:2003kq, Koide:2003rx} since that $\left|V_{\mu i}\right|\approx \left|V_{\tau i}\right|$ ($i=1, 2, 3.$). At present, 
the Long-baseline energy experiment NO$\nu$A has disfavored the exact 
$\mu \leftrightarrow \tau$ symmetry, some works have explored the breaking and other 
ideas on this appealing 
symmetry~\cite{Haba:2006hc,Xing:2006xa,GomezIzquierdo:2007vn,GomezIzquierdo:2009id,Xing:2010ez,He:2011kn,Araki:2011zg,Grimus:2012hu,Garg:2013xwa,Gupta:2013it,Luo:2014upa,Xing:2015fdg, Rivera-Agudelo:2015vza,Zhao:2016orh, Borgohain:2017inp, Garg:2017mjk, Borgohain:2018uhf,Samanta:2018hqm,Terrazas:2018pyl, Garg:2018rfz,Garg:2018jsg}. 

Along with this, $\mu\leftrightarrow \tau$ reflection symmetry has gained relevance since 
it predicts the CP violating Dirac phase ($\delta_{CP}=-90^{\circ}$), the atmospheric and the reactor angles are $45^{\circ}$ and 
non-zero respectively~\cite{Ahn:2008hy, Chen:2015siy, Chen:2016ica, Nishi:2016wki, Zhao:2017yvw, Liu:2017frs, Zhao:2018vxy, Nath:2018hjx}. 

Even though the quark and lepton sectors seem to obey different physics, we proposed a framework \cite{Gomez-Izquierdo:2017rxi} to simultaneously accommodate both sectors under the ${\bf S_{3}\otimes Z_{2} \otimes Z^{e}_{2}}$ discrete symmetry within the left-right theory. So that, we will recover the fermion mass matrices, that were obtained previously~\cite{Gomez-Izquierdo:2017rxi}, 
to make a complete study on fermion masses and mixings. In the present work, the quark sector 
will be studied in detail since this was only mentioned 
in~\cite{Gomez-Izquierdo:2017rxi}. 
As we will see, the up and down mass matrices possess the generalized Fritzsch 
textures~\cite{Fritzsch:1999ee} 
(which are not hierarchical~\cite{Verma:2015mgd}),
so that the CKM mixing matrix is parametrized 
by the quark masses and some free parameters that will be tuned by a $\chi^{2}$ analysis in order to 
fit the mixings. In the lepton sector, on the other hand, the mixing angles can be understood 
by a soft breaking of the $\lra$ 
symmetry
in the effective neutrino mass matrix that comes from type I see-saw mechanism. In the current analysis, we found a set of the free parameters that fit the PMNS mixing matrix for the inverted and degenerate hierarchy.
 
The paper is organized as follows: the fermion mass matrices will be introduced in Sec. II. The CKM and PMNS mixing matrices will be obtained in Sec. III and IV, respectively, besides of a $\chi^{2}$ analysis is presented to fit the free parameters in the relevant 
mixing matrices for the quark and lepton sectors separately. Finally, in Sec. V, we present our conclusions.

\section{Fermion Masses}
The following mass matrices were obtained in a particular 
model~\cite{Gomez-Izquierdo:2017rxi} 
where left-right theory~\cite{Pati:1974yy,Mohapatra:1974gc,Senjanovic:1975rk,Senjanovic:1978ev,Mohapatra:1979ia} 
and a ${\bf S}_{3}\otimes{\bf Z}_{2}\otimes {\bf Z}^{e}_{2}$ symmetry are the main ingredients.
\begin{itemize}
 \item {\bf Pseudomanisfest left-right theory (PLRT)}.
\begin{align}
{\bf M}_{q}=\begin{pmatrix}
a_{q}+b_{q} & b_{q} & c_{q} \\ 
b_{q} & a_{q}-b_{q} & c_{q} \\ 
c_{q} & c_{q} & g_{q}
\end{pmatrix},\quad {\bf M}_{\ell}=\begin{pmatrix}
a_{\ell} & 0 & 0 \\ 
0 & b_{\ell}+c_{\ell} & 0 \\ 
0 & 0 & b_{\ell}-c_{\ell}
\end{pmatrix}    ,\quad {\bf M}_{(L, R)}=\begin{pmatrix}
a_{(L, R)} & b_{(L, R)} & b_{(L, R)} \\ 
b_{(L, R)} & c_{(L, R)} & 0 \\ 
b_{(L, R)} & 0 & c_{(L, R)}
\end{pmatrix}. \label{eq10}
\end{align}

\item {\bf Manifest left-right theory (MLRT)}. 
\begin{align}
{\bf M}_{q}=\begin{pmatrix}
a_{q}+b_{q} & b_{q} & c_{q} \\ 
b_{q} & a_{q}-b_{q} & c_{q} \\ 
c^{\ast}_{q} & c^{\ast}_{q} & g_{q}
\end{pmatrix},\quad {\bf M}_{\ell}=\begin{pmatrix}
a_{\ell} & 0 & 0 \\ 
0 & b_{\ell}+c_{\ell} & 0 \\ 
0 & 0 & b_{\ell}-c_{\ell}
\end{pmatrix}    ,\quad {\bf M}_{(L, R)}=\begin{pmatrix}
a_{(L, R)} & b_{(L, R)} & b_{(L, R)} \\ 
b_{(L, R)} & c_{(L, R)} & 0 \\ 
b_{(L, R)} & 0 & c_{(L, R)}
\end{pmatrix}. \label{eq10.m}
\end{align}
\end{itemize}
where $q=u,~d$ stands for the label of up and down quark sector, and $\ell=e,~D$ for the 
charged leptons and Dirac neutrinos. 
On the other hand, as was stated in~\cite{Gomez-Izquierdo:2017rxi}, 
the fermion mass matrices are complex in the {\bf PLRT}. In {\bf MLRT}, 
the charged lepton and the Dirac neutrino mass matrices are reals and the Majorana neutrino is 
complex. 

Let us point out that an analytical study on the lepton mixing, in the {\bf PLRT}, was 
already made in detail in the particular case where the Majorana phases are CP parities, this 
means, these can be $0$ or $\pi$~\cite{Gomez-Izquierdo:2017rxi}. In what follows, the theoretical PMNS mass  mixing 
matrix is recovered but the Majorana phases can take any values, in general. 
At the same time, for the {\bf MLRT} the neutrino mass matrix is easily included in the above 
framework as we will se below.

\section{Quark Sector}

 In this model, the quark mass matrices can be rotated to a basis in which these mass matrices 
 acquire a form with some texture zeros. 
 Also, in the {\bf PLRT} and {\bf MLRT} framework the quark mass matrices can be expressed in 
 the following polar form 
 \begin{equation}\label{Eq:RotMats}
  {\bf M}_{q \texttt{j}} = 
  {\bf U}_{\pi/4}^{\top} {\bf Q}_{q \texttt{j} } 
  \left( 
   \mu_{q \texttt{j}} \, \mathbb{I}_{3 \times 3} 
   + {\cal M}_{q \texttt{j}} 
  \right) 
  {\bf P}_{q \texttt{j}}
  {\bf U}_{\pi/4} ,
 \end{equation}
 where 
 \begin{equation}\label{Eq:RotMats-1}
  {\bf U}_{\pi/4} =
  \frac{1}{ \sqrt{2} }
  \left( \begin{array}{ccc}
   1 & -1 & 0 \\
   1 &  1 & 0 \\
   0 &  0 & \sqrt{2}
  \end{array}  \right) 
  \quad \textrm{and} \quad
  {\cal M}_{q\texttt{j}} = 
  \left( \begin{array}{ccc}
   D_{q \texttt{j}} & B_{q \texttt{j}} & 0 \\
   B_{q \texttt{j}} & A_{q \texttt{j}} & C_{q \texttt{j}} \\
   0 & C_{q \texttt{j}} & 0
  \end{array}  \right).
 \end{equation}
 The ${\bf P}_{q \texttt{j}}$ and ${\bf Q}_{q \texttt{j}}$ are diagonal matrices, whose 
 explicit form depends on the theoretical framework in which we are working. 
 In the above expressions, the $\texttt{j}$ subscript denote to the {\bf PLRT} and {\bf MLRT} 
  frameworks. Concretely, $\texttt{j} = 1$ refers to the {\bf PLRT} framework, where we have 
 that $\mu_{q 1} = |g_{q}|$,
 \begin{equation}
  {\bf Q}_{q 1} = {\bf P}_{q 1}^{\top} ,
  \quad \textrm{and} \quad
  {\bf P}_{q 1} = 
  \textrm{diag} \left( 
  e^{ i \alpha_{q 1} }, e^{ i \beta_{q 1} }, e^{ i \gamma_{q 1} } 
  \right).
 \end{equation}
 The phase factors in the ${\bf P}_{q 1}$ matrix must satisfy the relations
 \begin{equation}\label{eq15} 
  \begin{array}{l}\vspace{2mm}
   2 \alpha_{q1} = \arg \left( a_{q} + b_{q} \right) , \quad 
   2 \beta_{q1}  = \arg \left( a_{q} - b_{q} \right) , \quad 
   2 \gamma_{q1} = \arg \left( g_{q} \right) , \\
   \alpha_{q1} + \beta_{q1} = \arg \left( b_{q} \right),\quad 
   \beta_{q1}  + \gamma_{q1} = \arg \left( c_{q} \right).
  \end{array}
 \end{equation}
 The entries of the ${\cal M}_{q1}$ matrix have the form 
 $A_{q 1} = |a_{q} + b_{q}| - |g_{q}|$, 
 $B_{q 1} = |b_{q}|$, $C_{1 q} = \sqrt{2} |c_{q}|$, and
 $D_{q 1} = |a_{q} - b_{q}| - |g_{q}|$.
 On the other hand, $\texttt{j} = 2$ refers to the {\bf MLRT} framework in which 
 $\mu_{q 2} = g_{q}$,
 \begin{equation}
  {\bf Q}_{q 2} = {\bf P}_{q 2}^{\dagger} ,
  \quad \textrm{and} \quad
  {\bf P}_{q 2} = 
  \textrm{diag} \left( 1, 1, e^{ i \gamma_{q 2} } \right) ,
 \end{equation}
 where $\gamma_{q2} = \arg \left( c_{q} \right)$. 
 The entries of the ${\cal M}_{q 2}$ matrix have the form $A_{q 2} = a_{q} + b_{q} - g_{q}$, 
 $B_{q 2} = b_{q}$, $C_{q 2} = \sqrt{2} |c_{q}|$, $D_{2 q} = a_{q} - b_{q} - g_{q}$.
 
 The real symmetric matrix ${\cal M}_{q \texttt{j}}$ in eq.~(\ref{Eq:RotMats-1}), with $\texttt{j} = 1, 2$, 
 can be brought to its diagonal shape by means of the following orthogonal transformation 
 
 \begin{equation}\label{Eq:RelM-1}
  {\cal M}_{q \texttt{j}} = 
   {\bf O}_{q \texttt{j}} \, 
   {\bf \Delta}_{q \texttt{j}} \, 
   {\bf O}_{q \texttt{j}}^{\top} ,
 \end{equation}
 where ${\bf O}_{q \texttt{j}}$ is a real orthogonal matrix, while
 \begin{equation}
  {\bf \Delta}_{q \texttt{j}} = 
  \textrm{diag} \left( 
   \sigma_{q1}^{^{(\texttt{j})}}, 
   \sigma_{q2}^{^{(\texttt{j})}}, 
   \sigma_{q3}^{^{(\texttt{j})}} 
  \right).
 \end{equation} 
 
 In the last matrix the $\sigma_{q\texttt{i}}^{^{(\texttt{j})}}$, with $\texttt{i}=1,2,3$, are the shifted quark masses~\cite{Canales:2013cga}.
 Now, it is easy conclude that quark mass matrices in both frameworks can be brought to 
 its diagonal shape by means of the following transformations 
\begin{equation}
  \begin{array}{ll}
   {\cal U}_{q1} \, {\bf M}_{q1} \, {\cal U}_{q1}^{\top}  =
	\textrm{diag} \left( m_{q1}, m_{q2}, m_{q3} \right), &
   \qquad \textrm{for {\bf PLRT}}, \\
   {\cal U}_{q2} \, {\bf M}_{q2} \, {\cal U}_{q2}^{\dagger} =
	\textrm{diag} \left( m_{q1}, m_{q2}, m_{q3} \right), &
	\qquad \textrm{for {\bf MLRT}}.
  \end{array}
 \end{equation}   
 In the above expressions the $m_{q i}$ are the physical quark masses, while
 \begin{equation}\label{Eq:Uu_Ud}
  {\cal U}_{q1} \equiv {\bf O}_{q}^{\top} \, {\bf P}_{q1}^{*} \, {\bf U}_{\pi/4}
  \quad \textrm{and} \quad  
  {\cal U}_{q2} \equiv {\bf O}_{q}^{\top} \, {\bf P}_{q2} \, {\bf U}_{\pi/4}. 
 \end{equation} 
 The relation between the physical quark masses and the shifted masses 
 is~\cite{Canales:2013cga,Canales:2012dr}: 
 \begin{equation}\label{Eq:Shifted-masses}
 \sigma_{q\texttt{i}}^{^{(\texttt{j})}} = m_{q\texttt{i}} - \mu_{q \texttt{j}} .
\end{equation}
	
 From the invariants of the real symmetric matrix ${\cal M}_{q \texttt{j}}$, 
 $\mathrm{tr}\left \{ {\cal M}_{q \texttt{j}} \right \}$, 
 $\mathrm{tr}\left \{ {\cal M}_{q \texttt{j}}^{2} \right \}$ and
 $\mathrm{det}\left \{ {\cal M}_{q \texttt{j}} \right \}$, 
 the parameters $A_{q \texttt{j}}$, $B_{q \texttt{j}}$, $C_{q \texttt{j}}$ and 
 $D_{q \texttt{j}}$ can be written in terms of the quark masses and two parameters.
 In this way, we get that the entries of the ${\cal M}_{q \texttt{j}}$ matrix take the form
\begin{equation}\label{Eq:Para_Mq_1}
  \begin{array}{ll}\vspace{2mm}
   \widetilde{A}_{q \texttt{j}} = 
	\frac{ A_{q \texttt{j} } }{ \sigma_{q 3}^{^{(\texttt{j})}} }
	= \widetilde{\sigma}_{q 1}^{^{(\texttt{j})}} 
	- \widetilde{\sigma}_{q 2}^{^{(\texttt{j})}} + \delta_{q} , & 
   \widetilde{B}_{q \texttt{j}} = 
	\frac{ B_{q \texttt{j} } }{ \sigma_{q 3}^{^{(\texttt{j})}} }
	= \sqrt{ \frac{ 
	 \delta_{q} }{ 1 - \delta_{q} } 
	 \xi_{q 1}^{^{(\texttt{j})}} \xi_{q 2 }^{^{(\texttt{j})} } 
	} , \\  
   \widetilde{C}_{q \texttt{j}} = 
	\frac{ C_{q \texttt{j} } }{ \sigma_{q 3}^{^{(\texttt{j})}} }
	= \sqrt{ \frac{ 
	 \widetilde{\sigma}_{q 1}^{^{(\texttt{j})}}
	 \widetilde{\sigma}_{q 2}^{^{(\texttt{j})}} 
	}{ 
	 1 - \delta_{q} } 
	} , & 
   \widetilde{D}_{q \texttt{j}} = 
	\frac{ D_{q \texttt{j} } }{ \sigma_{q 3}^{^{(\texttt{j})}} } 
	= 1 - \delta_{q} ,
  \end{array}
 \end{equation}
 where
 \begin{equation}\label{Eq:Para_Mq_2}
  \begin{array}{ll}\vspace{2mm}
   \xi_{q 1}^{^{(\texttt{j})}} = 
	1 - \widetilde{\sigma}_{q1}^{^{(\texttt{j})}} - \delta_{q} , &
   \xi_{q 2}^{^{(\texttt{j})}} = 
	1 + \widetilde{\sigma}_{q2}^{^{(\texttt{j})}} - \delta_{q} , \\ \vspace{2mm}
   \widetilde{\sigma}_{q1}^{^{(\texttt{j})}} = 
	\frac{ \sigma_{q1}^{^{(\texttt{j})}} }{ \sigma_{q3}^{^{(\texttt{j})}} } = 
	\frac{ 
	 \widetilde{m}_{q1} - \widetilde{\mu}_{q \texttt{j} } 
	}{
	 1 - \widetilde{\mu}_{q \texttt{j} 
	} }, &
   \widetilde{\sigma}_{q2}^{^{(\texttt{j})}} = 
    \frac{ | \sigma_{q2}^{^{(\texttt{j})}} | }{ \sigma_{q3}^{^{(\texttt{j})}} }
    = 
    \frac{ 
	 \left| \widetilde{m}_{q2} - \widetilde{\mu}_{q \texttt{j} } \right| 
    }{
     1 - \widetilde{\mu}_{q \texttt{j} }
    }, \\
   \widetilde{\mu}_{q \texttt{j}} = \frac{ \mu_{q \texttt{j}} }{ m_{q3} } , \quad
   \widetilde{m}_{q1} = \frac{ m_{q1} }{ m_{q3} }, &
   \widetilde{m}_{q2} = \frac{ m_{q2} }{ m_{q3} }.
  \end{array}
 \end{equation}
  
  In order to obtain the above parametrization we considered 
 $\sigma_{q2}^{^{(\texttt{j})}} = -| \sigma_{q2}^{^{(\texttt{j})}} |$. 
 With the aid of the expressions in eqs.~(\ref{Eq:Para_Mq_1}) and~(\ref{Eq:Para_Mq_2}), 
 we obtain that the parameters $\delta_{q}$ and $\widetilde{\mu}_{q \texttt{j}}$ must satisfy 
 the following relations
 \begin{equation}
  \widetilde{m}_{q1}  > \widetilde{\mu}_{q \texttt{j}} \geqslant 0
  \quad \textrm{and} \quad 
  \frac{ 1 - \widetilde{m}_{q1} }{ 1 - \widetilde{\mu}_{q \texttt{j} } } 
  > \delta_{q} > 0.
 \end{equation}
 
 From the  conditions above, we conclude that parameter $\widetilde{\mu}_{q \texttt{j}}$ must be
 positive and smaller than one.
 As $m_{q3} > 0$ and $\widetilde{\mu}_{q \texttt{j} } \geqslant 0$ we have 
 $|g_{q}| = g_{q}$ which implies that $\widetilde{\mu}_{q 1 } = \widetilde{\mu}_{q 2 }$.
 
 Therefore, in this parameterization the difference between the quark flavor mixing matrix 
 obtained in the {\bf PLRT} framework and that obtained in the {\bf MLRT} framework
 lies in the ${\bf P}_{q \texttt{j}}$ matrix, which is a diagonal matrix of phase factors.
 From here we will suppress the $\texttt{j}$ index in the expressions of 
 eqs.~(\ref{Eq:Para_Mq_1}) and~(\ref{Eq:Para_Mq_2}), whereby 
 $\sigma_{qi}^{\texttt{j}} \equiv \sigma_{qi}$, 
 $\widetilde{\sigma}_{q1,2}^{\texttt{j}} \equiv \widetilde{\sigma}_{q1,2}$, and
 $\widetilde{\mu}_{q 1 } = \widetilde{\mu}_{q 2 } \equiv \widetilde{\mu}_{q}$, thus
 $\xi_{q 1,2}^{^{(\texttt{j})}} \equiv \xi_{q 1,2}$. The real orthogonal matrix 
 ${\bf O}_{q\texttt{j}} \equiv {\bf O}_{q}$ in terms of the physical quark mass ratios has 
 the form:
 \begin{equation}\label{Eq:Matriz_Oq}
  {\bf O}_{q \texttt{j}} = 
  \left( \begin{array}{ccc}\vspace{2mm}
   \sqrt{ \frac{ 
	\widetilde{\sigma}_{q1} \delta_{q} \xi_{q2} 
   }{ 
    {\cal D}_{q1} } 
   } & 
   - \sqrt{ \frac{ 
	\widetilde{\sigma}_{q2} \delta_{q} \xi_{q1} 
   }{ 
	{\cal D}_{q2} } 
   } & 
   \sqrt{ \frac{ 
	\xi_{q1} \xi_{q2} 
   }{ 
    {\cal D}_{q3} } 
   } \\ \vspace{2mm} 
   - \sqrt{ \frac{ 
    \widetilde{\sigma}_{q1} \left(1 - \delta_{q} \right) \xi_{q1} 
   }{ 
	{\cal D}_{q1} } 
   } & 
   \sqrt{ \frac{  
	\widetilde{\sigma}_{q2} \left(1 - \delta_{q}\right) \xi_{q2} 
   }{ 
	{\cal D}_{q2} } 
   } & 
   \sqrt{ \frac{ 
	\delta_{q} \left(1 - \delta_{q}\right) 
   }{ 
	{\cal D}_{q3} } 
   } \\
   - \sqrt{ \frac{ 
	\widetilde{\sigma}_{q2} \xi_{q1} 
   }{ 
	{\cal D}_{q1} } 
   } & 
   - \sqrt{ \frac{ 
    \widetilde{\sigma}_{q1} \xi_{q2} 
   }{ 
	{\cal D}_{q2} } 
   } & 
   \sqrt{ \frac{ 
	\widetilde{\sigma}_{q1} \widetilde{\sigma}_{q2} \delta_{q} 
   }{ 
	{\cal D}_{q3} } 
   }   
  \end{array}  \right) ,
 \end{equation}
 where
 \begin{equation}
  \begin{array}{l}
   {\cal D}_{q1} = 
    \left( 1 - \widetilde{\sigma}_{q1} \right) 
    \left( \widetilde{\sigma}_{q1} + \widetilde{\sigma}_{q2} \right) 
    \left( 1 - \delta_{q}\right), \\
   {\cal D}_{q2} = 
    \left( 1 + \widetilde{\sigma}_{q2} \right) 
    \left( \widetilde{\sigma}_{q1} + \widetilde{\sigma}_{q2} \right) 
    \left( 1 - \delta_{q} \right),\\
   {\cal D}_{q3} = \left( 1 - \widetilde{\sigma}_{q1} \right) 
    \left( 1 + \widetilde{\sigma}_{q2} \right) 
    \left(1 - \delta_{q}\right).
  \end{array}
 \end{equation}

\subsection*{Quark Flavor Mixing Matrix}

The quark flavor mixing matrix CKM emerges from the mismatch between the diagonalization of $u$- 
 and $d$-type quark mass matrices. So, this mixing matrix is defined as 
 ${\bf V}_{\mathrm{CKM}} = {\bf U}_{u} {\bf U}_{d}^{\dagger}$, where ${\bf U}_{u}$ and 
 ${\bf U}_{d}$ are the unitary matrices that diagonalize to the u- and d-type quark mass 
 matrices, respectively.
 
 From eqs.~(\ref{Eq:Uu_Ud}) we obtain
 \begin{equation}\label{Eq:CKM}
  \begin{array}{ll}
   {\bf V}_{\mathrm{CKM}} = 
    {\bf O}_{u1}^{\top} \, {\bf P}_{u1}^{*} \, {\bf U}_{\pi/4}
    \left( {\bf O}_{d1}^{\top} \, {\bf P}_{d1}^{*} \, {\bf U}_{\pi/4} \right)^{\dagger}
    = e^{ i \zeta_{1} } \,
    {\bf O}_{u1}^{\top} {\bf P}_{1}^{( u-d )} {\bf O}_{d1}, &
   \qquad \textrm{for {\bf PLRT}}, \\   
   {\bf V}_{\mathrm{CKM}} = 
    {\bf O}_{u2}^{\top} \, {\bf P}_{u2} \, {\bf U}_{\pi/4}
    \left( {\bf O}_{d2}^{\top} \, {\bf P}_{d2} \, {\bf U}_{\pi/4} \right)^{\dagger}
    = {\bf O}_{u2}^{\top} {\bf P}_{2}^{( u-d )} {\bf O}_{d2}, &
    \qquad \textrm{for {\bf MLRT}},
  \end{array}
 \end{equation}
 where   
 \begin{equation}\label{Eq:Matrices.Pud}
  {\bf P}_{ \texttt{j} }^{ (u-d) } = 
   \textrm{diag} \left( 
    1 , e^{i \Theta_{ \texttt{j} } }, e^{ i \Gamma_{ \texttt{j} } }    
   \right) ,
   \qquad \texttt{j} = 1,2,
 \end{equation}
 with 
 \begin{equation}
  \begin{array}{ll}
   \Theta_{1} = - \left( \beta_{u1} - \beta_{d1} + \alpha_{d1} -\alpha_{u1} \right), &
   \Gamma_{1} = - \left( \gamma_{u1} - \gamma_{d1} + \alpha_{d1} - \alpha_{u1} \right) \\
   \Theta_{2} = 0, \quad 
   \Gamma_{2} = \gamma_{u2} - \gamma_{d2}, &
   \zeta_{1} = - \left( \alpha_{u1} -  \alpha_{d1} \right).
  \end{array}
 \end{equation}
 
 From eqs.~(\ref{Eq:Matriz_Oq}) and~(\ref{Eq:Matrices.Pud}) the explicit form of CKM mixing 
 matrix in both frameworks has the form:
 \begin{equation}\label{Eq:CKM-th-1}
  \begin{array}{ll} \vspace{2mm}
   V_{ud}^{^{(\texttt{j})}} = 
	\sqrt{ \frac{ 
	 \widetilde{\sigma}_{c} \widetilde{\sigma}_{s} \xi_{d1} \xi_{u1}
	}{ 
	 {\cal D}_{u1} {\cal D}_{d1} } 
    } e^{ i \Gamma_{\texttt{j}} } 
    + \sqrt{ \frac{ 
	 \widetilde{\sigma}_{d} \widetilde{\sigma}_{u} 
	}{ 
	 {\cal D}_{u1} {\cal D}_{d1} } 
    } \varepsilon_{11}^{^{(\texttt{j})}} , &  \vspace{2mm}
   V_{us}^{^{(\texttt{j})}} =   
    \sqrt{ \frac{ 
	 \widetilde{\sigma}_{c} \widetilde{\sigma}_{d} \xi_{d2} \xi_{u1}
    }{ 
	 {\cal D}_{u1} {\cal D}_{d2} } 
    } e^{ i \Gamma_{\texttt{j}} } 
    - \sqrt{ \frac{ 
	 \widetilde{\sigma}_{s} \widetilde{\sigma}_{u} 
    }{ 
	 {\cal D}_{u1} {\cal D}_{d2} } 
    } \varepsilon_{12}^{^{(\texttt{j})}} , \\  \vspace{2mm}
   V_{ub}^{^{(\texttt{j})}} =  
    - \sqrt{ \frac{ 
	 \widetilde{\sigma}_{c} \widetilde{\sigma}_{d} \widetilde{\sigma}_{s} \delta_{d} \xi_{u1}
	}{ 
	 {\cal D}_{u1} {\cal D}_{d3} } 
    } e^{ i \Gamma_{\texttt{j}} } 
    + \sqrt{ \frac{ 
	 \widetilde{\sigma}_{u} 
	}{ 
	 {\cal D}_{u1} {\cal D}_{d3} } 
    } \varepsilon_{13}^{^{(\texttt{j})}} ,  &  \vspace{2mm}
   V_{cd}^{^{(\texttt{j})}} =   
    \sqrt{ \frac{ 
	 \widetilde{\sigma}_{u} \widetilde{\sigma}_{s} \xi_{d1} \xi_{u2}
	}{ 
	 {\cal D}_{u2} {\cal D}_{d1} } 
    } e^{ i \Gamma_{\texttt{j}} } 
    - \sqrt{ \frac{ 
	 \widetilde{\sigma}_{c} \widetilde{\sigma}_{d} 
	}{ 
	 {\cal D}_{u2} {\cal D}_{d1} } 
    } \varepsilon_{21}^{^{(\texttt{j})}} , \\  \vspace{2mm}
   V_{cs}^{^{(\texttt{j})}} =  
    \sqrt{ \frac{ 
	 \widetilde{\sigma}_{d} \widetilde{\sigma}_{u} \xi_{d2} \xi_{u2}
	}{ 
	 {\cal D}_{u2} {\cal D}_{d2} } 
    } e^{ i \Gamma_{\texttt{j}} } 
    + \sqrt{ \frac{ 
	 \widetilde{\sigma}_{c} \widetilde{\sigma}_{s} 
	}{ 
	 {\cal D}_{u2} {\cal D}_{d2} } 
    } \varepsilon_{22}^{^{(\texttt{j})}} , &  \vspace{2mm}
   V_{cb}^{^{(\texttt{j})}} =   
    - \sqrt{ \frac{ 
     \widetilde{\sigma}_{d} \widetilde{\sigma}_{u} \widetilde{\sigma}_{s} \delta_{d} \xi_{u2}
	}{ 
	 {\cal D}_{u2} {\cal D}_{d3} } 
    } e^{ i \Gamma_{\texttt{j}} } 
    + \sqrt{ \frac{ 
	 \widetilde{\sigma}_{c}  
	}{ 
	 {\cal D}_{u2} {\cal D}_{d3} } 
    } \varepsilon_{23}^{^{(\texttt{j})}} , \\  \vspace{2mm}
   V_{td}^{^{(\texttt{j})}} =    
    - \sqrt{ \frac{ 
     \widetilde{\sigma}_{s} \widetilde{\sigma}_{c} \widetilde{\sigma}_{u} \delta_{u} \xi_{d1}
	}{ 
	 {\cal D}_{u3} {\cal D}_{d1} } 
    } e^{ i \Gamma_{\texttt{j}} } 
    + \sqrt{ \frac{ 
	 \widetilde{\sigma}_{d}  
	}{ 
	 {\cal D}_{u3} {\cal D}_{d1} } 
    } \varepsilon_{31}^{^{(\texttt{j})}} , &  \vspace{2mm}
   V_{ts}^{^{(\texttt{j})}} =    
    - \sqrt{ \frac{ 
	 \widetilde{\sigma}_{d} \widetilde{\sigma}_{c} \widetilde{\sigma}_{u} \delta_{u} \xi_{d2}
	}{ 
	 {\cal D}_{u3} {\cal D}_{d2} } 
    } e^{ i \Gamma_{\texttt{j}} } 
    - \sqrt{ \frac{ 
	 \widetilde{\sigma}_{s}  
	}{ 
	 {\cal D}_{u3} {\cal D}_{d2} } 
    } \varepsilon_{32}^{^{(\texttt{j})}} , \\  \vspace{2mm}
   V_{tb}^{^{(\texttt{j})}} =     
    \sqrt{ \frac{ 
	 \widetilde{\sigma}_{d} \widetilde{\sigma}_{s} 
	 \widetilde{\sigma}_{c} \widetilde{\sigma}_{u} \delta_{d} \delta_{u} 
	}{ 
	 {\cal D}_{u3} {\cal D}_{d3} } 
    } e^{ i \Gamma_{\texttt{j}} } 
    + \sqrt{ \frac{ 
	 1
	}{ 
	 {\cal D}_{u3} {\cal D}_{d3} } 
    } \varepsilon_{33}^{^{(\texttt{j})}} , \\  \vspace{2mm}  
  \end{array}  
 \end{equation}
where
 \begin{equation}\label{Eq:CKM-th-2}
  \begin{array}{ll}  \vspace{2mm}
   \varepsilon_{11}^{^{(\texttt{j})}} = 
    \sqrt{ \delta_{d} \delta_{u} \xi_{d2} \xi_{u2}  } 
    + \sqrt{ \left(1 - \delta_{d}\right) \left(1 - \delta_{u}\right) \xi_{d1} \xi_{u1} }
    e^{i \Theta_{\texttt{j}} } , &  \vspace{2mm}
   \varepsilon_{12}^{^{(\texttt{j})}} = 
    \sqrt{ \delta_{d} \delta_{u} \xi_{d1} \xi_{u2}  } 
    + \sqrt{ \left(1 - \delta_{d}\right) \left(1 - \delta_{u}\right) \xi_{d2} \xi_{u1} }
    e^{i \Theta_{\texttt{j}} } , \\  \vspace{2mm}
   \varepsilon_{13}^{^{(\texttt{j})}} =  
    \sqrt{ \delta_{u}  \xi_{d1}  \xi_{d2} \xi_{u2} } 
    - \sqrt{ \left(1 - \delta_{d}\right) \left(1 - \delta_{u}\right) \delta_{d}  \xi_{u1} }
    e^{i \Theta_{\texttt{j}} } ,  &  \vspace{2mm}
   \varepsilon_{21}^{^{(\texttt{j})}} = 
    \sqrt{ \delta_{d} \delta_{u} \xi_{d2} \xi_{u1}  } 
    + \sqrt{ \left(1 - \delta_{d}\right) \left(1 - \delta_{u}\right) \xi_{d1} \xi_{u2} }
    e^{i \Theta_{\texttt{j}} } ,  \\  \vspace{2mm}
   \varepsilon_{22}^{^{(\texttt{j})}} = 
    \sqrt{ \delta_{d} \delta_{u} \xi_{d1} \xi_{u1}  } 
    + \sqrt{ \left(1 - \delta_{d}\right) \left(1 - \delta_{u}\right) \xi_{d2} \xi_{u2} }
    e^{i \Theta_{\texttt{j}} } ,&  
   \varepsilon_{23}^{^{(\texttt{j})}} = 
    - \sqrt{ \delta_{u} \xi_{d1} \xi_{d2} \xi_{u1}  } 
    + \sqrt{ \left(1 - \delta_{d}\right) \left(1 - \delta_{u}\right) \delta_{d} \xi_{u2} }
    e^{i \Theta_{\texttt{j}} } , \\  \vspace{2mm}
   \varepsilon_{31}^{^{(\texttt{j})}} = 
    \sqrt{ \delta_{d} \xi_{d2} \xi_{u1} \xi_{u2}  } 
    - \sqrt{ \left(1 - \delta_{d}\right) \left(1 - \delta_{u}\right) \delta_{u} \xi_{d1} }
    e^{i \Theta_{\texttt{j}} }   ,&  \vspace{2mm}
   \varepsilon_{32}^{^{(\texttt{j})}} = 
    \sqrt{ \delta_{d} \xi_{d1} \xi_{u1} \xi_{u2}  } 
     - \sqrt{ \left(1 - \delta_{d}\right) \left(1 - \delta_{u}\right) \delta_{u} \xi_{d2} }
     e^{i \Theta_{\texttt{j}} } ,  \\  \vspace{2mm}
   \varepsilon_{33}^{^{(\texttt{j})}} = 
    \sqrt{ \xi_{d1} \xi_{d2} \xi_{u1} \xi_{u2}  } 
    + \sqrt{ \left(1 - \delta_{d}\right) \left(1 - \delta_{u}\right) \delta_{u} \delta_{d} }
    e^{i \Theta_{\texttt{j}} } .     
  \end{array}  
 \end{equation}
 
 The difference between the mixing matrices obtained in the frameworks of {\bf PLRT} and 
 {\bf MLRT} lies in the number of phase factors which each one contains. From the 
 model-independent point of view, the mixing matrix obtained in {\bf MLRT} is a particular 
 case of the matrix obtained in {\bf PLRT}, since we only need make zero the 
 $\Theta_{\texttt{j}}$ phase factor in eq.~(\ref{Eq:CKM-th-2}). 
  
\subsection*{Likelihood Test $\chi^{2}$}

In order to verify the viability of the model for describing the phenomenology associated 
 with quarks. The first issue that we need check is that experimental values for the masses 
 and flavor mixing in the quark sector are correctly reproduced by the model.
 To carry out the above, we perform a likelihood test $\chi^{2}$, in which we consider the 
values of the quark masses reported in Ref~\cite{Patrignani:2016xqp} 
and using the {\it RunDec} program~\cite{Chetyrkin:2000yt}, we obtain the following values 
for the quark mass ratios at the top quark mass scale:
 \begin{equation}\label{Eq:Quark-ratios}
  \begin{array}{ll}
   \widetilde{m}_{u} = \left( 1.33 \pm 0.73 \right) \times 10^{-5}, & 
   \widetilde{m}_{c} = \left( 3.91 \pm 0.42 \right) \times 10^{-3}, \\
   \widetilde{m}_{d} = \left( 1.49 \pm 0.39 \right) \times 10^{-3}, & 
   \widetilde{m}_{s} = \left( 2.19 \pm 0.53 \right) \times 10^{-2}. 
  \end{array}
 \end{equation}
For performing the likelihood test we define the $\chi^{2}$ function as:
 \begin{equation}
  \chi^{2} = \sum_{i=d,s,b} 
   \frac{ 
    \left( \left| V_{ui}^{th} \right| - \left| V_{ui}^{ex} \right| \right)^{2} 
   }{ 
    \sigma_{V_{ui}}^{2}
   }
   +
   \frac{ 
    \left( \left| V_{cb}^{th} \right| - \left| V_{cb}^{ex} \right| \right)^{2} 
   }{ 
    \sigma_{V_{cb}}^{2}
   }.
 \end{equation}
 
 In this expression the terms with superscript ``$ex$'' are the experimental data with 
 uncertainty $\sigma_{V_{kl}}$, whose values are~\cite{Patrignani:2016xqp}:
 \begin{equation}
  \begin{array}{ll}
   \left| V_{ud}^{ex} \right| = 0.97417 \pm 0.00021, &  
   \left| V_{us}^{ex} \right| = 0.2248  \pm 0.0006 , \\
   \left| V_{ub}^{ex} \right| = \left( 4.09 \pm 0.39 \right) \times 10^{-3}, &  
   \left| V_{cb}^{ex} \right| = \left( 40.5 \pm 1.5 \right) \times 10^{-3}.
  \end{array}
 \end{equation}
While the terms with superscript ``$th$'' in the same expression correspond to the 
 theoretical expressions for the magnitude of the entries of the quark mixing matrix CKM.
From eqs.~(\ref{Eq:Shifted-masses}), (\ref{Eq:CKM-th-1}) and~(\ref{Eq:CKM-th-2}) we have
 that the number of free parameters in $\chi^{2}$ function is six and five for the {\bf PLRT} 
 and {\bf MLRT}, respectively. 
However, the $\chi^{2}$ function depends only on four experimental data values, 
which correspond to the magnitude of the entries of the quark mixing matrix.
 In this numerical analysis, we consider the quark mixing matrix in the lower row of 
 eq.~(\ref{Eq:CKM}) as a particular case of the mixing matrix in the upper row of the same 
 equation.
So, in the {\bf PLRT} context, when we simultaneously consider to $\widetilde{\mu}_{d}$, 
$\widetilde{\mu}_{u}$, $\Theta_{1}$, $\Gamma_{1}$, $\delta_{d}$, and $\delta_{u}$ as free 
parameters in the likelihood test, we would only be able to determine the values of these 
parameters at best-fit point (BFP). 
Here, we perform a scan of the parameter space where we sought the BFP through the minimizing 
the $\chi^{2}$ function. 
In Table~\ref{tab:table1} we show the numerical values for the six free parameters  
obtained at the BFP. 
All these results were obtained considering the values in eq.~(\ref{Eq:Quark-ratios}) for 
quark mass ratios. 
The values in the first row of the table~\ref{tab:table1} are valid for the {\bf MLRT} and
{\bf PLRT} frameworks, since $\Theta_{1} = \Theta_{2} = 0$.

\begin{table}
  \begin{center}  
     \begin{tabular}{|c|c|c|c|c|c|c|}\hline \hline   
    $\Theta_{1}$ & $\Gamma_{1}$ & $\widetilde{\mu}_{d}$ & $\widetilde{\mu}_{u}$ &  
    $\delta_{d}$ & $\delta_{u}$ & $\chi^{2}_{\mathrm{min}}$ \\ \hline \hline   
    $0^{\circ}$  & $4.47^{\circ}$  & $4.978 \times 10^{-9}$ & $8.791 \times 10^{-9}$ & 
     $6.025 \times 10^{-2}$ & $4.163 \times 10^{-2}$ & $8.227 \times 10^{-1}$ \\  \hline  
    $3^{\circ}$  & $0.06^{\circ}$  & $5.797 \times 10^{-8}$ & $1.725 \times 10^{-8}$ & 
     $1.179 \times 10^{-1}$ & $9.320 \times 10^{-2}$ & $8.429 \times 10^{-1}$\\  \hline  
    $6^{\circ}$  & $2.17^{\circ}$  & $2.583 \times 10^{-9}$ & $8.838 \times 10^{-9}$ & 
     $8.262 \times 10^{-2}$ & $6.574 \times 10^{-2}$ & $8.219 \times 10^{-1}$\\  \hline  
    $9^{\circ}$  & $57.56^{\circ}$ & $1.009 \times 10^{-7}$ & $5.279 \times 10^{-8}$ & 
     $7.348 \times 10^{-2}$ & $7.317 \times 10^{-2}$ & $8.271 \times 10^{-1}$ \\  \hline  
    $12^{\circ}$ & $40.61^{\circ}$ & $7.174 \times 10^{-8}$ & $1.056 \times 10^{-8}$ & 
     $3.665 \times 10^{-2}$ & $3.064 \times 10^{-2}$ & $9.846 \times 10^{-1}$ \\ \hline \hline  
   \end{tabular}
  \caption{ Numerical values obtained for the six parameters in $\chi^{2}$ function at BFP. These 
   results were obtained by considering simultaneously to $\Theta_{1}$, $\Gamma_{1}$,
   $\widetilde{\mu}_{d}$, $\widetilde{\mu}_{u}$, $\delta_{d}$, and $\delta_{u}$ as free 
   parameters in the scan of the parameter space.  
  }\label{tab:table1}
 \end{center}
 \end{table} 

\begin{figure}
        \includegraphics[width=0.45\linewidth]{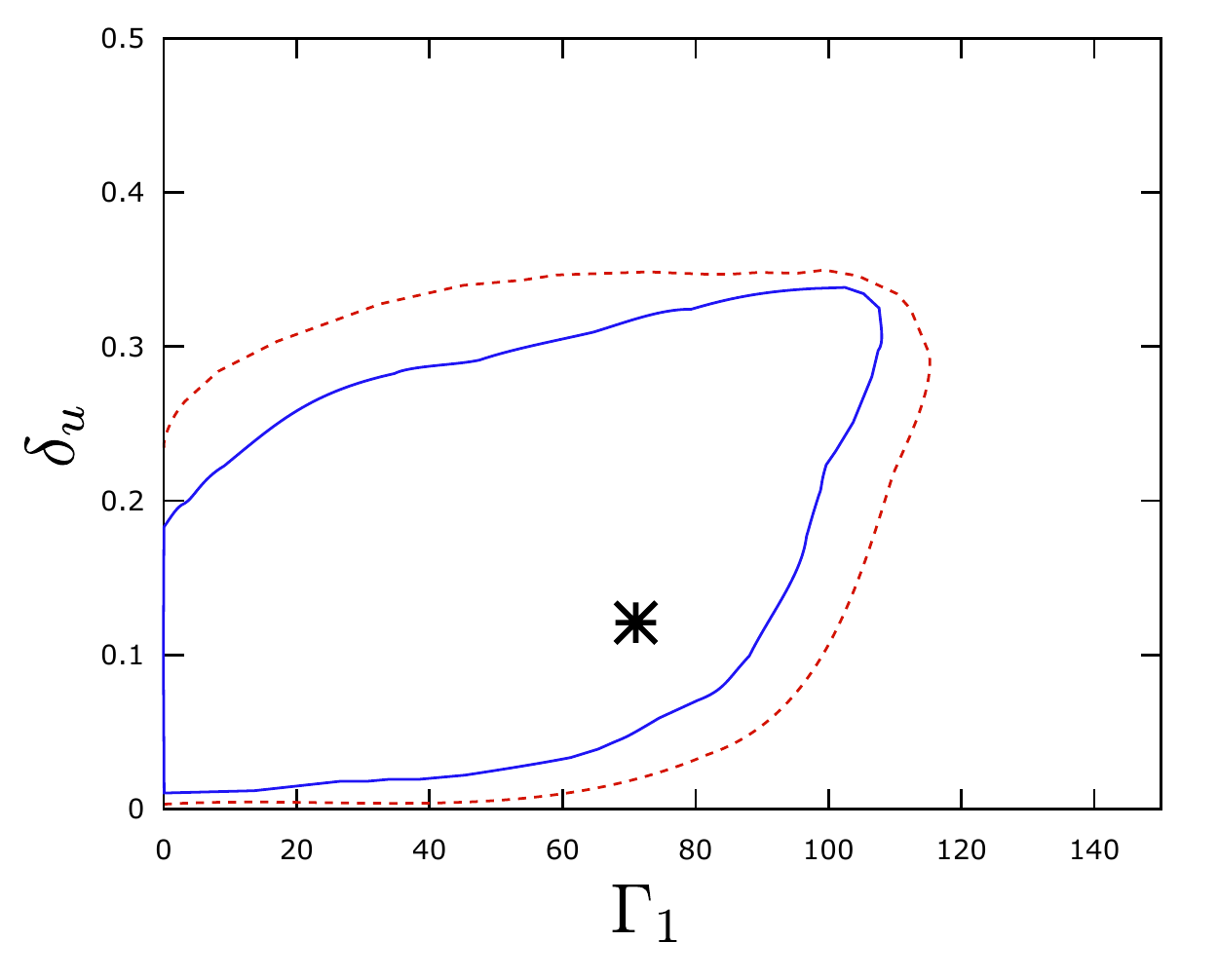}
         \includegraphics[width=0.45\linewidth]{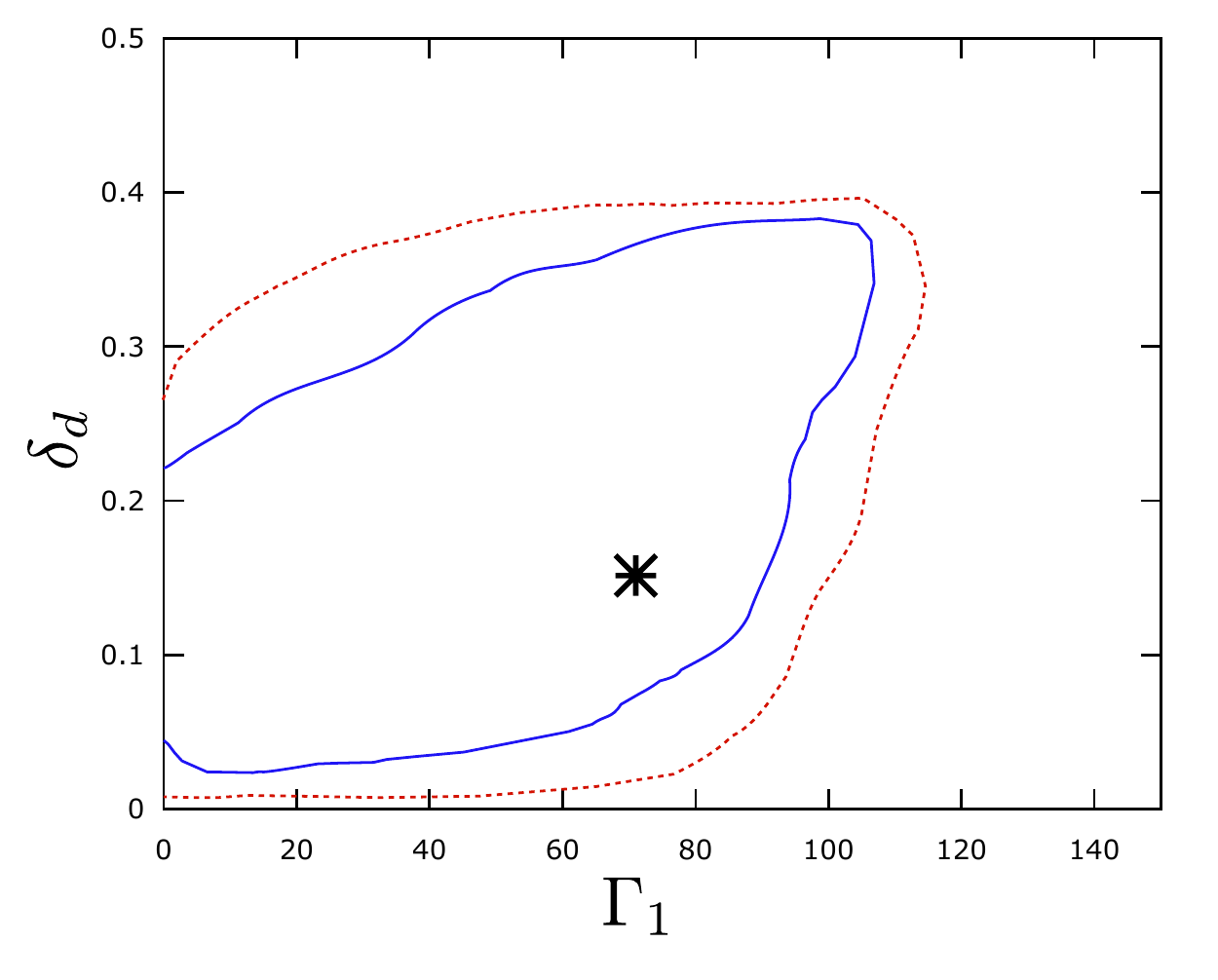}
\caption{\label{fig:quarks} 
     Allowed regions in the parameter space at 70\% CL(blue line) and 
     95\% CL(red dashed line). Here, the black asterisk correspond to the BFP, while the 
     $\Theta$, $\widetilde{\mu}_{u}$, and $\widetilde{\mu}_{d}$ parameters are fixed to the 
     values given in the first row of the table~\ref{tab:table1}.}
\end{figure}

 Now, the $\Theta_{2}$, $\widetilde{\mu}_{u}$, and $\widetilde{\mu}_{d}$ parameters are 
 fixed to the values given in the first row of the table~\ref{tab:table1}, thus the $\chi^{2}$ function
 has one degree of freedom. In Fig.~\ref{fig:quarks}, we show the allowed regions in the 
 parameter space at 70\% CL and 95\% CL, as well as the BFP which is denoted by black 
 asterisk. The resulting values for the free parameters $\Gamma_{1}$, $\delta_{d}$ and 
 $\delta_{u}$, at at 70\% (95\%) CL, are 
\begin{equation}
  \Gamma_{1} \left( ^{\circ} \right) = 71_{- 71 }^{+ 38 } \left(_{- 71 }^{+ 43 }  \right),
   \quad 
  \delta_{u} \left( 10^{-1}  \right) = 1.210_{-0.966}^{+2.146} 
   \left( _{-1.180}^{+2.270} \right),
   \quad
  \delta_{d} \left( 10^{-1}  \right) = 1.514_{-1.126}^{+2.303}
   \left( _{-1.422}^{+2.446} \right).
 \end{equation}
 In the BFP we obtain that $\chi^{2}_{\mathrm{min}} = 8.102 \times 10^{-1}$. 
 Form the likelihood test $\chi^{2}$ we obtain that the magnitudes of all quark mixing matrix 
 elements, at 95\% CL, are
 \begin{equation}
  \left( \begin{array}{ccc}
   0.97433 \pm 0.00018 & 
    0.22508_{-0.00078}^{+0.00080} &
    \left( 4.09_{-0.62}^{+0.60} \right) \times 10^{-3} \\
   0.22481_{-0.00083}^{+0.00076} &
    0.97356_{-0.00020}^{+0.00021} &
    \left( 4.053_{-0.241}^{+0.230} \right) \times 10^{-2} \\
   \left( 1.1942_{-0.1156}^{+0.1914} \right) \times 10^{-2} &
    \left( 3.8948_{-0.2714}^{+0.2176} \right) \times 10^{-2} &
    0.999170_{-0.000096}^{+0.000094}
  \end{array}   \right).
 \end{equation}  
The Jarlskog invariant is 
 \begin{equation}
  J_{\mathrm{CP}} = {\cal I}m \left( V_{ud} V_{cs} V_{us}^{*} V_{cd}^{*} \right)
  = \left( 2.92_{-0.29}^{+0.38} \right) \times 10^{-5} .
 \end{equation}
All these values are in good agreement with experimental data. Also, the results of the above 
 likelihood test can be considered as predictions of the {\bf PLRT} and {\bf MLRT} theoretical 
 frameworks. Because when $\Theta_{1} = \Theta_{2} = 0$ both schemes are equivalent.

\section{Lepton Sector}

As it can verified straightforward, the ${\bf M}_{e}$ charged lepton mass matrix is diagonalized by ${\bf U}_{e L}={\bf S}_{23}{\bf P}_{e}$ and ${\bf U}_{e R}={\bf S}_{23}{\bf P}^{\dg}_{e}$ in the case of {\bf PLRT} and ${\bf U}_{e L}={\bf S}_{23}$ and ${\bf U}_{e R}={\bf S}_{23}$ in the {\bf MLRT}
\begin{align}
 {\bf S}_{23}=\begin{pmatrix}
 1 & 0 & 0 \\ 
 0 & 0 & 1 \\ 
 0 & 1 & 0
\end{pmatrix}, \quad 
 {\bf P}_{e}= 
 \textrm{diag} \left( e^{i\eta_{e}/2}, e^{i\eta_{\mu}/2}, e^{i\eta_{\tau}/2} \right) 
 \label{eue}
\end{align}	
with $\vt m_{e}\vt=\vt a_{e}\vt$, $\vt m_{\mu}\vt=\vt b_{e}-c_{e}\vt$ and $\vt m_{\tau}\vt=\vt b_{e}+c_{e}\vt$ for the former framework and  $ m_{e}=a_{e}$, $m_{\mu}= b_{e}-c_{e}$ and $m_{\tau}=b_{e}+c_{e}$ in the second one.

The ${\bf M}_{\nu}$ neutrino mass matrix, that comes from the type I see-saw mechanism, is parametrized as
\begin{align}
{\bf \mc{M}_{\nu}}\approx\begin{pmatrix}
A_{\nu}& -B_{\nu}(1-\epsilon) & -B_{\nu}(1+\epsilon) \\ 
-B_{\nu}(1-\epsilon)& C_{\nu}(1-2\epsilon) & D_{\nu} \\ 
-B_{\nu}(1+\epsilon)& D_{\nu} & C_{\nu}(1+2\epsilon)
\end{pmatrix}\label{nvem}
\end{align}
where $A_{\nu}$, $B_{\nu}$, $C_{\nu}$ and $D_{\nu}$ are complex parameters; $\epsilon$ is a complex and real free parameter in the {\bf PLRT} and {\bf MLRT} frameworks, respectively. Along with this, the $\epsilon$ parameter was considered as a perturbation to the effective mass matrix such that $\vert \epsilon \vert\leq 0.3$ in order to break softly the $\mu\leftrightarrow\tau$  symmetry. So that, the $\left|\epsilon \right|^{2}$ quadratic terms were neglected in the above matrix. 
Let us remark that the above neutrino mass matrix has been already rotated by the ${\bf S}_{23}$ orthogonal matrix. 
As it was shown in~\cite{Gomez-Izquierdo:2017rxi}, 
the ${\bf {M}_{\nu}}$ effective neutrino mass matrix is diagonalized by ${\bf U}_{\nu}\approx {\bf S}_{23}{\bf \mc{U}_{\nu}}$ such that $\hat{\bf M}_{\nu}=\textrm{diag.}(m_{\nu_{1}}, m_{\nu_{2}}, m_{\nu_{3}})\approx {\bf U}^{\dg}_{\nu}{\bf M}_{\nu}{\bf U}^{\ast}_{\nu}={\bf \mc{U}^{\dg}_{\nu}}{\bf \mc{M}_{\nu}}{\bf \mc{U}^{\ast}_{\nu}}$ where ${\bf \mc U}_{\nu}\approx{\bf \mc{U}^{0}_{\nu}}{\bf \mc {U}^{\epsilon}}_{\nu}$. 
Here, 
${\bf \mc{U}^{0}_{\nu}}$ diagonalizes the ${\bf \mc{M}^{0}_{\nu}}$ neutrino mass matrix with exact $\mu\leftrightarrow\tau$ symmetry ($\left|\epsilon\right|=0$) this means  ${\bf \mc{U}^{0 T}_{\nu}}{\bf \mc{{M}}^{0}_{\nu}}$  ${\bf \mc{U}^{0}_{\nu}}=\hat{{\bf M}}^{0}_{\nu}=\textrm{diag}(m^{0}_{\nu_{1}}, m^{0}_{\nu_{2}}, m^{0}_{\nu_{3}})$. Along with this, the $\epsilon$ parameter breaks the $\mu\leftrightarrow\tau$ symmetry so that its contribution to the mixing matrix is contained in ${\bf \mc{U}^{\epsilon}_{\nu}}$.
\begin{align}
{\bf \mc{U}}^{0}_{\nu}=\begin{pmatrix}
\cos{\theta}_{\nu}~e^{i(\eta_{\nu}+\pi)} & \sin{\theta}_{\nu}~e^{i(\eta_{\nu}+\pi)}  & 0 \\ 
-\frac{\sin{\theta}_{\nu}}{\sqrt{2}}& \frac{\cos{\theta}_{\nu}}{\sqrt{2}} & -\frac{1}{\sqrt{2}} \\ 
-\frac{\sin{\theta}_{\nu}}{\sqrt{2}}& \frac{\cos{\theta}_{\nu}}{\sqrt{2}} & \frac{1}{\sqrt{2}}
\end{pmatrix},\qquad {\bf \mc{U}}^{\epsilon}_{\nu}&\approx\begin{pmatrix}
N_{1}&  0 & -N_{3}\sin{\theta}r_{1}\epsilon \\ 
0 & N_{2} & N_{3}\cos{\theta_{\nu}}r_{2}\epsilon \\ 
N_{1}\sin{\theta_{\nu}}r_{1}\epsilon & -N_{2}\cos{\theta_{\nu}}r_{2}\epsilon & N_{3},
\end{pmatrix}
\end{align}	
where $r_{(1, 2)}\equiv (m^{0}_{\nu_{3}}+m^{0}_{\nu_{(1, 2)}})/(m^{0}_{\nu_{3}}-m^{0}_{\nu_{(1, 2)}})$ and the $N_{i}$ the normalization factors are given as
\begin{align}
N_{1}=\left(1+\sin^{2}{\theta_{\nu}}\vert r_{1}\epsilon \vert^{2}\right )^{-1/2},\quad N_{2}=\left(1+\cos^{2}{\theta_{\nu}}\vert r_{2}\epsilon \vert^{2}\right)^{-1/2},\quad N_{3}=\left(1+\sin^{2}{\theta_{\nu}}\vert r_{1}\epsilon \vert^{2}+\cos^{2}{\theta_{\nu}}\vert r_{2}\epsilon \vert^{2}\right)^{-1/2}.\label{nofac}
\end{align}
Let us emphasize that two relative Majorana phases will be considered along this work in which the $m^{0}_{\nu_{3}}$ neutrino mass is kept positive. Explicitly, we have $\hat{{\bf M}}^{0}_{\nu}=\textrm{diag}(m^{0}_{\nu_{1}}, m^{0}_{\nu_{2}}, m^{0}_{\nu_{3}})= \textrm{diag}(\vert m^{0}_{\nu_{1}}\vert e^{i\alpha}, \vert m^{0}_{\nu_{2}}\vert e^{i\beta},\vert m^{0}_{\nu_{3}}\vert )$ where the associate Majorana phase of $m^{0}_{\nu_{3}}$ has been absorbed in the neutrino field.

\subsection*{Lepton Flavor Mixing Matrix}
In the {\bf PLRT} ({\bf MLRT}) case, we found that $V_{PMNS}\approx{\bf P}^{\dg}_{e}{\bf \mc{U}^{0}_{\nu}{\bf \mc{U}^{\epsilon}_{\nu}}}$ ($\approx {\bf \mc{U}^{0}_{\nu}{\bf \mc{U}^{\epsilon}_{\nu}}}$). Explicitly,
\begin{align}
{\bf V}_{PMNS}\approx{\bf P}^{\dagger}_{e}\begin{pmatrix}
\cos{\theta_{\nu}}N_{1}& \sin{\theta_{\nu}}N_{2}& \sin{2\theta_{\nu}}\frac{N_{3}}{2}(r_{2}-r_{1})\epsilon  \\ 
-\frac{\sin{\theta_{\nu}}}{\sqrt{2}}N_{1}(1+r_{1}\epsilon)& \frac{\cos{\theta_{\nu}}}{\sqrt{2}}N_{2}(1+r_{2}\epsilon) & -\frac{N_{3}}{\sqrt{2}}\left[1-\epsilon~r_{3}\right] \\ 
-\frac{\sin{\theta_{\nu}}}{\sqrt{2}}N_{1}(1-r_{1}\epsilon)& \frac{\cos{\theta_{\nu}}}{\sqrt{2}}N_{2}(1-r_{2}\epsilon)  & \frac{N_{3}}{\sqrt{2}}\left[1+\epsilon~ r_{3}\right]
\end{pmatrix} \label{pmma}
\end{align}	
with $r_{3}\equiv r_{2}\cos^{2}{\theta_{\nu}}+r_{1}\sin^{2}{\theta_{\nu}}$ and ${\bf P}^{\prime}_{e}=\textrm{diag}.(e^{i(\eta_{e}/2-\eta_{\nu}-\pi)}, e^{i\eta_{\mu}/2}, e^{i\eta_{\tau}/2})$. On the other hand,
comparing the magnitude of entries  ${\bf V}_{PMNS}$ with the mixing matrix in the 
standard parametrization of the PMNS, we obtain the following expressions for 
the lepton mixing angles
\begin{align}
\sin^{2}{\theta}_{13}&=\vert {\bf V}_{13}\vert^{2} =\frac{\sin^{2}{2\theta_{\nu}}}{4}N^{2}_{3}\vert \epsilon \vert^{2}~\vert r_{2}-r_{1} \vert^{2},\nonumber\\
\sin^{2}{\theta}_{23}&=\dfrac{\vert {\bf V}_{23}\vert^{2}}{1-\vert {\bf V}_{13}\vert^{2}}=\dfrac{N^{2}_{3}}{2}\frac{\vert 1-\epsilon~r_{3} \vert ^{2}}{1- \sin^{2}{\theta_{13}}},\nonumber\\
\sin^{2}{\theta_{12}}&=\dfrac{\vert {\bf V}_{12}\vert^{2}}{1-\vert {\bf V}_{13}\vert^{2}}=
\dfrac{N^{2}_{2}\sin^{2}{\theta_{\nu}}}{1-\sin^{2}{\theta}_{13}}.\label{mixan}
\end{align} 

In these mixing angles there are four free parameters namely, the absolute neutrino masses, two relative Majorana phase, the $\epsilon$ parameter and the $\theta_{\nu}$ angle. Some parameters could be reduced under certain considerations as follows: the $\theta_{\nu}$ parameter, in good approximation, coincides with the solar angle $\theta_{12}$ since we are in the limit of a soft breaking $\lra$ symmetry so the normalization factors, $N_{i}$, are expected to be of the order $1$, then $\theta_{12}=\theta_{\nu}$. Along with this, the mixing angles may be written in terms of one relative Majorana phase to do so we just have to observe that the reactor angle is non negligible when $\left|r_{2}-r_{1}\right|^{2}$ is large.
\begin{align}\label{r2r1}
\left|r_{2}-r_{1}\right|^{2}=\frac{4\left|m^{0}_{\nu_{3}}\right|^{2}\left|m^{0}_{\nu_{2}}-m^{0}_{\nu_{1}}\right|^{2}}{\left|m^{0}_{\nu_{3}}-m^{0}_{\nu_{1}}\right|^{2}\left|m^{0}_{\nu_{3}}-m^{0}_{\nu_{2}}\right|^{2}}.
\end{align}
 This happens if $\beta-\alpha=\pi$, then we have

\begin{align}
\left|m^{0}_{\nu_{2}}-m^{0}_{\nu_{1}}\right|^{2}&=\left[\left|m^{0}_{\nu_{2}}\right|+\left|m^{0}_{\nu_{1}}\right|\right]^{2},\nn\\
\left|m^{0}_{\nu_{3}}-m^{0}_{\nu_{1}}\right|^{2}&=\left|m^{0}_{\nu_{3}}\right|^{2}+\left|m^{0}_{\nu_{1}}\right|^{2}-2\left|m^{0}_{\nu_{1}}\right|\left|m^{0}_{\nu_{3}}\right|\cos{\alpha}\nonumber\\
\left|m^{0}_{\nu_{3}}-m^{0}_{\nu_{2}}\right|^{2}&=\left|m^{0}_{\nu_{3}}\right|^{2}+\left|m^{0}_{\nu_{2}}\right|^{2}+2\left|m^{0}_{\nu_{2}}\right|\left|m^{0}_{\nu_{3}}\right|\cos{\alpha}.
\end{align}
where the last two factors enhance the former one in order to get allowed values for the reactor angle. In addition, the factors $r_{2}$ and $r_{1}$ can be written in terms of the only relative Majorana phase, $\alpha$. Then, 
\begin{align}
r_{1}=\frac{\left|m^{0}_{\nu_{3}}\right|+\left|m^{0}_{\nu_{1}}\right|e^{i\alpha}}{\left|m^{0}_{\nu_{3}}\right|-\left|m^{0}_{\nu_{1}}\right|e^{i\alpha}},\qquad r_{2}=\frac{\left|m^{0}_{\nu_{3}}\right|-\left|m^{0}_{\nu_{2}}\right|e^{i\alpha}}{\left|m^{0}_{\nu_{3}}\right|+\left|m^{0}_{\nu_{2}}\right|e^{i\alpha}}.
\end{align}

In this way, the Majorana phases are related by the expression already mentioned, $\beta-\alpha=\pi$. This analysis is valid for {\bf PLRT} and {\bf MLRT}, however, in the latter framework the $\epsilon$ parameter is real. 

\subsection*{Likelihood Test $\chi^{2}$}
Once we fixed the $\theta_\nu$ parameter to the solar neutrino mixing angle $\theta_{12}$, the $\chi^{2}$ analysis is carried out to find allowed values of the three remaining free parameters $\epsilon$, the Majorana phase $\alpha$ and the mass of the lightest (common) neutrino $\left|m^{0}_{\nu_{3}}\right|$($m_0$). Two of the absolute neutrino masses can be written as a function of the the lightest mass and $\Delta m_{ij}^2$ as follows

\begin{eqnarray}
\vert m^{0}_{\nu_{2}}\vert&=&\sqrt{\Delta m^{2}_{13}+\Delta m^{2}_{21}+\vert m^{0 }_{\nu_{3}}\vert^{2}},\qquad\vert m^{0}_{\nu_{1}}\vert=\sqrt{\Delta m^{2}_{13}+\vert m^{0}_{\nu_{3}}\vert^{2}},\qquad \textrm{Inverted Hierarchy}\nonumber\\
\vert m^{0}_{\nu_{3}} \vert&=& \sqrt{\Delta m^{2}_{31}+m^{2}_{0}},\qquad \vert m^{0}_{\nu_{2}} \vert=\sqrt{\Delta m^{2}_{21}+m^{2}_{0}}. \qquad \textrm{Degenerate Hierarchy}\label{IDmass}
\end{eqnarray}
where $\vert m^{0 }_{\nu_{3}}\vert$ and $m_{0}$ ($\gtrsim 0.1~eV$) are the lightest and common neutrino masses for the inverted and degenerate ordering, respectively. 

In this analysis, the normal hierarchy will be left out since this was discarded in the previous analytical study~\cite{Gomez-Izquierdo:2017rxi}. The inverted and the degenerate hierarchies will be discussed next. 

The $\chi^{2}$ function is built as
\begin{equation}\label{chi}
\chi^{2}(\epsilon,\alpha,m_{0}(\left|m^{0}_{\nu_{3}}\right|))=\frac{\left(\sin^{2 }{\theta^{th}_{13}}-\sin^{2 }{\theta^{ex}_{13}}\right)^{2}}{\sigma^{2}_{13}}+\frac{\left(\sin^{2 }{\theta^{th}_{23}}-\sin^{2}{\theta^{ex}_{23}}\right)^{2}}{\sigma^{2}_{23}}.
\end{equation}
where the experimental data and theoretical expressions for the mixing angles are given in Eq. (\ref{ndata}) and Eq. (\ref{mixan}), respectively. 
We use the absolute neutrino masses in Eq. (\ref{IDmass}) as a function of $m_{0} (\left|m^{0}_{\nu_{3}}\right|)$, fixing $\Delta m^2_{ij}$ to  the central values of the global fit~\cite{deSalas:2017kay} and 
letting  $m_{0} (\left|m^{0}_{\nu_{3}}\right|)$ as a free parameter. For $\sigma_{13}$  and $\sigma_{23}$ we take the one sigma upper and lower uncertainties using summation in quadrature.

\begin{figure}
        \includegraphics[width=0.42\linewidth]{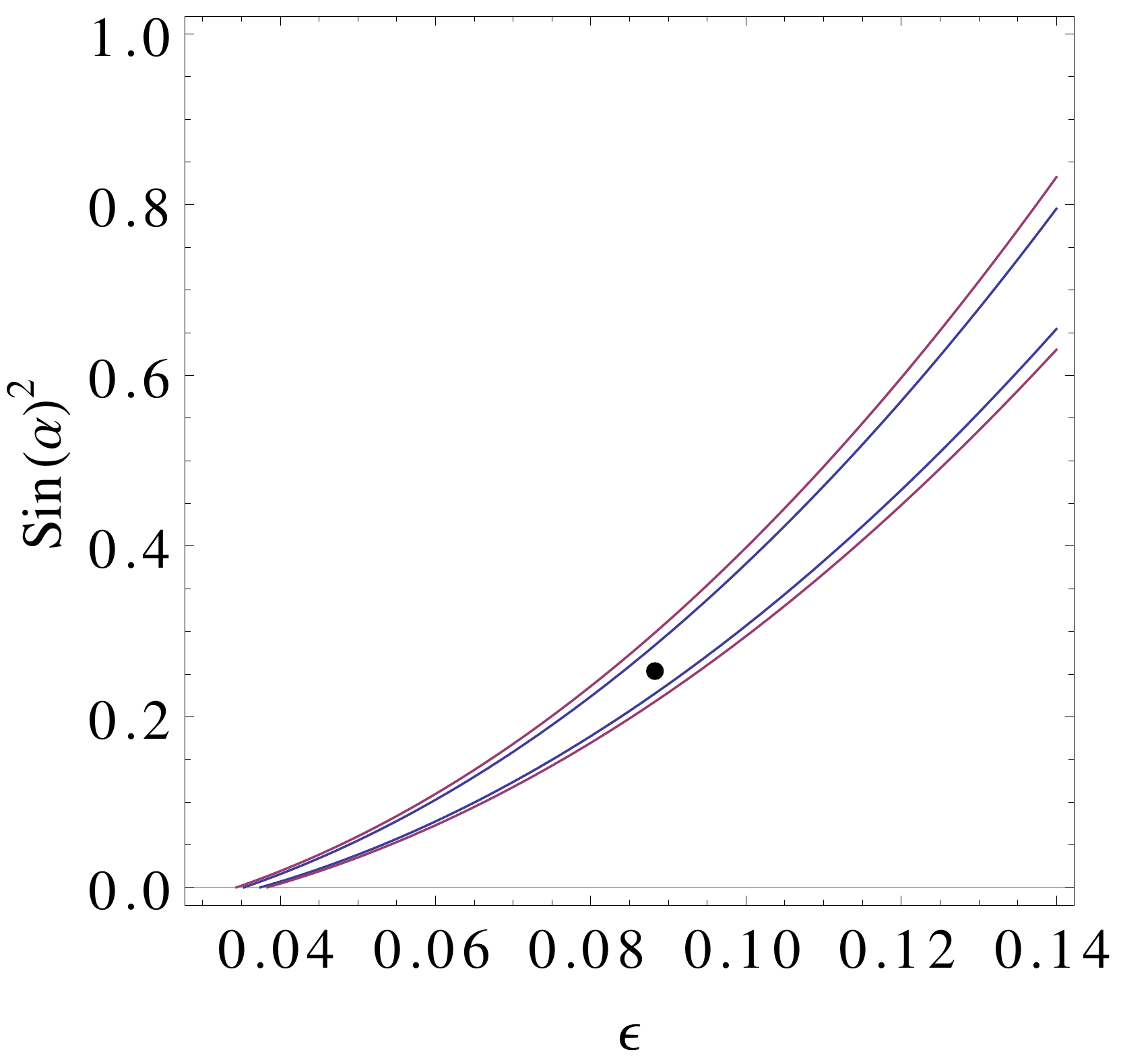}
         \includegraphics[width=0.44 \linewidth]{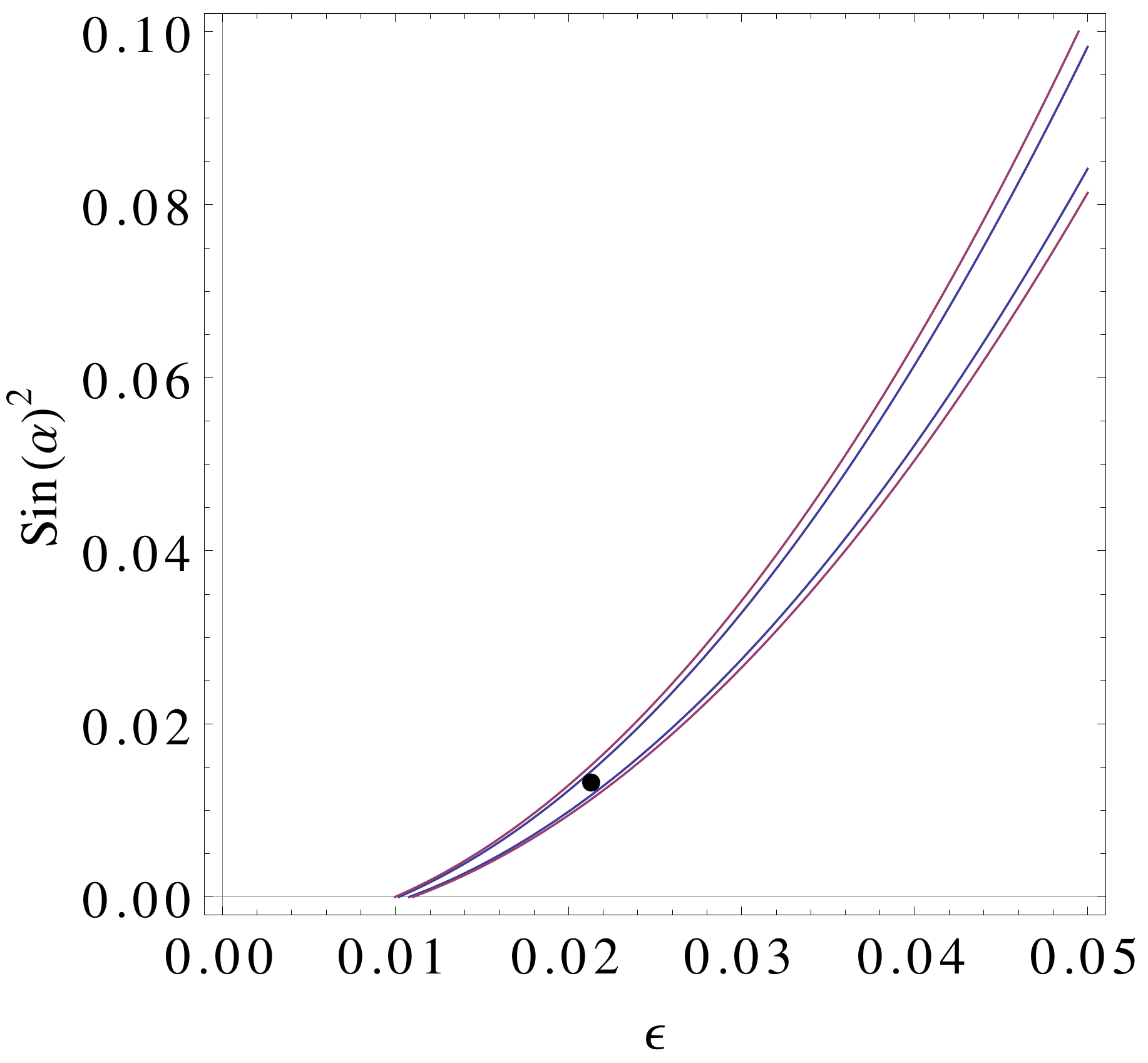}
    \caption{\label{fig:00} Allowed regions in the $\sin(\alpha)$-$\epsilon$ plane, at 90\%CL(blue) and 95\% CL(red) for degenerate (left) and inverted(right) hierarchy. In this case the $\theta_\nu$ parameter is fixed to the solar 
    angle, and $m_{0,3}$ is marginalized.}
\end{figure}

\begin{figure}
        \includegraphics[width=0.42\linewidth]{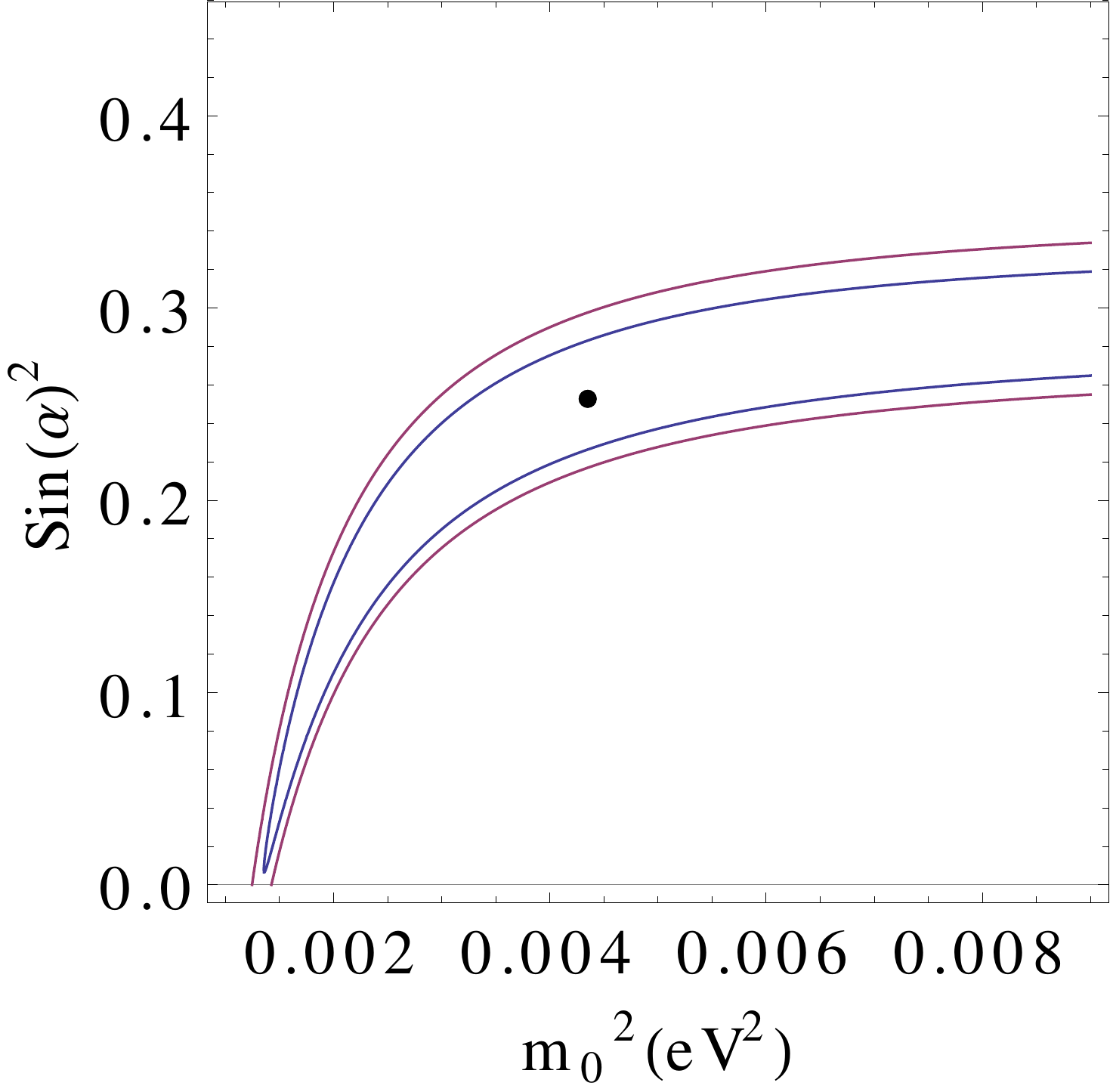}
              \includegraphics[width=0.46\linewidth]{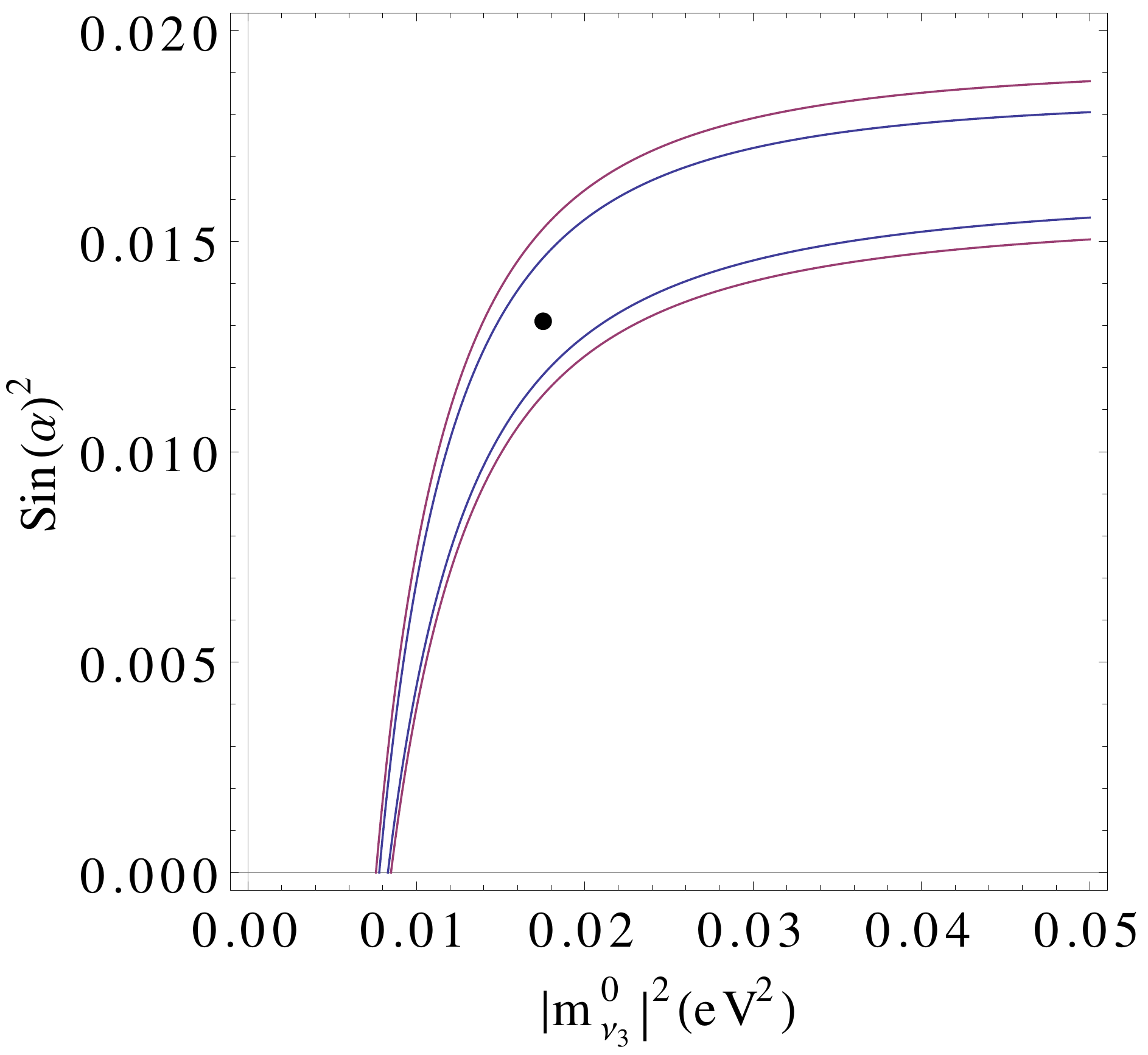}
          \caption{\label{fig:03} Allowed regions in the $\sin(\alpha)$-$\epsilon$ plane, at 90\%CL(blue) and 95\% CL(red) for degenerate (left) and inverted(right) hierarchy. The $\theta_\nu$ parameter is fixed to the solar angle and $\epsilon$ is marginalized.}
\end{figure}

\begin{figure}
           \includegraphics[width=0.41\linewidth]{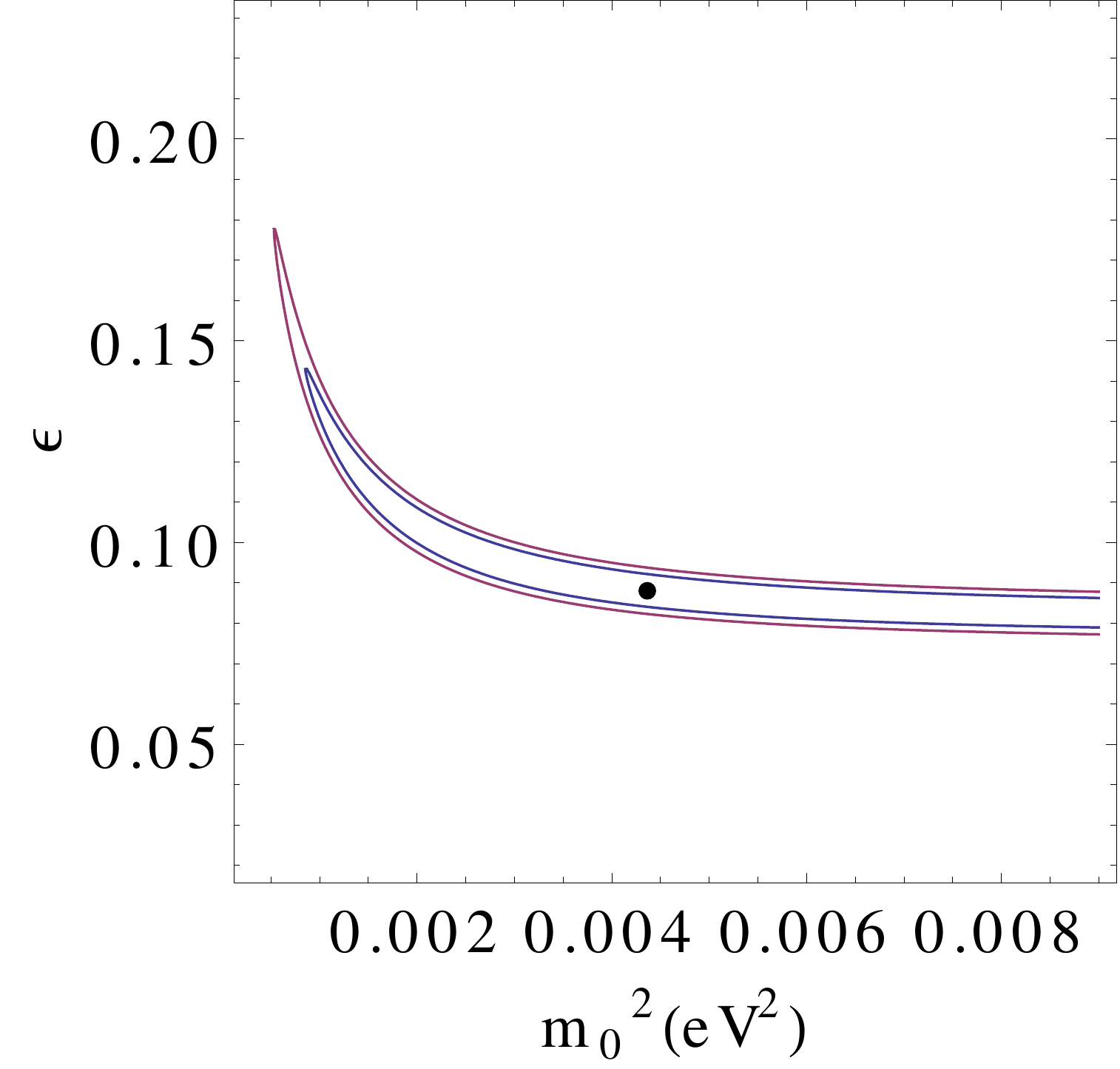}
             \includegraphics[width=0.42\linewidth]{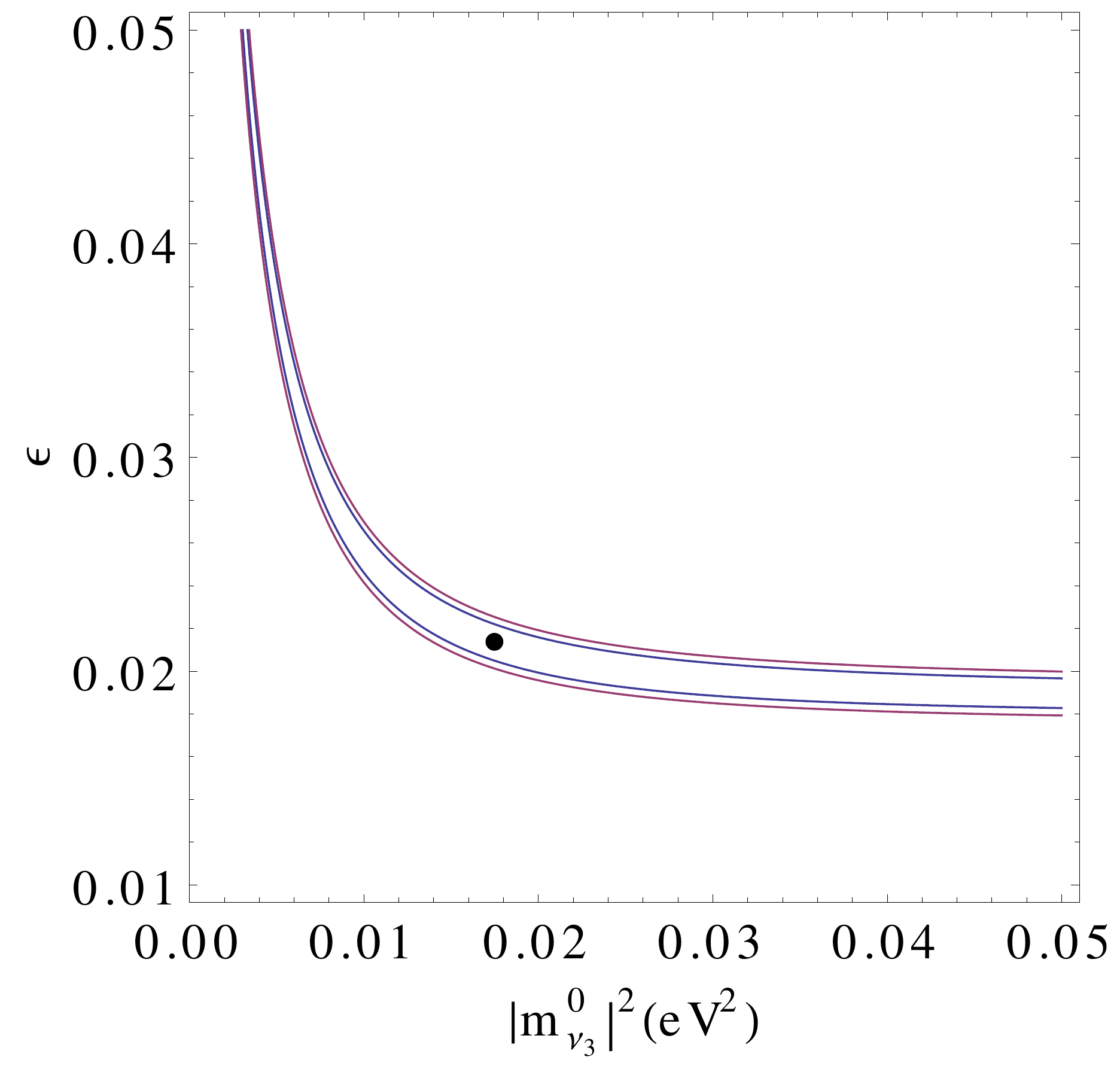}
    \caption{\label{fig:02}Allowed regions in the $m_0$-$\epsilon$ plane, at 90\%CL(blue) and 95\% CL(red) for degenerate (left) and inverted(right) hierarchy. Again, the $\theta_\nu$ parameter is fixed to the solar 
    angle and the $\alpha$ Majorana phase is marginalized.}
\end{figure}

The results of the minimization of the $\chi^2$ function are shown in Figures (\ref{fig:00}), (\ref{fig:03}) and (\ref{fig:02}),
we show the allowed regions at  90\% and 95\% CL in the plane of pairs of the three parameters marginalizing the $\chi^2$ function for the parameter not shown. In the left (right) panel is shown the case of degenerate (inverted) hierarchy for each figure. 
We can notice that 
the $\alpha$ parameter is more constrained in the case of inverted hierarchy than in the degenerate hierarchy case, and  that the fit prefers smaller values of the $\epsilon$ parameter in the case of inverted hierarchy.  For illustration purposes only we show the BFP in each case as a black dot. 

From comparison of our $\chi^2$ analysis with the qualitative analysis in~\cite{Gomez-Izquierdo:2017rxi} we find that a wide region of the parameter space is still
statistically compatible with experimental data.


\subsection*{Prediction on the Effective Majorana Mass of the Electron Neutrino}
%
From the neutrino oscillation experiments, we get information on the mass squared 
differences, but these experiments cannot say anything about the absolute neutrino mass scale. 
However, there are three processes that can address directly the determination of this 
important parameter: 
$i)$~analysis of CMB temperature fluctuations~\cite{Lesgourgues2006307},
$ii)$~the single $\beta$ decay~\cite{Otten086201} 
and 
$iii)$~neutrinoless double beta decay~($0\nu\beta\beta$)~\cite{RevModPhys.80.481}. 

\begin{figure}[h!] 
\begin{center}
\includegraphics[width=0.65\textwidth]{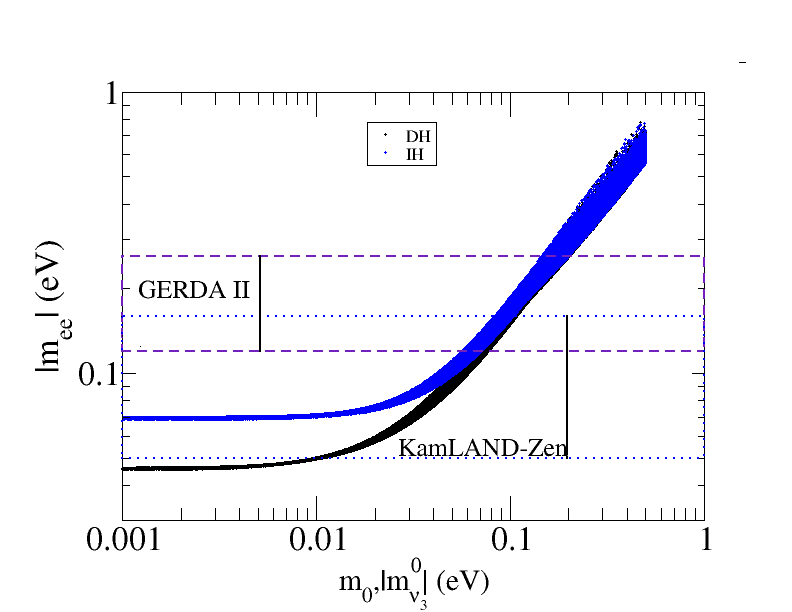}
\end{center}
\caption{\label{fig:mee} Effective mass $\vert m_{ee} \vert $  as a function of the common 
mass $ m_{0}$  in the case of Degenerate Hierarchy or of the lightest neutrino mass $\left|m^{0}_{\nu_{3}}\right|$ for Inverted Hierarchy. The horizontal regions defined 
by the blue dotted and purple dashed lines correspond to the limits by GERDA phase II~\cite{Agostini:2018tnm} and KamLAND-Zen~\cite{KamLAND-Zen:2016pfg} respectively.
}
\end{figure}

Here, we only focus on the last process which occurs 
if neutrinos are Majorana particles. With this decay process we can probe the absolute neutrino  mass scale by measuring of 
the effective Majorana mass of the electron neutrino, which is defined as:
\begin{equation}\label{mef}
 \vert m_{ee} \vert =
 \left| \sum_{i=1}^{3} m_{\nu_{i}} V_{ei}^{2}  \right|.
\end{equation}

The lowest upper bound on $\vert m_{ee}\vert < 0.22~eV$ was provided by GERDA phase-I 
data~\cite{Agostini:2013mzu}. That value has been significantly reduced by {\bf GERDA} phase-II 
data~\cite{Agostini:2017iyd}, see Fig. (\ref{fig:mee}).
According to our model, the above quantity can be performed directly using the fitted free parameters. Therefore, the plot in Fig.  (\ref{fig:mee}) shows the predicted regions for the effective Majorana mass of the electron neutrino.

\section{Conclusions}

We performed a complete study on the fermion masses and flavor mixing in the non-minimal left-right symmetric model  where the scalar sector was extended by three Higgs bidoublets, three right-handed (left-handed) triplets. The lepton sector has been previously studied in \cite{Gomez-Izquierdo:2017rxi}, where the Majorana phases were considered as CP parities ($0$ or $\pi$). In the present analysis we obtained precise formulas for the mixing angles with arbitrary Majorana phases and a chi squared statistical analysis was performed in order to fix the relevant free parameters using the updated neutrino oscillation data. Our results are in good agreement with ~\cite{Gomez-Izquierdo:2017rxi}, when 
fixed Majorana phases are considered.

On the other hand, we do this analysis for the first time in the  quark sector where the quark mass matrices come out being symmetric and hermitian in 
the {\bf PLRT} and {\bf MLRT} framework, respectively.
In the hadronic sector of {\bf PLRT} ({\bf MLRT}) framework, we write the quarks flavor mixing matrix, CKM, in terms of quark mass ratios, two shifted mass parameters 
$\widetilde{\mu}_{d}$ and $\widetilde{\mu}_{u}$, two parameters $\delta_{d}$ and 
$\delta_{u}$, two (one) phase factors. So, the difference between the CKM matrices obtained in the {\bf PLRT} and {\bf MLRT} 
framework lies in the number of phase factors, namely in {\bf PLRT} we have two phase factors, $\Gamma_{1}$ and $\Theta_{1}$, while in {\bf MLRT} only one, $\Theta_{2}$. 
		Whereby the quarks flavor mixing matrix in {\bf MRLT} is a particular case of the CKM matrix 
		obtained in {\bf PRLT}, since we only need take $\Theta_{2}=0$. We performed a likelihood 
		test $\chi^{2}$, in which the $\Theta_{2}$, $\widetilde{\mu}_{u}$, 
		and $\widetilde{\mu}_{d}$ parameters are fixed to the values given in the first row of the 
		table~\ref{tab:table1}, thus the $\chi^{2}$ function has one degree of freedom. 
		All values obtained in this $\chi^{2}$ analysis are in good agreement with experimental data.  
		Also, these values can be considered as predictions of the {\bf PLRT} and {\bf MLRT} 
		theoretical frameworks, because when $\Theta_{1} = \Theta_{2} = 0$ both schemes are 
		equivalent. The rich phenomenology of the model provides a region of the parameter space that is 
statistically compatible with experimental data.

\section*{Acknowledgements}

This work was partially supported by the Mexican grants 237004, PAPIIT IN111518 and 
Conacyt-32059. 
JCGI thanks the Department of Theoretical Physics at IFUNAM for the warm
hospitality. Also, JCGI would like to make an especial mention to Gabriela
Nabor, Marisol, Cecilia and Elizabeth G\'omez for financial and
moral support during this long time. 
FGC acknowledges the financial support from {\it CONACYT} and {\it PRODEP} under Grant 
No.~511-6/17-8017.

\bibliographystyle{bib_style_T1}
\bibliography{references.bib}

\begin{thebibliography}{100}
\providecommand{\url}[1]{\texttt{#1}}
\providecommand{\urlprefix}{URL }
\providecommand{\eprint}[2][]{\url{#2}}

\bibitem{Ishimori:2010au}
H.~Ishimori et~al., \emph{{Non-Abelian Discrete Symmetries in Particle
  Physics}}, \MYhref[journalLinks]{http://dx.doi.org/10.1143/PTPS.183.1}{Prog.
  Theor. Phys. Suppl.
  }\MYhref[journalLinks]{http://dx.doi.org/10.1143/PTPS.183.1}{\textbf{183}
  (2010) 1--163},
  \MYhref[eprintLinks]{http://arxiv.org/abs/1003.3552}{{\ttfamily
  arXiv:1003.3552 [hep-th]}}.

\bibitem{Grimus:2011fk}
W.~Grimus and P.~O. Ludl, \emph{{Finite flavour groups of fermions}},
  \MYhref[journalLinks]{http://dx.doi.org/10.1088/1751-8113/45/23/233001}{J.
  Phys.
  }\MYhref[journalLinks]{http://dx.doi.org/10.1088/1751-8113/45/23/233001}{\textbf{A45}
  (2012) 233001},
  \MYhref[eprintLinks]{http://arxiv.org/abs/1110.6376}{{\ttfamily
  arXiv:1110.6376 [hep-ph]}}.

\bibitem{Ishimori:2012zz}
H.~Ishimori et~al., \emph{{An introduction to non-Abelian discrete symmetries
  for particle physicists}},
  \MYhref[journalLinks]{http://dx.doi.org/10.1007/978-3-642-30805-5}{Lect.
  Notes Phys.
  }\MYhref[journalLinks]{http://dx.doi.org/10.1007/978-3-642-30805-5}{\textbf{858}
  (2012) 1--227}.

\bibitem{King:2013eh}
S.~F. King and C.~Luhn, \emph{{Neutrino Mass and Mixing with Discrete
  Symmetry}},
  \MYhref[journalLinks]{http://dx.doi.org/10.1088/0034-4885/76/5/056201}{Rept.
  Prog. Phys.
  }\MYhref[journalLinks]{http://dx.doi.org/10.1088/0034-4885/76/5/056201}{\textbf{76}
  (2013) 056201},
  \MYhref[eprintLinks]{http://arxiv.org/abs/1301.1340}{{\ttfamily
  arXiv:1301.1340 [hep-ph]}}.

\bibitem{Cabibbo:1963yz}
N.~Cabibbo, \emph{{Unitary Symmetry and Leptonic Decays}},
  \MYhref[journalLinks]{http://dx.doi.org/10.1103/PhysRevLett.10.531}{Phys.Rev.Lett.
  }\MYhref[journalLinks]{http://dx.doi.org/10.1103/PhysRevLett.10.531}{\textbf{10}
  (1963) 531--533}.

\bibitem{Kobayashi:1973fv}
M.~Kobayashi and T.~Maskawa, \emph{{CP Violation in the Renormalizable Theory
  of Weak Interaction}},
  \MYhref[journalLinks]{http://dx.doi.org/10.1143/PTP.49.652}{Prog. Theor.
  Phys.
  }\MYhref[journalLinks]{http://dx.doi.org/10.1143/PTP.49.652}{\textbf{49}
  (1973) 652--657}.

\bibitem{Maki:1962mu}
Z.~Maki, M.~Nakagawa and S.~Sakata, \emph{{Remarks on the unified model of
  elementary particles}},
  \MYhref[journalLinks]{http://dx.doi.org/10.1143/PTP.28.870}{Prog.Theor.Phys.
  }\MYhref[journalLinks]{http://dx.doi.org/10.1143/PTP.28.870}{\textbf{28}
  (1962) 870--880}.

\bibitem{Pontecorvo:1967fh}
B.~Pontecorvo, \emph{{Neutrino Experiments and the Problem of Conservation of
  Leptonic Charge}}, Sov. Phys. JETP \textbf{26} (1968) 984--988, [Zh. Eksp.
  Teor. Fiz.53,1717(1967)].

\bibitem{Patrignani:2016xqp}
C.~Patrignani et~al. (Particle Data Group), \emph{{Review of Particle
  Physics}},
  \MYhref[journalLinks]{http://dx.doi.org/10.1088/1674-1137/40/10/100001}{Chin.
  Phys.
  }\MYhref[journalLinks]{http://dx.doi.org/10.1088/1674-1137/40/10/100001}{\textbf{C40}
  (2016) 10 100001}.

\bibitem{Minkowski:1977sc}
P.~Minkowski, \emph{{$\mu \to e \gamma$ at a Rate of One Out of 1-Billion Muon
  Decays?}},
  \MYhref[journalLinks]{http://dx.doi.org/10.1016/0370-2693(77)90435-X}{Phys.
  Lett.
  }\MYhref[journalLinks]{http://dx.doi.org/10.1016/0370-2693(77)90435-X}{\textbf{B67}
  (1977) 421}.

\bibitem{GellMann:1980vs}
M.~Gell-Mann, P.~Ramond and R.~Slansky, \emph{{Complex Spinors and Unified
  Theories}}, Conf.Proc. \textbf{C790927} (1979) 315--321,
  \MYhref[eprintLinks]{http://arxiv.org/abs/1306.4669}{{\ttfamily
  arXiv:1306.4669 [hep-th]}}.

\bibitem{Mohapatra:1979ia}
R.~N. Mohapatra and G.~Senjanovic, \emph{{Neutrino Mass and Spontaneous Parity
  Violation}},
  \MYhref[journalLinks]{http://dx.doi.org/10.1103/PhysRevLett.44.912}{Phys.Rev.Lett.
  }\MYhref[journalLinks]{http://dx.doi.org/10.1103/PhysRevLett.44.912}{\textbf{44}
  (1980) 912}.

\bibitem{Schechter:1980gr}
J.~Schechter and J.~W.~F. Valle, \emph{{Neutrino Masses in SU(2) x U(1)
  Theories}},
  \MYhref[journalLinks]{http://dx.doi.org/10.1103/PhysRevD.22.2227}{Phys. Rev.
  }\MYhref[journalLinks]{http://dx.doi.org/10.1103/PhysRevD.22.2227}{\textbf{D22}
  (1980) 2227}.

\bibitem{Mohapatra:1980yp}
R.~N. Mohapatra and G.~Senjanovic, \emph{{Neutrino Masses and Mixings in Gauge
  Models with Spontaneous Parity Violation}},
  \MYhref[journalLinks]{http://dx.doi.org/10.1103/PhysRevD.23.165}{Phys. Rev.
  }\MYhref[journalLinks]{http://dx.doi.org/10.1103/PhysRevD.23.165}{\textbf{D23}
  (1981) 165}.

\bibitem{Schechter:1981cv}
J.~Schechter and J.~W.~F. Valle, \emph{{Neutrino Decay and Spontaneous
  Violation of Lepton Number}},
  \MYhref[journalLinks]{http://dx.doi.org/10.1103/PhysRevD.25.774}{Phys. Rev.
  }\MYhref[journalLinks]{http://dx.doi.org/10.1103/PhysRevD.25.774}{\textbf{D25}
  (1982) 774}.

\bibitem{deSalas:2017kay}
P.~F. de~Salas et~al., \emph{{Status of neutrino oscillations 2018: 3$\sigma$
  hint for normal mass ordering and improved CP sensitivity}},
  \MYhref[journalLinks]{http://dx.doi.org/10.1016/j.physletb.2018.06.019}{Phys.
  Lett.
  }\MYhref[journalLinks]{http://dx.doi.org/10.1016/j.physletb.2018.06.019}{\textbf{B782}
  (2018) 633--640},
  \MYhref[eprintLinks]{http://arxiv.org/abs/1708.01186}{{\ttfamily
  arXiv:1708.01186 [hep-ph]}}.

\bibitem{Esteban:2016qun}
I.~Esteban et~al., \emph{{Updated fit to three neutrino mixing: exploring the
  accelerator-reactor complementarity}},
  \MYhref[journalLinks]{http://dx.doi.org/10.1007/JHEP01(2017)087}{JHEP
  }\MYhref[journalLinks]{http://dx.doi.org/10.1007/JHEP01(2017)087}{\textbf{01}
  (2017) 087}, \MYhref[eprintLinks]{http://arxiv.org/abs/1611.01514}{{\ttfamily
  arXiv:1611.01514 [hep-ph]}}.

\bibitem{Capozzi:2018ubv}
F.~Capozzi, E.~Lisi, A.~Marrone and A.~Palazzo, \emph{{Current unknowns in the
  three neutrino framework}}  (2018),
  \MYhref[eprintLinks]{http://arxiv.org/abs/1804.09678}{{\ttfamily
  arXiv:1804.09678 [hep-ph]}}.

\bibitem{globalfit}
Valencia-Globalfit, \url{http://globalfit.astroparticles.es/} (2018).

\bibitem{Pakvasa:1977in}
S.~Pakvasa and H.~Sugawara, \emph{{Discrete Symmetry and Cabibbo Angle}},
  \MYhref[journalLinks]{http://dx.doi.org/10.1016/0370-2693(78)90172-7}{Phys.
  Lett.
  }\MYhref[journalLinks]{http://dx.doi.org/10.1016/0370-2693(78)90172-7}{\textbf{73B}
  (1978) 61--64}.

\bibitem{Kubo:2003iw}
J.~Kubo, A.~Mondragon, M.~Mondragon and E.~Rodriguez-Jauregui, \emph{{The
  Flavor symmetry}},
  \MYhref[journalLinks]{http://dx.doi.org/10.1143/PTP.109.795}{Prog. Theor.
  Phys.
  }\MYhref[journalLinks]{http://dx.doi.org/10.1143/PTP.109.795}{\textbf{109}
  (2003) 795--807}, [Erratum: Prog. Theor. Phys.114,287(2005)],
  \MYhref[eprintLinks]{http://arxiv.org/abs/hep-ph/0302196}{{\ttfamily
  arXiv:hep-ph/0302196 [hep-ph]}}.

\bibitem{Kubo:2003pd}
J.~Kubo, \emph{{Majorana phase in minimal S(3) invariant extension of the
  standard model}},
  \MYhref[journalLinks]{http://dx.doi.org/10.1016/j.physletb.2005.06.013,
  10.1016/j.physletb.2003.10.048}{Phys. Lett.
  }\MYhref[journalLinks]{http://dx.doi.org/10.1016/j.physletb.2005.06.013,
  10.1016/j.physletb.2003.10.048}{\textbf{B578} (2004) 156--164}, [Erratum:
  Phys. Lett.B619,387(2005)],
  \MYhref[eprintLinks]{http://arxiv.org/abs/hep-ph/0309167}{{\ttfamily
  arXiv:hep-ph/0309167 [hep-ph]}}.

\bibitem{Kobayashi:2003fh}
T.~Kobayashi, J.~Kubo and H.~Terao, \emph{{Exact S(3) symmetry solving the
  supersymmetric flavor problem}},
  \MYhref[journalLinks]{http://dx.doi.org/10.1016/j.physletb.2003.03.002}{Phys.
  Lett.
  }\MYhref[journalLinks]{http://dx.doi.org/10.1016/j.physletb.2003.03.002}{\textbf{B568}
  (2003) 83--91},
  \MYhref[eprintLinks]{http://arxiv.org/abs/hep-ph/0303084}{{\ttfamily
  arXiv:hep-ph/0303084 [hep-ph]}}.

\bibitem{Chen:2004rr}
S.-L. Chen, M.~Frigerio and E.~Ma, \emph{Large neutrino mixing and normal mass
  hierarchy: A discrete understanding},
  \MYhref[journalLinks]{http://dx.doi.org/10.1103/PhysRevD.70.079905,
  10.1103/PhysRevD.70.073008}{Phys. Rev.
  }\MYhref[journalLinks]{http://dx.doi.org/10.1103/PhysRevD.70.079905,
  10.1103/PhysRevD.70.073008}{\textbf{D70} (2004) 073008}, [Erratum: Phys.
  Rev.D70,079905(2004)],
  \MYhref[eprintLinks]{http://arxiv.org/abs/hep-ph/0404084}{{\ttfamily
  arXiv:hep-ph/0404084 [hep-ph]}}.

\bibitem{Kubo:2005sr}
J.~Kubo et~al., \emph{{A minimal S(3)-invariant extension of the standard
  model}},
  \MYhref[journalLinks]{http://dx.doi.org/10.1088/1742-6596/18/1/013}{J. Phys.
  Conf. Ser.
  }\MYhref[journalLinks]{http://dx.doi.org/10.1088/1742-6596/18/1/013}{\textbf{18}
  (2005) 380--384}.

\bibitem{Mondragon:2006hi}
A.~Mondragon, \emph{{Models of flavour with discrete symmetries}},
  \MYhref[journalLinks]{http://dx.doi.org/10.1063/1.2359411}{AIP Conf. Proc.
  }\MYhref[journalLinks]{http://dx.doi.org/10.1063/1.2359411}{\textbf{857}
  (2006) 2 266},
  \MYhref[eprintLinks]{http://arxiv.org/abs/hep-ph/0609243}{{\ttfamily
  arXiv:hep-ph/0609243 [hep-ph]}}.

\bibitem{Felix:2006pn}
O.~Felix, A.~Mondragon, M.~Mondragon and E.~Peinado, \emph{{Neutrino masses and
  mixings in a minimal S(3)-invariant extension of the standard model}},
  \MYhref[journalLinks]{http://dx.doi.org/10.1063/1.2751980}{AIP Conf. Proc.
  }\MYhref[journalLinks]{http://dx.doi.org/10.1063/1.2751980}{\textbf{917}
  (2007) 383--389},
  \MYhref[eprintLinks]{http://arxiv.org/abs/hep-ph/0610061}{{\ttfamily
  arXiv:hep-ph/0610061}}.

\bibitem{Mondragon:2007af}
A.~Mondragon, M.~Mondragon and E.~Peinado, \emph{{Lepton masses, mixings and
  FCNC in a minimal $S_3$-invariant extension of the Standard Model}},
  \MYhref[journalLinks]{http://dx.doi.org/10.1103/PhysRevD.76.076003}{Phys.
  Rev.
  }\MYhref[journalLinks]{http://dx.doi.org/10.1103/PhysRevD.76.076003}{\textbf{D76}
  (2007) 076003},
  \MYhref[eprintLinks]{http://arxiv.org/abs/0706.0354}{{\ttfamily
  arXiv:0706.0354 [hep-ph]}}.

\bibitem{Mondragon:2007nk}
A.~Mondragon, M.~Mondragon and E.~Peinado, \emph{{S(3)-flavour symmetry as
  realized in lepton flavour violating processes}},
  \MYhref[journalLinks]{http://dx.doi.org/10.1088/1751-8113/41/30/304035}{J.
  Phys.
  }\MYhref[journalLinks]{http://dx.doi.org/10.1088/1751-8113/41/30/304035}{\textbf{A41}
  (2008) 304035},
  \MYhref[eprintLinks]{http://arxiv.org/abs/0712.1799}{{\ttfamily
  arXiv:0712.1799 [hep-ph]}}.

\bibitem{Mondragon:2007jx}
A.~Mondragon, M.~Mondragon and E.~Peinado, \emph{{Nearly tri-bimaximal mixing
  in the S(3) flavour symmetry}},
  \MYhref[journalLinks]{http://dx.doi.org/10.1063/1.2965040}{AIP Conf. Proc.
  }\MYhref[journalLinks]{http://dx.doi.org/10.1063/1.2965040}{\textbf{1026}
  (2008) 164--169},
  \MYhref[eprintLinks]{http://arxiv.org/abs/0712.2488}{{\ttfamily
  arXiv:0712.2488 [hep-ph]}}.

\bibitem{Meloni:2010aw}
D.~Meloni, S.~Morisi and E.~Peinado, \emph{{Fritzsch neutrino mass matrix from
  $S_3$ symmetry}},
  \MYhref[journalLinks]{http://dx.doi.org/10.1088/0954-3899/38/1/015003}{J.
  Phys.
  }\MYhref[journalLinks]{http://dx.doi.org/10.1088/0954-3899/38/1/015003}{\textbf{G38}
  (2011) 015003},
  \MYhref[eprintLinks]{http://arxiv.org/abs/1005.3482}{{\ttfamily
  arXiv:1005.3482 [hep-ph]}}.

\bibitem{Dicus:2010iq}
D.~A. Dicus, S.-F. Ge and W.~W. Repko, \emph{{Neutrino mixing with broken $S_3$
  symmetry}},
  \MYhref[journalLinks]{http://dx.doi.org/10.1103/PhysRevD.82.033005}{Phys.
  Rev.
  }\MYhref[journalLinks]{http://dx.doi.org/10.1103/PhysRevD.82.033005}{\textbf{D82}
  (2010) 033005},
  \MYhref[eprintLinks]{http://arxiv.org/abs/1004.3266}{{\ttfamily
  arXiv:1004.3266 [hep-ph]}}.

\bibitem{Bhattacharyya:2010hp}
G.~Bhattacharyya, P.~Leser and H.~Pas, \emph{{Exotic Higgs boson decay modes as
  a harbinger of $S_3$ flavor symmetry}},
  \MYhref[journalLinks]{http://dx.doi.org/10.1103/PhysRevD.83.011701}{Phys.
  Rev.
  }\MYhref[journalLinks]{http://dx.doi.org/10.1103/PhysRevD.83.011701}{\textbf{D83}
  (2011) 011701},
  \MYhref[eprintLinks]{http://arxiv.org/abs/1006.5597}{{\ttfamily
  arXiv:1006.5597 [hep-ph]}}.

\bibitem{Canales:2011ug}
F.~Gonzalez~Canales and A.~Mondragon, \emph{{The $S_{3}$ symmetry: Flavour and
  texture zeroes}},
  \MYhref[journalLinks]{http://dx.doi.org/10.1088/1742-6596/287/1/012015}{J.
  Phys. Conf. Ser.
  }\MYhref[journalLinks]{http://dx.doi.org/10.1088/1742-6596/287/1/012015}{\textbf{287}
  (2011) 012015},
  \MYhref[eprintLinks]{http://arxiv.org/abs/1101.3807}{{\ttfamily
  arXiv:1101.3807 [hep-ph]}}.

\bibitem{Dong:2011vb}
P.~V. Dong, H.~N. Long, C.~H. Nam and V.~V. Vien, \emph{{The $S_3$ flavor
  symmetry in 3-3-1 models}},
  \MYhref[journalLinks]{http://dx.doi.org/10.1103/PhysRevD.85.053001}{Phys.
  Rev.
  }\MYhref[journalLinks]{http://dx.doi.org/10.1103/PhysRevD.85.053001}{\textbf{D85}
  (2012) 053001},
  \MYhref[eprintLinks]{http://arxiv.org/abs/1111.6360}{{\ttfamily
  arXiv:1111.6360 [hep-ph]}}.

\bibitem{Dias:2012bh}
A.~G. Dias, A.~C.~B. Machado and C.~C. Nishi, \emph{{An $S_3$ Model for Lepton
  Mass Matrices with Nearly Minimal Texture}},
  \MYhref[journalLinks]{http://dx.doi.org/10.1103/PhysRevD.86.093005}{Phys.
  Rev.
  }\MYhref[journalLinks]{http://dx.doi.org/10.1103/PhysRevD.86.093005}{\textbf{D86}
  (2012) 093005},
  \MYhref[eprintLinks]{http://arxiv.org/abs/1206.6362}{{\ttfamily
  arXiv:1206.6362 [hep-ph]}}.

\bibitem{Canales:2012ix}
F.~G. Canales, A.~Mondragon, U.~S. Salazar and L.~Velasco-Sevilla, \emph{{$S_3$
  as a unified family theory for quarks and leptons}}, arXiv:1210.0288  (2012),
  \MYhref[eprintLinks]{http://arxiv.org/abs/1210.0288}{{\ttfamily
  arXiv:1210.0288 [hep-ph]}}.

\bibitem{Canales:2012dr}
F.~Gonzalez~Canales, A.~Mondragon and M.~Mondragon, \emph{{The $S_3$ Flavour
  Symmetry: Neutrino Masses and Mixings}},
  \MYhref[journalLinks]{http://dx.doi.org/10.1002/prop.201200121}{Fortsch.Phys.
  }\MYhref[journalLinks]{http://dx.doi.org/10.1002/prop.201200121}{\textbf{61}
  (2013) 546--570},
  \MYhref[eprintLinks]{http://arxiv.org/abs/1205.4755}{{\ttfamily
  arXiv:1205.4755 [hep-ph]}}.

\bibitem{GonzalezCanales:2012za}
F.~Gonzalez~Canales and A.~Mondragon, \emph{{The flavour symmetry S(3) and the
  neutrino mass matrix with two texture zeroes}},
  \MYhref[journalLinks]{http://dx.doi.org/10.1088/1742-6596/378/1/012014}{J.Phys.Conf.Ser.
  }\MYhref[journalLinks]{http://dx.doi.org/10.1088/1742-6596/378/1/012014}{\textbf{378}
  (2012) 012014}.

\bibitem{Canales:2013ura}
F.~G. Canales et~al., \emph{{Fermion mixing in an $S_{3}$ model with three
  Higgs doublets}},
  \MYhref[journalLinks]{http://dx.doi.org/10.1088/1742-6596/447/1/012053}{J.Phys.Conf.Ser.
  }\MYhref[journalLinks]{http://dx.doi.org/10.1088/1742-6596/447/1/012053}{\textbf{447}
  (2013) 012053}.

\bibitem{Canales:2013cga}
F.~González~Canales et~al., \emph{{Quark sector of S3 models: classification
  and comparison with experimental data}},
  \MYhref[journalLinks]{http://dx.doi.org/10.1103/PhysRevD.88.096004}{Phys.Rev.
  }\MYhref[journalLinks]{http://dx.doi.org/10.1103/PhysRevD.88.096004}{\textbf{D88}
  (2013) 096004},
  \MYhref[eprintLinks]{http://arxiv.org/abs/1304.6644}{{\ttfamily
  arXiv:1304.6644 [hep-ph]}}.

\bibitem{Ma:2013zca}
E.~Ma and B.~Melic, \emph{{Updated $S_{3}$ model of quarks}},
  \MYhref[journalLinks]{http://dx.doi.org/10.1016/j.physletb.2013.07.015}{Phys.
  Lett.
  }\MYhref[journalLinks]{http://dx.doi.org/10.1016/j.physletb.2013.07.015}{\textbf{B725}
  (2013) 402--406},
  \MYhref[eprintLinks]{http://arxiv.org/abs/1303.6928}{{\ttfamily
  arXiv:1303.6928 [hep-ph]}}.

\bibitem{Kajiyama:2013sza}
Y.~Kajiyama, H.~Okada and K.~Yagyu, \emph{{Electron/Muon Specific Two Higgs
  Doublet Model}},
  \MYhref[journalLinks]{http://dx.doi.org/10.1016/j.nuclphysb.2014.08.009}{Nucl.
  Phys.
  }\MYhref[journalLinks]{http://dx.doi.org/10.1016/j.nuclphysb.2014.08.009}{\textbf{B887}
  (2014) 358--370},
  \MYhref[eprintLinks]{http://arxiv.org/abs/1309.6234}{{\ttfamily
  arXiv:1309.6234 [hep-ph]}}.

\bibitem{Hernandez:2013hea}
A.~E. C\'arcamo~Hern\'andez, R.~Martinez and F.~Ochoa, \emph{{Fermion masses
  and mixings in the 3-3-1 model with right-handed neutrinos based on the $S_3$
  flavor symmetry}},
  \MYhref[journalLinks]{http://dx.doi.org/10.1140/epjc/s10052-016-4480-3}{Eur.
  Phys. J.
  }\MYhref[journalLinks]{http://dx.doi.org/10.1140/epjc/s10052-016-4480-3}{\textbf{C76}
  (2016) 11 634},
  \MYhref[eprintLinks]{http://arxiv.org/abs/1309.6567}{{\ttfamily
  arXiv:1309.6567 [hep-ph]}}.

\bibitem{Das:2014fea}
D.~Das and U.~K. Dey, \emph{{Analysis of an extended scalar sector with $S_3$
  symmetry}},
  \MYhref[journalLinks]{http://dx.doi.org/10.1103/PhysRevD.91.039905,
  10.1103/PhysRevD.89.095025}{Phys. Rev.
  }\MYhref[journalLinks]{http://dx.doi.org/10.1103/PhysRevD.91.039905,
  10.1103/PhysRevD.89.095025}{\textbf{D89} (2014) 9 095025}, [Erratum: Phys.
  Rev.D91,no.3,039905(2015)],
  \MYhref[eprintLinks]{http://arxiv.org/abs/1404.2491}{{\ttfamily
  arXiv:1404.2491 [hep-ph]}}.

\bibitem{Ma:2014qra}
E.~Ma and R.~Srivastava, \emph{{Dirac or inverse seesaw neutrino masses with
  $B-L$ gauge symmetry and $S_3$ flavor symmetry}},
  \MYhref[journalLinks]{http://dx.doi.org/10.1016/j.physletb.2014.12.049}{Phys.
  Lett.
  }\MYhref[journalLinks]{http://dx.doi.org/10.1016/j.physletb.2014.12.049}{\textbf{B741}
  (2015) 217--222},
  \MYhref[eprintLinks]{http://arxiv.org/abs/1411.5042}{{\ttfamily
  arXiv:1411.5042 [hep-ph]}}.

\bibitem{Hernandez:2014vta}
A.~E. C\'arcamo~Hern\'andez, R.~Martinez and J.~Nisperuza, \emph{{$S_3$
  discrete group as a source of the quark mass and mixing pattern in $331$
  models}},
  \MYhref[journalLinks]{http://dx.doi.org/10.1140/epjc/s10052-015-3278-z}{Eur.
  Phys. J.
  }\MYhref[journalLinks]{http://dx.doi.org/10.1140/epjc/s10052-015-3278-z}{\textbf{C75}
  (2015) 2 72}, \MYhref[eprintLinks]{http://arxiv.org/abs/1401.0937}{{\ttfamily
  arXiv:1401.0937 [hep-ph]}}.

\bibitem{Hernandez:2014lpa}
A.~E. C\'arcamo~Hern\'andez, E.~Cata\~no Mur and R.~Martinez, \emph{{Lepton
  masses and mixing in $SU(3)_{C}\otimes SU(3)_{L}\otimes U(1)_{X}$ models with
  a $S_3$ flavor symmetry}},
  \MYhref[journalLinks]{http://dx.doi.org/10.1103/PhysRevD.90.073001}{Phys.
  Rev.
  }\MYhref[journalLinks]{http://dx.doi.org/10.1103/PhysRevD.90.073001}{\textbf{D90}
  (2014) 7 073001},
  \MYhref[eprintLinks]{http://arxiv.org/abs/1407.5217}{{\ttfamily
  arXiv:1407.5217 [hep-ph]}}.

\bibitem{Gupta:2014nba}
S.~Gupta, C.~S. Kim and P.~Sharma, \emph{{Radiative and seesaw threshold
  corrections to the $S_3$ symmetric neutrino mass matrix}},
  \MYhref[journalLinks]{http://dx.doi.org/10.1016/j.physletb.2014.12.005}{Phys.
  Lett.
  }\MYhref[journalLinks]{http://dx.doi.org/10.1016/j.physletb.2014.12.005}{\textbf{B740}
  (2015) 353--358},
  \MYhref[eprintLinks]{http://arxiv.org/abs/1408.0172}{{\ttfamily
  arXiv:1408.0172 [hep-ph]}}.

\bibitem{Das:2015sca}
D.~Das, U.~K. Dey and P.~B. Pal, \emph{{$S_3$ symmetry and the quark mixing
  matrix}},
  \MYhref[journalLinks]{http://dx.doi.org/10.1016/j.physletb.2015.12.038}{Phys.
  Lett.
  }\MYhref[journalLinks]{http://dx.doi.org/10.1016/j.physletb.2015.12.038}{\textbf{B753}
  (2016) 315--318},
  \MYhref[eprintLinks]{http://arxiv.org/abs/1507.06509}{{\ttfamily
  arXiv:1507.06509 [hep-ph]}}.

\bibitem{Hernandez:2015dga}
A.~E. C\'arcamo~Hern\'andez, I.~de~Medeiros~Varzielas and E.~Schumacher,
  \emph{{Fermion and scalar phenomenology of a two-Higgs-doublet model with
  $S_3$}},
  \MYhref[journalLinks]{http://dx.doi.org/10.1103/PhysRevD.93.016003}{Phys.
  Rev.
  }\MYhref[journalLinks]{http://dx.doi.org/10.1103/PhysRevD.93.016003}{\textbf{D93}
  (2016) 1 016003},
  \MYhref[eprintLinks]{http://arxiv.org/abs/1509.02083}{{\ttfamily
  arXiv:1509.02083 [hep-ph]}}.

\bibitem{Hernandez:2015zeh}
A.~E. C\'arcamo~Hern\'andez, I.~de~Medeiros~Varzielas and N.~A. Neill,
  \emph{{Novel Randall-Sundrum model with $S_{3}$ flavor symmetry}},
  \MYhref[journalLinks]{http://dx.doi.org/10.1103/PhysRevD.94.033011}{Phys.
  Rev.
  }\MYhref[journalLinks]{http://dx.doi.org/10.1103/PhysRevD.94.033011}{\textbf{D94}
  (2016) 3 033011},
  \MYhref[eprintLinks]{http://arxiv.org/abs/1511.07420}{{\ttfamily
  arXiv:1511.07420 [hep-ph]}}.

\bibitem{Arbelaez:2016mhg}
C.~Arbel\'aez, A.~E. C\'arcamo~Hern\'andez, S.~Kovalenko and I.~Schmidt,
  \emph{{Radiative Seesaw-type Mechanism of Fermion Masses and Non-trivial
  Quark Mixing}},
  \MYhref[journalLinks]{http://dx.doi.org/10.1140/epjc/s10052-017-4948-9}{Eur.
  Phys. J.
  }\MYhref[journalLinks]{http://dx.doi.org/10.1140/epjc/s10052-017-4948-9}{\textbf{C77}
  (2017) 6 422},
  \MYhref[eprintLinks]{http://arxiv.org/abs/1602.03607}{{\ttfamily
  arXiv:1602.03607 [hep-ph]}}.

\bibitem{Hernandez:2015hrt}
A.~E. Cárcamo~Hernández, \emph{{A novel and economical explanation for SM
  fermion masses and mixings}},
  \MYhref[journalLinks]{http://dx.doi.org/10.1140/epjc/s10052-016-4351-y}{Eur.
  Phys. J.
  }\MYhref[journalLinks]{http://dx.doi.org/10.1140/epjc/s10052-016-4351-y}{\textbf{C76}
  (2016) 9 503},
  \MYhref[eprintLinks]{http://arxiv.org/abs/1512.09092}{{\ttfamily
  arXiv:1512.09092 [hep-ph]}}.

\bibitem{CarcamoHernandez:2016pdu}
A.~E. C\'arcamo~Hern\'andez, S.~Kovalenko and I.~Schmidt, \emph{{Radiatively
  generated hierarchy of lepton and quark masses}},
  \MYhref[journalLinks]{http://dx.doi.org/10.1007/JHEP02(2017)125}{JHEP
  }\MYhref[journalLinks]{http://dx.doi.org/10.1007/JHEP02(2017)125}{\textbf{02}
  (2017) 125}, \MYhref[eprintLinks]{http://arxiv.org/abs/1611.09797}{{\ttfamily
  arXiv:1611.09797 [hep-ph]}}.

\bibitem{Pramanick:2016mdp}
S.~Pramanick and A.~Raychaudhuri, \emph{{Neutrino mass model with $S_3$
  symmetry and seesaw interplay}},
  \MYhref[journalLinks]{http://dx.doi.org/10.1103/PhysRevD.94.115028}{Phys.
  Rev.
  }\MYhref[journalLinks]{http://dx.doi.org/10.1103/PhysRevD.94.115028}{\textbf{D94}
  (2016) 11 115028},
  \MYhref[eprintLinks]{http://arxiv.org/abs/1609.06103}{{\ttfamily
  arXiv:1609.06103 [hep-ph]}}.

\bibitem{Gomez-Izquierdo:2017rxi}
J.~C. G\'omez-Izquierdo, \emph{{Non-minimal flavored ${S}_{3}\otimes {Z}_{2}$
  left-right symmetric model}},
  \MYhref[journalLinks]{http://dx.doi.org/10.1140/epjc/s10052-017-5094-0}{Eur.
  Phys. J.
  }\MYhref[journalLinks]{http://dx.doi.org/10.1140/epjc/s10052-017-5094-0}{\textbf{C77}
  (2017) 8 551},
  \MYhref[eprintLinks]{http://arxiv.org/abs/1701.01747}{{\ttfamily
  arXiv:1701.01747 [hep-ph]}}.

\bibitem{Barradas-Guevara:2017iyt}
E.~Barradas-Guevara, O.~Felix-Beltran, F.~Gonzalez-Canales and M.~Zeleny-Mora,
  \emph{{Lepton CP violation in a $\nu$2HDM with flavor}},
  \MYhref[journalLinks]{http://dx.doi.org/10.1103/PhysRevD.97.035003}{Phys.
  Rev.
  }\MYhref[journalLinks]{http://dx.doi.org/10.1103/PhysRevD.97.035003}{\textbf{D97}
  (2018) 3 035003},
  \MYhref[eprintLinks]{http://arxiv.org/abs/1704.03474}{{\ttfamily
  arXiv:1704.03474 [hep-ph]}}.

\bibitem{Cruz:2017add}
A.~A. Cruz and M.~Mondragón, \emph{{Neutrino masses, mixing, and leptogenesis
  in an S3 model}}  (2017),
  \MYhref[eprintLinks]{http://arxiv.org/abs/1701.07929}{{\ttfamily
  arXiv:1701.07929 [hep-ph]}}.

\bibitem{Das:2017zrm}
D.~Das, U.~K. Dey and P.~B. Pal, \emph{{Quark mixing in an $S_3$ symmetric
  model with two Higgs doublets}},
  \MYhref[journalLinks]{http://dx.doi.org/10.1103/PhysRevD.96.031701}{Phys.
  Rev.
  }\MYhref[journalLinks]{http://dx.doi.org/10.1103/PhysRevD.96.031701}{\textbf{D96}
  (2017) 3 031701},
  \MYhref[eprintLinks]{http://arxiv.org/abs/1705.07784}{{\ttfamily
  arXiv:1705.07784 [hep-ph]}}.

\bibitem{Espinoza:2018itz}
C.~Espinoza, E.~A. Garc\'es, M.~Mondragon and H.~Reyes-González, \emph{{The
  $S3$ Symmetric Model with a Dark Scalar}}  (2018),
  \MYhref[eprintLinks]{http://arxiv.org/abs/1804.01879}{{\ttfamily
  arXiv:1804.01879 [hep-ph]}}.

\bibitem{Ge:2018ofp}
S.-F. Ge, A.~Kusenko and T.~T. Yanagida,
  \MYhref[journalLinks]{http://dx.doi.org/10.1016/j.physletb.2018.04.040}{\emph{{Large
  Leptonic Dirac CP Phase from Broken Democracy with Random Perturbations}}
  }\MYhref[journalLinks]{http://dx.doi.org/10.1016/j.physletb.2018.04.040}{
  (2018)}, \MYhref[eprintLinks]{http://arxiv.org/abs/1803.03888}{{\ttfamily
  arXiv:1803.03888 [hep-ph]}}.

\bibitem{Gomez-Izquierdo:2018jrx}
J.~C. Gómez-Izquierdo and M.~Mondragón, \emph{{B-L Model with ${\bf S}_{3}$
  Symmetry: Nearest Neighbor Interaction Textures and Broken
  $\mu\leftrightarrow\tau$ Symmetry}}  (2018),
  \MYhref[eprintLinks]{http://arxiv.org/abs/1804.08746}{{\ttfamily
  arXiv:1804.08746 [hep-ph]}}.

\bibitem{Ma:2001dn}
E.~Ma and G.~Rajasekaran, \emph{{Softly broken A(4) symmetry for nearly
  degenerate neutrino masses}},
  \MYhref[journalLinks]{http://dx.doi.org/10.1103/PhysRevD.64.113012}{Phys.
  Rev.
  }\MYhref[journalLinks]{http://dx.doi.org/10.1103/PhysRevD.64.113012}{\textbf{D64}
  (2001) 113012},
  \MYhref[eprintLinks]{http://arxiv.org/abs/hep-ph/0106291}{{\ttfamily
  arXiv:hep-ph/0106291 [hep-ph]}}.

\bibitem{He:2006dk}
X.-G. He, Y.-Y. Keum and R.~R. Volkas, \emph{{A(4) flavor symmetry breaking
  scheme for understanding quark and neutrino mixing angles}},
  \MYhref[journalLinks]{http://dx.doi.org/10.1088/1126-6708/2006/04/039}{JHEP
  }\MYhref[journalLinks]{http://dx.doi.org/10.1088/1126-6708/2006/04/039}{\textbf{04}
  (2006) 039},
  \MYhref[eprintLinks]{http://arxiv.org/abs/hep-ph/0601001}{{\ttfamily
  arXiv:hep-ph/0601001 [hep-ph]}}.

\bibitem{Chen:2009um}
M.-C. Chen and S.~F. King, \emph{{A4 See-Saw Models and Form Dominance}},
  \MYhref[journalLinks]{http://dx.doi.org/10.1088/1126-6708/2009/06/072}{JHEP
  }\MYhref[journalLinks]{http://dx.doi.org/10.1088/1126-6708/2009/06/072}{\textbf{06}
  (2009) 072}, \MYhref[eprintLinks]{http://arxiv.org/abs/0903.0125}{{\ttfamily
  arXiv:0903.0125 [hep-ph]}}.

\bibitem{Ahn:2012tv}
Y.~H. Ahn and S.~K. Kang, \emph{{Non-zero $\theta_{13}$ and CP violation in a
  model with $A_4$ flavor symmetry}},
  \MYhref[journalLinks]{http://dx.doi.org/10.1103/PhysRevD.86.093003}{Phys.
  Rev.
  }\MYhref[journalLinks]{http://dx.doi.org/10.1103/PhysRevD.86.093003}{\textbf{D86}
  (2012) 093003},
  \MYhref[eprintLinks]{http://arxiv.org/abs/1203.4185}{{\ttfamily
  arXiv:1203.4185 [hep-ph]}}.

\bibitem{Memenga:2013vc}
N.~Memenga, W.~Rodejohann and H.~Zhang, \emph{{$A_4$ flavor symmetry model for
  Dirac neutrinos and sizable $U_{e3}$}},
  \MYhref[journalLinks]{http://dx.doi.org/10.1103/PhysRevD.87.053021}{Phys.
  Rev.
  }\MYhref[journalLinks]{http://dx.doi.org/10.1103/PhysRevD.87.053021}{\textbf{D87}
  (2013) 5 053021},
  \MYhref[eprintLinks]{http://arxiv.org/abs/1301.2963}{{\ttfamily
  arXiv:1301.2963 [hep-ph]}}.

\bibitem{Felipe:2013vwa}
R.~Gonzalez~Felipe, H.~Serodio and J.~P. Silva, \emph{{Neutrino masses and
  mixing in A4 models with three Higgs doublets}},
  \MYhref[journalLinks]{http://dx.doi.org/10.1103/PhysRevD.88.015015}{Phys.
  Rev.
  }\MYhref[journalLinks]{http://dx.doi.org/10.1103/PhysRevD.88.015015}{\textbf{D88}
  (2013) 1 015015},
  \MYhref[eprintLinks]{http://arxiv.org/abs/1304.3468}{{\ttfamily
  arXiv:1304.3468 [hep-ph]}}.

\bibitem{Varzielas:2012ai}
I.~de~Medeiros~Varzielas and D.~Pidt, \emph{{UV completions of flavour models
  and large $\theta_{13}$}},
  \MYhref[journalLinks]{http://dx.doi.org/10.1007/JHEP03(2013)065}{JHEP
  }\MYhref[journalLinks]{http://dx.doi.org/10.1007/JHEP03(2013)065}{\textbf{03}
  (2013) 065}, \MYhref[eprintLinks]{http://arxiv.org/abs/1211.5370}{{\ttfamily
  arXiv:1211.5370 [hep-ph]}}.

\bibitem{Ishimori:2012fg}
H.~Ishimori and E.~Ma, \emph{{New Simple $A_4$ Neutrino Model for Nonzero
  $\theta_{13}$ and Large $\delta_{CP}$}},
  \MYhref[journalLinks]{http://dx.doi.org/10.1103/PhysRevD.86.045030}{Phys.
  Rev.
  }\MYhref[journalLinks]{http://dx.doi.org/10.1103/PhysRevD.86.045030}{\textbf{D86}
  (2012) 045030},
  \MYhref[eprintLinks]{http://arxiv.org/abs/1205.0075}{{\ttfamily
  arXiv:1205.0075 [hep-ph]}}.

\bibitem{Hernandez:2013dta}
A.~E. Carcamo~Hernandez et~al., \emph{{Lepton masses and mixings in an $A_4$
  multi-Higgs model with a radiative seesaw mechanism}},
  \MYhref[journalLinks]{http://dx.doi.org/10.1103/PhysRevD.88.076014}{Phys.
  Rev.
  }\MYhref[journalLinks]{http://dx.doi.org/10.1103/PhysRevD.88.076014}{\textbf{D88}
  (2013) 7 076014},
  \MYhref[eprintLinks]{http://arxiv.org/abs/1307.6499}{{\ttfamily
  arXiv:1307.6499 [hep-ph]}}.

\bibitem{Babu:2002dz}
K.~S. Babu, E.~Ma and J.~W.~F. Valle, \emph{{Underlying A(4) symmetry for the
  neutrino mass matrix and the quark mixing matrix}},
  \MYhref[journalLinks]{http://dx.doi.org/10.1016/S0370-2693(02)03153-2}{Phys.
  Lett.
  }\MYhref[journalLinks]{http://dx.doi.org/10.1016/S0370-2693(02)03153-2}{\textbf{B552}
  (2003) 207--213},
  \MYhref[eprintLinks]{http://arxiv.org/abs/hep-ph/0206292}{{\ttfamily
  arXiv:hep-ph/0206292 [hep-ph]}}.

\bibitem{Altarelli:2005yx}
G.~Altarelli and F.~Feruglio, \emph{{Tri-bimaximal neutrino mixing, A(4) and
  the modular symmetry}},
  \MYhref[journalLinks]{http://dx.doi.org/10.1016/j.nuclphysb.2006.02.015}{Nucl.
  Phys.
  }\MYhref[journalLinks]{http://dx.doi.org/10.1016/j.nuclphysb.2006.02.015}{\textbf{B741}
  (2006) 215--235},
  \MYhref[eprintLinks]{http://arxiv.org/abs/hep-ph/0512103}{{\ttfamily
  arXiv:hep-ph/0512103 [hep-ph]}}.

\bibitem{Gupta:2011ct}
S.~Gupta, A.~S. Joshipura and K.~M. Patel, \emph{{Minimal extension of
  tri-bimaximal mixing and generalized $Z_2 X Z_2$ symmetries}},
  \MYhref[journalLinks]{http://dx.doi.org/10.1103/PhysRevD.85.031903}{Phys.
  Rev.
  }\MYhref[journalLinks]{http://dx.doi.org/10.1103/PhysRevD.85.031903}{\textbf{D85}
  (2012) 031903},
  \MYhref[eprintLinks]{http://arxiv.org/abs/1112.6113}{{\ttfamily
  arXiv:1112.6113 [hep-ph]}}.

\bibitem{Altarelli:2005yp}
G.~Altarelli and F.~Feruglio, \emph{{Tri-bimaximal neutrino mixing from
  discrete symmetry in extra dimensions}},
  \MYhref[journalLinks]{http://dx.doi.org/10.1016/j.nuclphysb.2005.05.005}{Nucl.
  Phys.
  }\MYhref[journalLinks]{http://dx.doi.org/10.1016/j.nuclphysb.2005.05.005}{\textbf{B720}
  (2005) 64--88},
  \MYhref[eprintLinks]{http://arxiv.org/abs/hep-ph/0504165}{{\ttfamily
  arXiv:hep-ph/0504165 [hep-ph]}}.

\bibitem{Kadosh:2010rm}
A.~Kadosh and E.~Pallante, \emph{{An A(4) flavor model for quarks and leptons
  in warped geometry}},
  \MYhref[journalLinks]{http://dx.doi.org/10.1007/JHEP08(2010)115}{JHEP
  }\MYhref[journalLinks]{http://dx.doi.org/10.1007/JHEP08(2010)115}{\textbf{08}
  (2010) 115}, \MYhref[eprintLinks]{http://arxiv.org/abs/1004.0321}{{\ttfamily
  arXiv:1004.0321 [hep-ph]}}.

\bibitem{Kadosh:2013nra}
A.~Kadosh, \emph{{$\Theta_{13}$ and charged Lepton Flavor Violation in "warped"
  $A_4$ models}},
  \MYhref[journalLinks]{http://dx.doi.org/10.1007/JHEP06(2013)114}{JHEP
  }\MYhref[journalLinks]{http://dx.doi.org/10.1007/JHEP06(2013)114}{\textbf{06}
  (2013) 114}, \MYhref[eprintLinks]{http://arxiv.org/abs/1303.2645}{{\ttfamily
  arXiv:1303.2645 [hep-ph]}}.

\bibitem{delAguila:2010vg}
F.~del Aguila, A.~Carmona and J.~Santiago, \emph{{Neutrino Masses from an A4
  Symmetry in Holographic Composite Higgs Models}},
  \MYhref[journalLinks]{http://dx.doi.org/10.1007/JHEP08(2010)127}{JHEP
  }\MYhref[journalLinks]{http://dx.doi.org/10.1007/JHEP08(2010)127}{\textbf{08}
  (2010) 127}, \MYhref[eprintLinks]{http://arxiv.org/abs/1001.5151}{{\ttfamily
  arXiv:1001.5151 [hep-ph]}}.

\bibitem{Campos:2014lla}
M.~D. Campos et~al., \emph{{Fermion masses and mixings in an $SU(5)$ grand
  unified model with an extra flavor symmetry}},
  \MYhref[journalLinks]{http://dx.doi.org/10.1103/PhysRevD.90.016006}{Phys.
  Rev.
  }\MYhref[journalLinks]{http://dx.doi.org/10.1103/PhysRevD.90.016006}{\textbf{D90}
  (2014) 1 016006},
  \MYhref[eprintLinks]{http://arxiv.org/abs/1403.2525}{{\ttfamily
  arXiv:1403.2525 [hep-ph]}}.

\bibitem{Vien:2014pta}
V.~V. Vien and H.~N. Long, \emph{{Neutrino mixing with nonzero $\theta_{13}$
  and CP violation in the 3-3-1 model based on $A_4$ flavor symmetry}},
  \MYhref[journalLinks]{http://dx.doi.org/10.1142/S0217751X15501171}{Int. J.
  Mod. Phys.
  }\MYhref[journalLinks]{http://dx.doi.org/10.1142/S0217751X15501171}{\textbf{A30}
  (2015) 21 1550117},
  \MYhref[eprintLinks]{http://arxiv.org/abs/1405.4665}{{\ttfamily
  arXiv:1405.4665 [hep-ph]}}.

\bibitem{Karmakar:2014dva}
B.~Karmakar and A.~Sil, \emph{{Nonzero $?_{13}$ and leptogenesis in a type-I
  seesaw model with $A_4$ symmetry}},
  \MYhref[journalLinks]{http://dx.doi.org/10.1103/PhysRevD.91.013004}{Phys.
  Rev.
  }\MYhref[journalLinks]{http://dx.doi.org/10.1103/PhysRevD.91.013004}{\textbf{D91}
  (2015) 013004},
  \MYhref[eprintLinks]{http://arxiv.org/abs/1407.5826}{{\ttfamily
  arXiv:1407.5826 [hep-ph]}}.

\bibitem{Karmakar:2015jza}
B.~Karmakar and A.~Sil, \emph{{Spontaneous CP violation in lepton-sector: A
  common origin for $\theta_{13}$, the Dirac CP phase, and leptogenesis}},
  \MYhref[journalLinks]{http://dx.doi.org/10.1103/PhysRevD.93.013006}{Phys.
  Rev.
  }\MYhref[journalLinks]{http://dx.doi.org/10.1103/PhysRevD.93.013006}{\textbf{D93}
  (2016) 1 013006},
  \MYhref[eprintLinks]{http://arxiv.org/abs/1509.07090}{{\ttfamily
  arXiv:1509.07090 [hep-ph]}}.

\bibitem{Joshipura:2015dsa}
A.~S. Joshipura and K.~M. Patel, \emph{{Generalized $\mu-\tau$ symmetry and
  discrete subgroups of $O(3)$}},
  \MYhref[journalLinks]{http://dx.doi.org/10.1016/j.physletb.2015.07.062}{Phys.
  Lett.
  }\MYhref[journalLinks]{http://dx.doi.org/10.1016/j.physletb.2015.07.062}{\textbf{B749}
  (2015) 159--166},
  \MYhref[eprintLinks]{http://arxiv.org/abs/1507.01235}{{\ttfamily
  arXiv:1507.01235 [hep-ph]}}.

\bibitem{Hernandez:2015tna}
A.~E. C\'arcamo~Hern\'andez and R.~Martinez, \emph{{A predictive 3-3-1 model
  with $A_4$ flavor symmetry}},
  \MYhref[journalLinks]{http://dx.doi.org/10.1016/j.nuclphysb.2016.02.025}{Nucl.
  Phys.
  }\MYhref[journalLinks]{http://dx.doi.org/10.1016/j.nuclphysb.2016.02.025}{\textbf{B905}
  (2016) 337--358},
  \MYhref[eprintLinks]{http://arxiv.org/abs/1501.05937}{{\ttfamily
  arXiv:1501.05937 [hep-ph]}}.

\bibitem{Bhattacharya:2016lts}
S.~Bhattacharya, B.~Karmakar, N.~Sahu and A.~Sil, \emph{{Unifying the flavor
  origin of dark matter with leptonic nonzero $\theta_{13}$}},
  \MYhref[journalLinks]{http://dx.doi.org/10.1103/PhysRevD.93.115041}{Phys.
  Rev.
  }\MYhref[journalLinks]{http://dx.doi.org/10.1103/PhysRevD.93.115041}{\textbf{D93}
  (2016) 11 115041},
  \MYhref[eprintLinks]{http://arxiv.org/abs/1603.04776}{{\ttfamily
  arXiv:1603.04776 [hep-ph]}}.

\bibitem{Karmakar:2016cvb}
B.~Karmakar and A.~Sil, \emph{{An $A_4$ realization of inverse seesaw: neutrino
  masses, $\theta_{13}$ and leptonic non-unitarity}},
  \MYhref[journalLinks]{http://dx.doi.org/10.1103/PhysRevD.96.015007}{Phys.
  Rev.
  }\MYhref[journalLinks]{http://dx.doi.org/10.1103/PhysRevD.96.015007}{\textbf{D96}
  (2017) 1 015007},
  \MYhref[eprintLinks]{http://arxiv.org/abs/1610.01909}{{\ttfamily
  arXiv:1610.01909 [hep-ph]}}.

\bibitem{Bhattacharya:2016rqj}
S.~Bhattacharya, B.~Karmakar, N.~Sahu and A.~Sil, \emph{{Flavor origin of dark
  matter and its relation with leptonic nonzero $\theta_{13}$ and Dirac CP
  phase $\delta$}},
  \MYhref[journalLinks]{http://dx.doi.org/10.1007/JHEP05(2017)068}{JHEP
  }\MYhref[journalLinks]{http://dx.doi.org/10.1007/JHEP05(2017)068}{\textbf{05}
  (2017) 068}, \MYhref[eprintLinks]{http://arxiv.org/abs/1611.07419}{{\ttfamily
  arXiv:1611.07419 [hep-ph]}}.

\bibitem{Chattopadhyay:2017zvs}
P.~Chattopadhyay and K.~M. Patel, \emph{{Discrete symmetries for electroweak
  natural type-I seesaw mechanism}},
  \MYhref[journalLinks]{http://dx.doi.org/10.1016/j.nuclphysb.2017.06.008}{Nucl.
  Phys.
  }\MYhref[journalLinks]{http://dx.doi.org/10.1016/j.nuclphysb.2017.06.008}{\textbf{B921}
  (2017) 487--506},
  \MYhref[eprintLinks]{http://arxiv.org/abs/1703.09541}{{\ttfamily
  arXiv:1703.09541 [hep-ph]}}.

\bibitem{CarcamoHernandez:2017kra}
A.~E. C\'arcamo~Hern\'andez and H.~N. Long, \emph{{A highly predictive $A_{4}$
  flavour 3-3-1 model with radiative inverse seesaw mechanism}},
  \MYhref[journalLinks]{http://dx.doi.org/10.1088/1361-6471/aaace7}{J. Phys.
  }\MYhref[journalLinks]{http://dx.doi.org/10.1088/1361-6471/aaace7}{\textbf{G45}
  (2018) 4 045001},
  \MYhref[eprintLinks]{http://arxiv.org/abs/1705.05246}{{\ttfamily
  arXiv:1705.05246 [hep-ph]}}.

\bibitem{CentellesChulia:2017koy}
S.~Centelles~Chuliá, R.~Srivastava and J.~W.~F. Valle, \emph{{Generalized
  Bottom-Tau unification, neutrino oscillations and dark matter: predictions
  from a lepton quarticity flavor approach}},
  \MYhref[journalLinks]{http://dx.doi.org/10.1016/j.physletb.2017.07.065}{Phys.
  Lett.
  }\MYhref[journalLinks]{http://dx.doi.org/10.1016/j.physletb.2017.07.065}{\textbf{B773}
  (2017) 26--33},
  \MYhref[eprintLinks]{http://arxiv.org/abs/1706.00210}{{\ttfamily
  arXiv:1706.00210 [hep-ph]}}.

\bibitem{Bjorkeroth:2017tsz}
F.~Bjorkeroth, E.~J. Chun and S.~F. King, \emph{{Accidental Peccei-Quinn
  symmetry from discrete flavour symmetry and Pati-Salam}},
  \MYhref[journalLinks]{http://dx.doi.org/10.1016/j.physletb.2017.12.058}{Phys.
  Lett.
  }\MYhref[journalLinks]{http://dx.doi.org/10.1016/j.physletb.2017.12.058}{\textbf{B777}
  (2018) 428--434},
  \MYhref[eprintLinks]{http://arxiv.org/abs/1711.05741}{{\ttfamily
  arXiv:1711.05741 [hep-ph]}}.

\bibitem{CarcamoHernandez:2018aon}
A.~E. Cárcamo~Hernández and S.~F. King, \emph{{Muon anomalies and the $SU(5)$
  Yukawa relations}}  (2018),
  \MYhref[eprintLinks]{http://arxiv.org/abs/1803.07367}{{\ttfamily
  arXiv:1803.07367 [hep-ph]}}.

\bibitem{Patel:2010hr}
K.~M. Patel, \emph{{An SO(10)XS4 Model of Quark-Lepton Complementarity}},
  \MYhref[journalLinks]{http://dx.doi.org/10.1016/j.physletb.2010.11.024}{Phys.
  Lett.
  }\MYhref[journalLinks]{http://dx.doi.org/10.1016/j.physletb.2010.11.024}{\textbf{B695}
  (2011) 225--230},
  \MYhref[eprintLinks]{http://arxiv.org/abs/1008.5061}{{\ttfamily
  arXiv:1008.5061 [hep-ph]}}.

\bibitem{Mohapatra:2012tb}
R.~N. Mohapatra and C.~C. Nishi, \emph{{$S_4$ Flavored CP Symmetry for
  Neutrinos}},
  \MYhref[journalLinks]{http://dx.doi.org/10.1103/PhysRevD.86.073007}{Phys.
  Rev.
  }\MYhref[journalLinks]{http://dx.doi.org/10.1103/PhysRevD.86.073007}{\textbf{D86}
  (2012) 073007},
  \MYhref[eprintLinks]{http://arxiv.org/abs/1208.2875}{{\ttfamily
  arXiv:1208.2875 [hep-ph]}}.

\bibitem{BhupalDev:2012nm}
P.~S. Bhupal~Dev, B.~Dutta, R.~N. Mohapatra and M.~Severson,
  \emph{{$\theta_{13}$ and Proton Decay in a Minimal $SO(10) \times S_4$ model
  of Flavor}},
  \MYhref[journalLinks]{http://dx.doi.org/10.1103/PhysRevD.86.035002}{Phys.
  Rev.
  }\MYhref[journalLinks]{http://dx.doi.org/10.1103/PhysRevD.86.035002}{\textbf{D86}
  (2012) 035002},
  \MYhref[eprintLinks]{http://arxiv.org/abs/1202.4012}{{\ttfamily
  arXiv:1202.4012 [hep-ph]}}.

\bibitem{Varzielas:2012pa}
I.~de~Medeiros~Varzielas and L.~Lavoura, \emph{{Flavour models for $TM_{1}$
  lepton mixing}},
  \MYhref[journalLinks]{http://dx.doi.org/10.1088/0954-3899/40/8/085002}{J.
  Phys.
  }\MYhref[journalLinks]{http://dx.doi.org/10.1088/0954-3899/40/8/085002}{\textbf{G40}
  (2013) 085002},
  \MYhref[eprintLinks]{http://arxiv.org/abs/1212.3247}{{\ttfamily
  arXiv:1212.3247 [hep-ph]}}.

\bibitem{Ding:2013hpa}
G.-J. Ding, S.~F. King, C.~Luhn and A.~J. Stuart, \emph{{Spontaneous CP
  violation from vacuum alignment in $S_4$ models of leptons}},
  \MYhref[journalLinks]{http://dx.doi.org/10.1007/JHEP05(2013)084}{JHEP
  }\MYhref[journalLinks]{http://dx.doi.org/10.1007/JHEP05(2013)084}{\textbf{05}
  (2013) 084}, \MYhref[eprintLinks]{http://arxiv.org/abs/1303.6180}{{\ttfamily
  arXiv:1303.6180 [hep-ph]}}.

\bibitem{Ishimori:2010fs}
H.~Ishimori, Y.~Shimizu, M.~Tanimoto and A.~Watanabe, \emph{{Neutrino masses
  and mixing from $S_{4}$ flavor twisting}},
  \MYhref[journalLinks]{http://dx.doi.org/10.1103/PhysRevD.83.033004}{Phys.
  Rev.
  }\MYhref[journalLinks]{http://dx.doi.org/10.1103/PhysRevD.83.033004}{\textbf{D83}
  (2011) 033004},
  \MYhref[eprintLinks]{http://arxiv.org/abs/1010.3805}{{\ttfamily
  arXiv:1010.3805 [hep-ph]}}.

\bibitem{Ding:2013eca}
G.-J. Ding and Y.-L. Zhou, \emph{{Dirac Neutrinos with $S_4$ Flavor Symmetry in
  Warped Extra Dimensions}},
  \MYhref[journalLinks]{http://dx.doi.org/10.1016/j.nuclphysb.2013.08.011}{Nucl.
  Phys.
  }\MYhref[journalLinks]{http://dx.doi.org/10.1016/j.nuclphysb.2013.08.011}{\textbf{B876}
  (2013) 418--452},
  \MYhref[eprintLinks]{http://arxiv.org/abs/1304.2645}{{\ttfamily
  arXiv:1304.2645 [hep-ph]}}.

\bibitem{Hagedorn:2011un}
C.~Hagedorn and M.~Serone, \emph{{Leptons in Holographic Composite Higgs Models
  with Non-Abelian Discrete Symmetries}},
  \MYhref[journalLinks]{http://dx.doi.org/10.1007/JHEP10(2011)083}{JHEP
  }\MYhref[journalLinks]{http://dx.doi.org/10.1007/JHEP10(2011)083}{\textbf{10}
  (2011) 083}, \MYhref[eprintLinks]{http://arxiv.org/abs/1106.4021}{{\ttfamily
  arXiv:1106.4021 [hep-ph]}}.

\bibitem{Campos:2014zaa}
M.~D. Campos, A.~E. Cárcamo~Hernández, H.~Pas and E.~Schumacher, \emph{{Higgs
  $\rightarrow$ $\mu\tau$ as an indication for $S_4$ flavor symmetry}},
  \MYhref[journalLinks]{http://dx.doi.org/10.1103/PhysRevD.91.116011}{Phys.
  Rev.
  }\MYhref[journalLinks]{http://dx.doi.org/10.1103/PhysRevD.91.116011}{\textbf{D91}
  (2015) 11 116011},
  \MYhref[eprintLinks]{http://arxiv.org/abs/1408.1652}{{\ttfamily
  arXiv:1408.1652 [hep-ph]}}.

\bibitem{Dong:2010zu}
P.~V. Dong, H.~N. Long, D.~V. Soa and V.~V. Vien, \emph{{The 3-3-1 model with
  $S_4$ flavor symmetry}},
  \MYhref[journalLinks]{http://dx.doi.org/10.1140/epjc/s10052-011-1544-2}{Eur.
  Phys. J.
  }\MYhref[journalLinks]{http://dx.doi.org/10.1140/epjc/s10052-011-1544-2}{\textbf{C71}
  (2011) 1544}, \MYhref[eprintLinks]{http://arxiv.org/abs/1009.2328}{{\ttfamily
  arXiv:1009.2328 [hep-ph]}}.

\bibitem{VanVien:2015xha}
V.~V. Vien, H.~N. Long and D.~P. Khoi, \emph{{Neutrino Mixing with Non-Zero
  $\theta_{13}$ and CP Violation in the 3-3-1 Model Based on $S_4$ Flavor
  Symmetry}},
  \MYhref[journalLinks]{http://dx.doi.org/10.1142/S0217751X1550102X}{Int. J.
  Mod. Phys.
  }\MYhref[journalLinks]{http://dx.doi.org/10.1142/S0217751X1550102X}{\textbf{A30}
  (2015) 17 1550102},
  \MYhref[eprintLinks]{http://arxiv.org/abs/1506.06063}{{\ttfamily
  arXiv:1506.06063 [hep-ph]}}.

\bibitem{deAnda:2017yeb}
F.~J. de~Anda, S.~F. King and E.~Perdomo, \emph{{$\mathbf{SO(10)}\times
  \mathbf{S_4}$ grand unified theory of flavour and leptogenesis}},
  \MYhref[journalLinks]{http://dx.doi.org/10.1007/JHEP12(2017)075}{JHEP
  }\MYhref[journalLinks]{http://dx.doi.org/10.1007/JHEP12(2017)075}{\textbf{12}
  (2017) 075}, \MYhref[eprintLinks]{http://arxiv.org/abs/1710.03229}{{\ttfamily
  arXiv:1710.03229 [hep-ph]}}.

\bibitem{Frampton:1994rk}
P.~H. Frampton and T.~W. Kephart, \emph{{Simple nonAbelian finite flavor groups
  and fermion masses}},
  \MYhref[journalLinks]{http://dx.doi.org/10.1142/S0217751X95002187}{Int. J.
  Mod. Phys.
  }\MYhref[journalLinks]{http://dx.doi.org/10.1142/S0217751X95002187}{\textbf{A10}
  (1995) 4689--4704},
  \MYhref[eprintLinks]{http://arxiv.org/abs/hep-ph/9409330}{{\ttfamily
  arXiv:hep-ph/9409330 [hep-ph]}}.

\bibitem{Grimus:2003kq}
W.~Grimus and L.~Lavoura, \emph{{A Discrete symmetry group for maximal
  atmospheric neutrino mixing}},
  \MYhref[journalLinks]{http://dx.doi.org/10.1016/j.physletb.2003.08.032}{Phys.
  Lett.
  }\MYhref[journalLinks]{http://dx.doi.org/10.1016/j.physletb.2003.08.032}{\textbf{B572}
  (2003) 189--195},
  \MYhref[eprintLinks]{http://arxiv.org/abs/hep-ph/0305046}{{\ttfamily
  arXiv:hep-ph/0305046 [hep-ph]}}.

\bibitem{Grimus:2004rj}
W.~Grimus et~al., \emph{{Lepton mixing angle $\theta_{13} = 0$ with a
  horizontal symmetry $D_4$}},
  \MYhref[journalLinks]{http://dx.doi.org/10.1088/1126-6708/2004/07/078}{JHEP
  }\MYhref[journalLinks]{http://dx.doi.org/10.1088/1126-6708/2004/07/078}{\textbf{07}
  (2004) 078},
  \MYhref[eprintLinks]{http://arxiv.org/abs/hep-ph/0407112}{{\ttfamily
  arXiv:hep-ph/0407112 [hep-ph]}}.

\bibitem{Frigerio:2004jg}
M.~Frigerio, S.~Kaneko, E.~Ma and M.~Tanimoto, \emph{{Quaternion family
  symmetry of quarks and leptons}},
  \MYhref[journalLinks]{http://dx.doi.org/10.1103/PhysRevD.71.011901}{Phys.
  Rev.
  }\MYhref[journalLinks]{http://dx.doi.org/10.1103/PhysRevD.71.011901}{\textbf{D71}
  (2005) 011901},
  \MYhref[eprintLinks]{http://arxiv.org/abs/hep-ph/0409187}{{\ttfamily
  arXiv:hep-ph/0409187 [hep-ph]}}.

\bibitem{Adulpravitchai:2008yp}
A.~Adulpravitchai, A.~Blum and C.~Hagedorn, \emph{{A Supersymmetric D4 Model
  for $\mu-\tau$ Symmetry}},
  \MYhref[journalLinks]{http://dx.doi.org/10.1088/1126-6708/2009/03/046}{JHEP
  }\MYhref[journalLinks]{http://dx.doi.org/10.1088/1126-6708/2009/03/046}{\textbf{03}
  (2009) 046}, \MYhref[eprintLinks]{http://arxiv.org/abs/0812.3799}{{\ttfamily
  arXiv:0812.3799 [hep-ph]}}.

\bibitem{Ishimori:2008gp}
H.~Ishimori et~al., \emph{{$D(4)$ Flavor Symmetry for Neutrino Masses and
  Mixing}},
  \MYhref[journalLinks]{http://dx.doi.org/10.1016/j.physletb.2008.03.007}{Phys.
  Lett.
  }\MYhref[journalLinks]{http://dx.doi.org/10.1016/j.physletb.2008.03.007}{\textbf{B662}
  (2008) 178--184},
  \MYhref[eprintLinks]{http://arxiv.org/abs/0802.2310}{{\ttfamily
  arXiv:0802.2310 [hep-ph]}}.

\bibitem{Hagedorn:2010mq}
C.~Hagedorn and R.~Ziegler, \emph{{$\mu-\tau$ Symmetry and Charged Lepton Mass
  Hierarchy in a Supersymmetric $D_4$ Model}},
  \MYhref[journalLinks]{http://dx.doi.org/10.1103/PhysRevD.82.053011}{Phys.
  Rev.
  }\MYhref[journalLinks]{http://dx.doi.org/10.1103/PhysRevD.82.053011}{\textbf{D82}
  (2010) 053011},
  \MYhref[eprintLinks]{http://arxiv.org/abs/1007.1888}{{\ttfamily
  arXiv:1007.1888 [hep-ph]}}.

\bibitem{Vien:2013zra}
V.~V. Vien and H.~N. Long, \emph{{The $D_4$ flavor symmery in 3-3-1 model with
  neutral leptons}},
  \MYhref[journalLinks]{http://dx.doi.org/10.1142/S0217751X13501595}{Int. J.
  Mod. Phys.
  }\MYhref[journalLinks]{http://dx.doi.org/10.1142/S0217751X13501595}{\textbf{A28}
  (2013) 1350159},
  \MYhref[eprintLinks]{http://arxiv.org/abs/1312.5034}{{\ttfamily
  arXiv:1312.5034 [hep-ph]}}.

\bibitem{Babu:2004tn}
K.~S. Babu and J.~Kubo, \emph{{Dihedral families of quarks, leptons and
  Higgses}},
  \MYhref[journalLinks]{http://dx.doi.org/10.1103/PhysRevD.71.056006}{Phys.Rev.
  }\MYhref[journalLinks]{http://dx.doi.org/10.1103/PhysRevD.71.056006}{\textbf{D71}
  (2005) 056006},
  \MYhref[eprintLinks]{http://arxiv.org/abs/hep-ph/0411226}{{\ttfamily
  arXiv:hep-ph/0411226 [hep-ph]}}.

\bibitem{Kajiyama:2005rk}
Y.~Kajiyama, E.~Itou and J.~Kubo, \emph{{Nonabelian discrete family symmetry to
  soften the SUSY flavor problem and to suppress proton decay}},
  \MYhref[journalLinks]{http://dx.doi.org/10.1016/j.nuclphysb.2006.02.042}{Nucl.
  Phys.
  }\MYhref[journalLinks]{http://dx.doi.org/10.1016/j.nuclphysb.2006.02.042}{\textbf{B743}
  (2006) 74--103},
  \MYhref[eprintLinks]{http://arxiv.org/abs/hep-ph/0511268}{{\ttfamily
  arXiv:hep-ph/0511268}}.

\bibitem{Kajiyama:2007pr}
Y.~Kajiyama, \emph{{R-parity violation and non-Abelian discrete family
  symmetry}},
  \MYhref[journalLinks]{http://dx.doi.org/10.1088/1126-6708/2007/04/007}{JHEP
  }\MYhref[journalLinks]{http://dx.doi.org/10.1088/1126-6708/2007/04/007}{\textbf{04}
  (2007) 007},
  \MYhref[eprintLinks]{http://arxiv.org/abs/hep-ph/0702056}{{\ttfamily
  arXiv:hep-ph/0702056}}.

\bibitem{Kifune:2007fj}
N.~Kifune, J.~Kubo and A.~Lenz, \emph{{Flavor Changing Neutral Higgs Bosons in
  a Supersymmetric Extension based on a $Q_{6}$ Family Symmetry}},
  \MYhref[journalLinks]{http://dx.doi.org/10.1103/PhysRevD.77.076010}{Phys.Rev.
  }\MYhref[journalLinks]{http://dx.doi.org/10.1103/PhysRevD.77.076010}{\textbf{D77}
  (2008) 076010},
  \MYhref[eprintLinks]{http://arxiv.org/abs/0712.0503}{{\ttfamily
  arXiv:0712.0503 [hep-ph]}}.

\bibitem{Babu:2009nn}
K.~Babu and Y.~Meng, \emph{{Flavor Violation in Supersymmetric Q(6) Model}},
  \MYhref[journalLinks]{http://dx.doi.org/10.1103/PhysRevD.80.075003}{Phys.Rev.
  }\MYhref[journalLinks]{http://dx.doi.org/10.1103/PhysRevD.80.075003}{\textbf{D80}
  (2009) 075003},
  \MYhref[eprintLinks]{http://arxiv.org/abs/0907.4231}{{\ttfamily
  arXiv:0907.4231 [hep-ph]}}.

\bibitem{Kawashima:2009jv}
K.~Kawashima, J.~Kubo and A.~Lenz, \emph{{Testing the new CP phase in a
  Supersymmetric Model with Q(6) Family Symmetry by B(s) Mixing}},
  \MYhref[journalLinks]{http://dx.doi.org/10.1016/j.physletb.2009.09.064}{Phys.Lett.
  }\MYhref[journalLinks]{http://dx.doi.org/10.1016/j.physletb.2009.09.064}{\textbf{B681}
  (2009) 60--67},
  \MYhref[eprintLinks]{http://arxiv.org/abs/0907.2302}{{\ttfamily
  arXiv:0907.2302 [hep-ph]}}.

\bibitem{Kaburaki:2010xc}
Y.~Kaburaki, K.~Konya, J.~Kubo and A.~Lenz, \emph{{Triangle Relation of Dark
  Matter, EDM and CP Violation in B0 Mixing in a Supersymmetric $Q_{6}$
  Model}},
  \MYhref[journalLinks]{http://dx.doi.org/10.1103/PhysRevD.84.016007}{Phys.
  Rev.
  }\MYhref[journalLinks]{http://dx.doi.org/10.1103/PhysRevD.84.016007}{\textbf{D84}
  (2011) 016007},
  \MYhref[eprintLinks]{http://arxiv.org/abs/1012.2435}{{\ttfamily
  arXiv:1012.2435 [hep-ph]}}.

\bibitem{Babu:2011mv}
K.~Babu, K.~Kawashima and J.~Kubo, \emph{{Variations on the Supersymmetric
  $Q_6$ Model of Flavor}},
  \MYhref[journalLinks]{http://dx.doi.org/10.1103/PhysRevD.83.095008}{Phys.Rev.
  }\MYhref[journalLinks]{http://dx.doi.org/10.1103/PhysRevD.83.095008}{\textbf{D83}
  (2011) 095008},
  \MYhref[eprintLinks]{http://arxiv.org/abs/1103.1664}{{\ttfamily
  arXiv:1103.1664 [hep-ph]}}.

\bibitem{Araki:2011zg}
T.~Araki and Y.~Li, \emph{{$Q_{6}$ flavor symmetry model for the extension of
  the minimal standard model by three right-handed sterile neutrinos}},
  \MYhref[journalLinks]{http://dx.doi.org/10.1103/PhysRevD.85.065016}{Phys.Rev.
  }\MYhref[journalLinks]{http://dx.doi.org/10.1103/PhysRevD.85.065016}{\textbf{D85}
  (2012) 065016},
  \MYhref[eprintLinks]{http://arxiv.org/abs/1112.5819}{{\ttfamily
  arXiv:1112.5819 [hep-ph]}}.

\bibitem{Gomez-Izquierdo:2013uaa}
J.~C. G\'omez-Izquierdo, F.~Gonz\'alez-Canales and M.~Mondragon, \emph{{$Q_{6}$
  as the flavor symmetry in a non-minimal SUSY $SU(5)$ model}},
  \MYhref[journalLinks]{http://dx.doi.org/10.1140/epjc/s10052-015-3440-7}{Eur.
  Phys. J.
  }\MYhref[journalLinks]{http://dx.doi.org/10.1140/epjc/s10052-015-3440-7}{\textbf{C75}
  (2015) 5 221},
  \MYhref[eprintLinks]{http://arxiv.org/abs/1312.7385}{{\ttfamily
  arXiv:1312.7385 [hep-ph]}}.

\bibitem{Gomez-Izquierdo:2017med}
J.~C. Gómez-Izquierdo, F.~Gonzalez-Canales and M.~Mondragón, \emph{{On ${\bf
  Q}_{6}$ flavor symmetry and the breaking of $\mu \leftrightarrow \tau$
  symmetry}},
  \MYhref[journalLinks]{http://dx.doi.org/10.1142/S0217751X17501718}{Int. J.
  Mod. Phys.
  }\MYhref[journalLinks]{http://dx.doi.org/10.1142/S0217751X17501718}{\textbf{A32}
  (2017) 28-29 1750171},
  \MYhref[eprintLinks]{http://arxiv.org/abs/1705.06324}{{\ttfamily
  arXiv:1705.06324 [hep-ph]}}.

\bibitem{Luhn:2007sy}
C.~Luhn, S.~Nasri and P.~Ramond, \emph{{Tri-bimaximal neutrino mixing and the
  family symmetry semidirect product of Z(7) and Z(3)}},
  \MYhref[journalLinks]{http://dx.doi.org/10.1016/j.physletb.2007.06.059}{Phys.
  Lett.
  }\MYhref[journalLinks]{http://dx.doi.org/10.1016/j.physletb.2007.06.059}{\textbf{B652}
  (2007) 27--33},
  \MYhref[eprintLinks]{http://arxiv.org/abs/0706.2341}{{\ttfamily
  arXiv:0706.2341 [hep-ph]}}.

\bibitem{Hagedorn:2008bc}
C.~Hagedorn, M.~A. Schmidt and A.~{\relax Yu}. Smirnov, \emph{{Lepton Mixing
  and Cancellation of the Dirac Mass Hierarchy in $SO(10)$ GUTs with Flavor
  Symmetries $T(7)$ and $\Sigma(81)$}},
  \MYhref[journalLinks]{http://dx.doi.org/10.1103/PhysRevD.79.036002}{Phys.
  Rev.
  }\MYhref[journalLinks]{http://dx.doi.org/10.1103/PhysRevD.79.036002}{\textbf{D79}
  (2009) 036002},
  \MYhref[eprintLinks]{http://arxiv.org/abs/0811.2955}{{\ttfamily
  arXiv:0811.2955 [hep-ph]}}.

\bibitem{Cao:2010mp}
Q.-H. Cao, S.~Khalil, E.~Ma and H.~Okada, \emph{{Observable $T_7$ Lepton Flavor
  Symmetry at the Large Hadron Collider}},
  \MYhref[journalLinks]{http://dx.doi.org/10.1103/PhysRevLett.106.131801}{Phys.
  Rev. Lett.
  }\MYhref[journalLinks]{http://dx.doi.org/10.1103/PhysRevLett.106.131801}{\textbf{106}
  (2011) 131801},
  \MYhref[eprintLinks]{http://arxiv.org/abs/1009.5415}{{\ttfamily
  arXiv:1009.5415 [hep-ph]}}.

\bibitem{Luhn:2012bc}
C.~Luhn, K.~M. Parattu and A.~Wingerter, \emph{{A Minimal Model of Neutrino
  Flavor}},
  \MYhref[journalLinks]{http://dx.doi.org/10.1007/JHEP12(2012)096}{JHEP
  }\MYhref[journalLinks]{http://dx.doi.org/10.1007/JHEP12(2012)096}{\textbf{12}
  (2012) 096}, \MYhref[eprintLinks]{http://arxiv.org/abs/1210.1197}{{\ttfamily
  arXiv:1210.1197 [hep-ph]}}.

\bibitem{Kajiyama:2013lja}
Y.~Kajiyama, H.~Okada and K.~Yagyu, \emph{{$T_7$ Flavor Model in Three Loop
  Seesaw and Higgs Phenomenology}},
  \MYhref[journalLinks]{http://dx.doi.org/10.1007/JHEP10(2013)196}{JHEP
  }\MYhref[journalLinks]{http://dx.doi.org/10.1007/JHEP10(2013)196}{\textbf{10}
  (2013) 196}, \MYhref[eprintLinks]{http://arxiv.org/abs/1307.0480}{{\ttfamily
  arXiv:1307.0480 [hep-ph]}}.

\bibitem{Vien:2014gza}
V.~V. Vien and H.~N. Long, \emph{{The $T_7$ flavor symmetry in 3-3-1 model with
  neutral leptons}},
  \MYhref[journalLinks]{http://dx.doi.org/10.1007/JHEP04(2014)133}{JHEP
  }\MYhref[journalLinks]{http://dx.doi.org/10.1007/JHEP04(2014)133}{\textbf{04}
  (2014) 133}, \MYhref[eprintLinks]{http://arxiv.org/abs/1402.1256}{{\ttfamily
  arXiv:1402.1256 [hep-ph]}}.

\bibitem{Vien:2015koa}
V.~V. Vien, \emph{{$T_7$ flavor symmetry scheme for understanding neutrino mass
  and mixing in 3-3-1 model with neutral leptons}},
  \MYhref[journalLinks]{http://dx.doi.org/10.1142/S0217732314501399}{Mod. Phys.
  Lett.
  }\MYhref[journalLinks]{http://dx.doi.org/10.1142/S0217732314501399}{\textbf{A29}
  (2014) 28}, \MYhref[eprintLinks]{http://arxiv.org/abs/1508.02585}{{\ttfamily
  arXiv:1508.02585 [hep-ph]}}.

\bibitem{Hernandez:2015cra}
A.~E. C\'arcamo~Hern\'andez and R.~Martinez, \emph{{Fermion mass and mixing
  pattern in a minimal $T_{7}$ flavor 331 model}},
  \MYhref[journalLinks]{http://dx.doi.org/10.1088/0954-3899/43/4/045003}{J.
  Phys.
  }\MYhref[journalLinks]{http://dx.doi.org/10.1088/0954-3899/43/4/045003}{\textbf{G43}
  (2016) 4 045003},
  \MYhref[eprintLinks]{http://arxiv.org/abs/1501.07261}{{\ttfamily
  arXiv:1501.07261 [hep-ph]}}.

\bibitem{Arbelaez:2015toa}
C.~Arbel\'aez, A.~E. C\'arcamo~Hern\'andez, S.~Kovalenko and I.~Schmidt,
  \emph{{Adjoint $SU(5)$ GUT model with $T_{7}$ flavor symmetry}},
  \MYhref[journalLinks]{http://dx.doi.org/10.1103/PhysRevD.92.115015}{Phys.
  Rev.
  }\MYhref[journalLinks]{http://dx.doi.org/10.1103/PhysRevD.92.115015}{\textbf{D92}
  (2015) 11 115015},
  \MYhref[eprintLinks]{http://arxiv.org/abs/1507.03852}{{\ttfamily
  arXiv:1507.03852 [hep-ph]}}.

\bibitem{Ding:2011qt}
G.-J. Ding, \emph{{Tri-Bimaximal Neutrino Mixing and the $T_{13}$ Flavor
  Symmetry}},
  \MYhref[journalLinks]{http://dx.doi.org/10.1016/j.nuclphysb.2011.08.012}{Nucl.
  Phys.
  }\MYhref[journalLinks]{http://dx.doi.org/10.1016/j.nuclphysb.2011.08.012}{\textbf{B853}
  (2011) 635--662},
  \MYhref[eprintLinks]{http://arxiv.org/abs/1105.5879}{{\ttfamily
  arXiv:1105.5879 [hep-ph]}}.

\bibitem{Hartmann:2011dn}
C.~Hartmann, \emph{{The Frobenius group T13 and the canonical see-saw mechanism
  applied to neutrino mixing}},
  \MYhref[journalLinks]{http://dx.doi.org/10.1103/PhysRevD.85.013012}{Phys.
  Rev.
  }\MYhref[journalLinks]{http://dx.doi.org/10.1103/PhysRevD.85.013012}{\textbf{D85}
  (2012) 013012},
  \MYhref[eprintLinks]{http://arxiv.org/abs/1109.5143}{{\ttfamily
  arXiv:1109.5143 [hep-ph]}}.

\bibitem{Hartmann:2011pq}
C.~Hartmann and A.~Zee, \emph{{Neutrino Mixing and the Frobenius Group $T13$}},
  \MYhref[journalLinks]{http://dx.doi.org/10.1016/j.nuclphysb.2011.07.023}{Nucl.
  Phys.
  }\MYhref[journalLinks]{http://dx.doi.org/10.1016/j.nuclphysb.2011.07.023}{\textbf{B853}
  (2011) 105--124},
  \MYhref[eprintLinks]{http://arxiv.org/abs/1106.0333}{{\ttfamily
  arXiv:1106.0333 [hep-ph]}}.

\bibitem{Kajiyama:2010sb}
Y.~Kajiyama and H.~Okada, \emph{{$T(13)$ Flavor Symmetry and Decaying Dark
  Matter}},
  \MYhref[journalLinks]{http://dx.doi.org/10.1016/j.nuclphysb.2011.02.020}{Nucl.
  Phys.
  }\MYhref[journalLinks]{http://dx.doi.org/10.1016/j.nuclphysb.2011.02.020}{\textbf{B848}
  (2011) 303--313},
  \MYhref[eprintLinks]{http://arxiv.org/abs/1011.5753}{{\ttfamily
  arXiv:1011.5753 [hep-ph]}}.

\bibitem{Sen:2007vx}
S.~Sen, \emph{{Quark masses in supersymmetric $SU(3)$(color)$\times SU(3)(W)
  \times U(1)(X)$ model with discrete $T$-prime flavor symmetry}},
  \MYhref[journalLinks]{http://dx.doi.org/10.1103/PhysRevD.76.115020}{Phys.
  Rev.
  }\MYhref[journalLinks]{http://dx.doi.org/10.1103/PhysRevD.76.115020}{\textbf{D76}
  (2007) 115020},
  \MYhref[eprintLinks]{http://arxiv.org/abs/0710.2734}{{\ttfamily
  arXiv:0710.2734 [hep-ph]}}.

\bibitem{Chen:2007afa}
M.-C. Chen and K.~T. Mahanthappa, \emph{{CKM and Tri-bimaximal MNS Matrices in
  a $SU(5) \times ^{(d)}T$ Model}},
  \MYhref[journalLinks]{http://dx.doi.org/10.1016/j.physletb.2007.06.064}{Phys.
  Lett.
  }\MYhref[journalLinks]{http://dx.doi.org/10.1016/j.physletb.2007.06.064}{\textbf{B652}
  (2007) 34--39},
  \MYhref[eprintLinks]{http://arxiv.org/abs/0705.0714}{{\ttfamily
  arXiv:0705.0714 [hep-ph]}}.

\bibitem{Frampton:2008bz}
P.~H. Frampton, T.~W. Kephart and S.~Matsuzaki, \emph{{Simplified
  Renormalizable T-prime Model for Tribimaximal Mixing and Cabibbo Angle}},
  \MYhref[journalLinks]{http://dx.doi.org/10.1103/PhysRevD.78.073004}{Phys.
  Rev.
  }\MYhref[journalLinks]{http://dx.doi.org/10.1103/PhysRevD.78.073004}{\textbf{D78}
  (2008) 073004},
  \MYhref[eprintLinks]{http://arxiv.org/abs/0807.4713}{{\ttfamily
  arXiv:0807.4713 [hep-ph]}}.

\bibitem{Eby:2011ph}
D.~A. Eby, P.~H. Frampton, X.-G. He and T.~W. Kephart, \emph{{Quartification
  with T' Flavor}},
  \MYhref[journalLinks]{http://dx.doi.org/10.1103/PhysRevD.84.037302}{Phys.
  Rev.
  }\MYhref[journalLinks]{http://dx.doi.org/10.1103/PhysRevD.84.037302}{\textbf{D84}
  (2011) 037302},
  \MYhref[eprintLinks]{http://arxiv.org/abs/1103.5737}{{\ttfamily
  arXiv:1103.5737 [hep-ph]}}.

\bibitem{Frampton:2013lva}
P.~H. Frampton, C.~M. Ho and T.~W. Kephart, \emph{{Heterotic discrete flavor
  model}},
  \MYhref[journalLinks]{http://dx.doi.org/10.1103/PhysRevD.89.027701}{Phys.
  Rev.
  }\MYhref[journalLinks]{http://dx.doi.org/10.1103/PhysRevD.89.027701}{\textbf{D89}
  (2014) 2 027701},
  \MYhref[eprintLinks]{http://arxiv.org/abs/1305.4402}{{\ttfamily
  arXiv:1305.4402 [hep-ph]}}.

\bibitem{Chen:2013wba}
M.-C. Chen, J.~Huang, K.~Mahanthappa and A.~M. Wijangco, \emph{{Large
  $\theta_{13}$ in a SUSY $SU(5) \times T'$ Model}},
  \MYhref[journalLinks]{http://dx.doi.org/10.1007/JHEP10(2013)112}{JHEP
  }\MYhref[journalLinks]{http://dx.doi.org/10.1007/JHEP10(2013)112}{\textbf{1310}
  (2013) 112}, \MYhref[eprintLinks]{http://arxiv.org/abs/1307.7711}{{\ttfamily
  arXiv:1307.7711}}.

\bibitem{Ma:2007wu}
E.~Ma, \emph{{Near tribimaximal neutrino mixing with $\Delta(27)$ symmetry}},
  \MYhref[journalLinks]{http://dx.doi.org/10.1016/j.physletb.2007.12.060}{Phys.
  Lett.
  }\MYhref[journalLinks]{http://dx.doi.org/10.1016/j.physletb.2007.12.060}{\textbf{B660}
  (2008) 505--507},
  \MYhref[eprintLinks]{http://arxiv.org/abs/0709.0507}{{\ttfamily
  arXiv:0709.0507 [hep-ph]}}.

\bibitem{Varzielas:2012nn}
I.~de~Medeiros~Varzielas, D.~Emmanuel-Costa and P.~Leser, \emph{{Geometrical CP
  Violation from Non-Renormalisable Scalar Potentials}},
  \MYhref[journalLinks]{http://dx.doi.org/10.1016/j.physletb.2012.08.008}{Phys.
  Lett.
  }\MYhref[journalLinks]{http://dx.doi.org/10.1016/j.physletb.2012.08.008}{\textbf{B716}
  (2012) 193--196},
  \MYhref[eprintLinks]{http://arxiv.org/abs/1204.3633}{{\ttfamily
  arXiv:1204.3633 [hep-ph]}}.

\bibitem{Bhattacharyya:2012pi}
G.~Bhattacharyya, I.~de~Medeiros~Varzielas and P.~Leser, \emph{{A common origin
  of fermion mixing and geometrical CP violation, and its test through Higgs
  physics at the LHC}},
  \MYhref[journalLinks]{http://dx.doi.org/10.1103/PhysRevLett.109.241603}{Phys.
  Rev. Lett.
  }\MYhref[journalLinks]{http://dx.doi.org/10.1103/PhysRevLett.109.241603}{\textbf{109}
  (2012) 241603},
  \MYhref[eprintLinks]{http://arxiv.org/abs/1210.0545}{{\ttfamily
  arXiv:1210.0545 [hep-ph]}}.

\bibitem{Ma:2013xqa}
E.~Ma, \emph{{Neutrino Mixing and Geometric CP Violation with $\Delta(27)$
  Symmetry}},
  \MYhref[journalLinks]{http://dx.doi.org/10.1016/j.physletb.2013.05.011}{Phys.
  Lett.
  }\MYhref[journalLinks]{http://dx.doi.org/10.1016/j.physletb.2013.05.011}{\textbf{B723}
  (2013) 161--163},
  \MYhref[eprintLinks]{http://arxiv.org/abs/1304.1603}{{\ttfamily
  arXiv:1304.1603 [hep-ph]}}.

\bibitem{Nishi:2013jqa}
C.~C. Nishi, \emph{{Generalized $CP$ symmetries in $\Delta(27)$ flavor
  models}},
  \MYhref[journalLinks]{http://dx.doi.org/10.1103/PhysRevD.88.033010}{Phys.
  Rev.
  }\MYhref[journalLinks]{http://dx.doi.org/10.1103/PhysRevD.88.033010}{\textbf{D88}
  (2013) 3 033010},
  \MYhref[eprintLinks]{http://arxiv.org/abs/1306.0877}{{\ttfamily
  arXiv:1306.0877 [hep-ph]}}.

\bibitem{Varzielas:2013sla}
I.~de~Medeiros~Varzielas and D.~Pidt, \emph{{Towards realistic models of quark
  masses with geometrical CP violation}},
  \MYhref[journalLinks]{http://dx.doi.org/10.1088/0954-3899/41/2/025004}{J.
  Phys.
  }\MYhref[journalLinks]{http://dx.doi.org/10.1088/0954-3899/41/2/025004}{\textbf{G41}
  (2014) 025004},
  \MYhref[eprintLinks]{http://arxiv.org/abs/1307.0711}{{\ttfamily
  arXiv:1307.0711 [hep-ph]}}.

\bibitem{Ma:2014eka}
E.~Ma and A.~Natale, \emph{{Scotogenic $Z_2$ or $U(1)_D$ Model of Neutrino Mass
  with $\Delta(27)$ Symmetry}},
  \MYhref[journalLinks]{http://dx.doi.org/10.1016/j.physletb.2014.05.070}{Phys.
  Lett.
  }\MYhref[journalLinks]{http://dx.doi.org/10.1016/j.physletb.2014.05.070}{\textbf{B734}
  (2014) 403--405},
  \MYhref[eprintLinks]{http://arxiv.org/abs/1403.6772}{{\ttfamily
  arXiv:1403.6772 [hep-ph]}}.

\bibitem{Abbas:2014ewa}
M.~Abbas and S.~Khalil, \emph{{Fermion masses and mixing in $\Delta(27)$
  flavour model}},
  \MYhref[journalLinks]{http://dx.doi.org/10.1103/PhysRevD.91.053003}{Phys.
  Rev.
  }\MYhref[journalLinks]{http://dx.doi.org/10.1103/PhysRevD.91.053003}{\textbf{D91}
  (2015) 5 053003},
  \MYhref[eprintLinks]{http://arxiv.org/abs/1406.6716}{{\ttfamily
  arXiv:1406.6716 [hep-ph]}}.

\bibitem{Abbas:2015zna}
M.~Abbas, S.~Khalil, A.~Rashed and A.~Sil, \emph{{Neutrino masses and deviation
  from tribimaximal mixing in $\Delta(27)$ model with inverse seesaw
  mechanism}},
  \MYhref[journalLinks]{http://dx.doi.org/10.1103/PhysRevD.93.013018}{Phys.
  Rev.
  }\MYhref[journalLinks]{http://dx.doi.org/10.1103/PhysRevD.93.013018}{\textbf{D93}
  (2016) 1 013018},
  \MYhref[eprintLinks]{http://arxiv.org/abs/1508.03727}{{\ttfamily
  arXiv:1508.03727 [hep-ph]}}.

\bibitem{Varzielas:2015aua}
I.~de~Medeiros~Varzielas, \emph{{$\Delta(27)$ family symmetry and neutrino
  mixing}},
  \MYhref[journalLinks]{http://dx.doi.org/10.1007/JHEP08(2015)157}{JHEP
  }\MYhref[journalLinks]{http://dx.doi.org/10.1007/JHEP08(2015)157}{\textbf{08}
  (2015) 157}, \MYhref[eprintLinks]{http://arxiv.org/abs/1507.00338}{{\ttfamily
  arXiv:1507.00338 [hep-ph]}}.

\bibitem{Bjorkeroth:2015uou}
F.~Bjorkeroth, F.~J. de~Anda, I.~de~Medeiros~Varzielas and S.~F. King,
  \emph{{Towards a complete $\Delta(27) \times SO(10)$ SUSY GUT}},
  \MYhref[journalLinks]{http://dx.doi.org/10.1103/PhysRevD.94.016006}{Phys.
  Rev.
  }\MYhref[journalLinks]{http://dx.doi.org/10.1103/PhysRevD.94.016006}{\textbf{D94}
  (2016) 1 016006},
  \MYhref[eprintLinks]{http://arxiv.org/abs/1512.00850}{{\ttfamily
  arXiv:1512.00850 [hep-ph]}}.

\bibitem{Chen:2015jta}
P.~Chen et~al., \emph{{Warped flavor symmetry predictions for neutrino
  physics}},
  \MYhref[journalLinks]{http://dx.doi.org/10.1007/JHEP01(2016)007}{JHEP
  }\MYhref[journalLinks]{http://dx.doi.org/10.1007/JHEP01(2016)007}{\textbf{01}
  (2016) 007}, \MYhref[eprintLinks]{http://arxiv.org/abs/1509.06683}{{\ttfamily
  arXiv:1509.06683 [hep-ph]}}.

\bibitem{Vien:2016tmh}
V.~V. Vien, A.~E. Cárcamo~Hernández and H.~N. Long, \emph{{The $\Delta(27)$
  flavor 3-3-1 model with neutral leptons}},
  \MYhref[journalLinks]{http://dx.doi.org/10.1016/j.nuclphysb.2016.10.010}{Nucl.
  Phys.
  }\MYhref[journalLinks]{http://dx.doi.org/10.1016/j.nuclphysb.2016.10.010}{\textbf{B913}
  (2016) 792--814},
  \MYhref[eprintLinks]{http://arxiv.org/abs/1601.03300}{{\ttfamily
  arXiv:1601.03300 [hep-ph]}}.

\bibitem{Hernandez:2016eod}
A.~E. C\'arcamo~Hern\'andez, H.~N. Long and V.~V. Vien, \emph{{A 3-3-1 model
  with right-handed neutrinos based on the $\Delta \left( 27 \right)$ family
  symmetry}},
  \MYhref[journalLinks]{http://dx.doi.org/10.1140/epjc/s10052-016-4074-0}{Eur.
  Phys. J.
  }\MYhref[journalLinks]{http://dx.doi.org/10.1140/epjc/s10052-016-4074-0}{\textbf{C76}
  (2016) 5 242},
  \MYhref[eprintLinks]{http://arxiv.org/abs/1601.05062}{{\ttfamily
  arXiv:1601.05062 [hep-ph]}}.

\bibitem{CarcamoHernandez:2017owh}
A.~E. C\'arcamo~Hern\'andez, S.~Kovalenko, J.~W.~F. Valle and C.~A.
  Vaquera-Araujo, \emph{{Predictive Pati-Salam theory of fermion masses and
  mixing}},
  \MYhref[journalLinks]{http://dx.doi.org/10.1007/JHEP07(2017)118}{JHEP
  }\MYhref[journalLinks]{http://dx.doi.org/10.1007/JHEP07(2017)118}{\textbf{07}
  (2017) 118}, \MYhref[eprintLinks]{http://arxiv.org/abs/1705.06320}{{\ttfamily
  arXiv:1705.06320 [hep-ph]}}.

\bibitem{deMedeirosVarzielas:2017sdv}
I.~de~Medeiros~Varzielas, G.~G. Ross and J.~Talbert, \emph{{A Unified Model of
  Quarks and Leptons with a Universal Texture Zero}},
  \MYhref[journalLinks]{http://dx.doi.org/10.1007/JHEP03(2018)007}{JHEP
  }\MYhref[journalLinks]{http://dx.doi.org/10.1007/JHEP03(2018)007}{\textbf{03}
  (2018) 007}, \MYhref[eprintLinks]{http://arxiv.org/abs/1710.01741}{{\ttfamily
  arXiv:1710.01741 [hep-ph]}}.

\bibitem{Bernal:2017xat}
N.~Bernal, A.~E. C\'arcamo~Hern\'andez, I.~de~Medeiros~Varzielas and
  S.~Kovalenko, \emph{{Fermion masses and mixings and dark matter constraints
  in a model with radiative seesaw mechanism}},
  \MYhref[journalLinks]{http://dx.doi.org/10.1007/JHEP05(2018)053}{JHEP
  }\MYhref[journalLinks]{http://dx.doi.org/10.1007/JHEP05(2018)053}{\textbf{05}
  (2018) 053}, \MYhref[eprintLinks]{http://arxiv.org/abs/1712.02792}{{\ttfamily
  arXiv:1712.02792 [hep-ph]}}.

\bibitem{CarcamoHernandez:2018iel}
A.~E. Cárcamo~Hernández, H.~N. Long and V.~V. Vien, \emph{{Fermion masses and
  mixings in a 3-3-1 model with $\Delta\left(27\right)$ family symmetry and
  inverse seesaw mechanism}}  (2018),
  \MYhref[eprintLinks]{http://arxiv.org/abs/1803.01636}{{\ttfamily
  arXiv:1803.01636 [hep-ph]}}.

\bibitem{Everett:2008et}
L.~L. Everett and A.~J. Stuart, \emph{{Icosahedral (A(5)) Family Symmetry and
  the Golden Ratio Prediction for Solar Neutrino Mixing}},
  \MYhref[journalLinks]{http://dx.doi.org/10.1103/PhysRevD.79.085005}{Phys.
  Rev.
  }\MYhref[journalLinks]{http://dx.doi.org/10.1103/PhysRevD.79.085005}{\textbf{D79}
  (2009) 085005},
  \MYhref[eprintLinks]{http://arxiv.org/abs/0812.1057}{{\ttfamily
  arXiv:0812.1057 [hep-ph]}}.

\bibitem{Feruglio:2011qq}
F.~Feruglio and A.~Paris, \emph{{The Golden Ratio Prediction for the Solar
  Angle from a Natural Model with $A_{5}$ Flavour Symmetry}},
  \MYhref[journalLinks]{http://dx.doi.org/10.1007/JHEP03(2011)101}{JHEP
  }\MYhref[journalLinks]{http://dx.doi.org/10.1007/JHEP03(2011)101}{\textbf{03}
  (2011) 101}, \MYhref[eprintLinks]{http://arxiv.org/abs/1101.0393}{{\ttfamily
  arXiv:1101.0393 [hep-ph]}}.

\bibitem{Cooper:2012bd}
I.~K. Cooper, S.~F. King and A.~J. Stuart, \emph{{A Golden $A_5$ Model of
  Leptons with a Minimal NLO Correction}},
  \MYhref[journalLinks]{http://dx.doi.org/10.1016/j.nuclphysb.2013.07.027}{Nucl.
  Phys.
  }\MYhref[journalLinks]{http://dx.doi.org/10.1016/j.nuclphysb.2013.07.027}{\textbf{B875}
  (2013) 650--677},
  \MYhref[eprintLinks]{http://arxiv.org/abs/1212.1066}{{\ttfamily
  arXiv:1212.1066 [hep-ph]}}.

\bibitem{Varzielas:2013hga}
I.~de~Medeiros~Varzielas and L.~Lavoura, \emph{{Golden ratio lepton mixing and
  nonzero reactor angle with $A_5$}},
  \MYhref[journalLinks]{http://dx.doi.org/10.1088/0954-3899/41/5/055005}{J.
  Phys.
  }\MYhref[journalLinks]{http://dx.doi.org/10.1088/0954-3899/41/5/055005}{\textbf{G41}
  (2014) 055005},
  \MYhref[eprintLinks]{http://arxiv.org/abs/1312.0215}{{\ttfamily
  arXiv:1312.0215 [hep-ph]}}.

\bibitem{Gehrlein:2014wda}
J.~Gehrlein, J.~P. Oppermann, D.~Schäfer and M.~Spinrath, \emph{{An $SU(5)
  \times A_5$ golden ratio flavour model}},
  \MYhref[journalLinks]{http://dx.doi.org/10.1016/j.nuclphysb.2014.11.023}{Nucl.
  Phys.
  }\MYhref[journalLinks]{http://dx.doi.org/10.1016/j.nuclphysb.2014.11.023}{\textbf{B890}
  (2014) 539--568},
  \MYhref[eprintLinks]{http://arxiv.org/abs/1410.2057}{{\ttfamily
  arXiv:1410.2057 [hep-ph]}}.

\bibitem{Gehrlein:2015dxa}
J.~Gehrlein, S.~T. Petcov, M.~Spinrath and X.~Zhang, \emph{{Leptogenesis in an
  $SU(5) \times A_5$ Golden Ratio Flavour Model}},
  \MYhref[journalLinks]{http://dx.doi.org/10.1016/j.nuclphysb.2015.04.019}{Nucl.
  Phys.
  }\MYhref[journalLinks]{http://dx.doi.org/10.1016/j.nuclphysb.2015.04.019}{\textbf{B896}
  (2015) 311--329},
  \MYhref[eprintLinks]{http://arxiv.org/abs/1502.00110}{{\ttfamily
  arXiv:1502.00110 [hep-ph]}}.

\bibitem{DiIura:2015kfa}
A.~Di~Iura, C.~Hagedorn and D.~Meloni, \emph{{Lepton mixing from the interplay
  of the alternating group A$_{5}$ and CP}},
  \MYhref[journalLinks]{http://dx.doi.org/10.1007/JHEP08(2015)037}{JHEP
  }\MYhref[journalLinks]{http://dx.doi.org/10.1007/JHEP08(2015)037}{\textbf{08}
  (2015) 037}, \MYhref[eprintLinks]{http://arxiv.org/abs/1503.04140}{{\ttfamily
  arXiv:1503.04140 [hep-ph]}}.

\bibitem{Ballett:2015wia}
P.~Ballett, S.~Pascoli and J.~Turner, \emph{{Mixing angle and phase
  correlations from $A_{5}$ with generalized CP and their prospects for
  discovery}},
  \MYhref[journalLinks]{http://dx.doi.org/10.1103/PhysRevD.92.093008}{Phys.
  Rev.
  }\MYhref[journalLinks]{http://dx.doi.org/10.1103/PhysRevD.92.093008}{\textbf{D92}
  (2015) 9 093008},
  \MYhref[eprintLinks]{http://arxiv.org/abs/1503.07543}{{\ttfamily
  arXiv:1503.07543 [hep-ph]}}.

\bibitem{Gehrlein:2015dza}
J.~Gehrlein, S.~T. Petcov, M.~Spinrath and X.~Zhang, \emph{{Leptogenesis in an
  SU(5) x A5 Golden Ratio Flavour Model: Addendum}},
  \MYhref[journalLinks]{http://dx.doi.org/10.1016/j.nuclphysb.2015.08.019}{Nucl.
  Phys.
  }\MYhref[journalLinks]{http://dx.doi.org/10.1016/j.nuclphysb.2015.08.019}{\textbf{B899}
  (2015) 617--630},
  \MYhref[eprintLinks]{http://arxiv.org/abs/1508.07930}{{\ttfamily
  arXiv:1508.07930 [hep-ph]}}.

\bibitem{Turner:2015uta}
J.~Turner, \emph{{Predictions for leptonic mixing angle correlations and
  nontrivial Dirac CP violation from A$_5$ with generalized CP symmetry}},
  \MYhref[journalLinks]{http://dx.doi.org/10.1103/PhysRevD.92.116007}{Phys.
  Rev.
  }\MYhref[journalLinks]{http://dx.doi.org/10.1103/PhysRevD.92.116007}{\textbf{D92}
  (2015) 11 116007},
  \MYhref[eprintLinks]{http://arxiv.org/abs/1507.06224}{{\ttfamily
  arXiv:1507.06224 [hep-ph]}}.

\bibitem{Li:2015jxa}
C.-C. Li and G.-J. Ding, \emph{{Lepton Mixing in $A_5$ Family Symmetry and
  Generalized CP}},
  \MYhref[journalLinks]{http://dx.doi.org/10.1007/JHEP05(2015)100}{JHEP
  }\MYhref[journalLinks]{http://dx.doi.org/10.1007/JHEP05(2015)100}{\textbf{05}
  (2015) 100}, \MYhref[eprintLinks]{http://arxiv.org/abs/1503.03711}{{\ttfamily
  arXiv:1503.03711 [hep-ph]}}.

\bibitem{Fritzsch:1977vd}
H.~Fritzsch, \emph{{Weak Interaction Mixing in the Six - Quark Theory}},
  \MYhref[journalLinks]{http://dx.doi.org/10.1016/0370-2693(78)90524-5}{Phys.Lett.
  }\MYhref[journalLinks]{http://dx.doi.org/10.1016/0370-2693(78)90524-5}{\textbf{B73}
  (1978) 317--322}.

\bibitem{Fritzsch:1979zq}
H.~Fritzsch, \emph{{Quark Masses and Flavor Mixing}},
  \MYhref[journalLinks]{http://dx.doi.org/10.1016/0550-3213(79)90362-6}{Nucl.
  Phys.
  }\MYhref[journalLinks]{http://dx.doi.org/10.1016/0550-3213(79)90362-6}{\textbf{B155}
  (1979) 189--207}.

\bibitem{Fritzsch:1985eg}
H.~Fritzsch, \emph{{Flavor Mixing and the Internal Structure of the Quark Mass
  Matrix}},
  \MYhref[journalLinks]{http://dx.doi.org/10.1016/0370-2693(86)91592-3}{Phys.
  Lett.
  }\MYhref[journalLinks]{http://dx.doi.org/10.1016/0370-2693(86)91592-3}{\textbf{166B}
  (1986) 423--428}.

\bibitem{Branco:1988iq}
G.~C. Branco, L.~Lavoura and F.~Mota, \emph{{Nearest Neighbor Interactions and
  the Physical Content of Fritzsch Mass Matrices}},
  \MYhref[journalLinks]{http://dx.doi.org/10.1103/PhysRevD.39.3443}{Phys. Rev.
  }\MYhref[journalLinks]{http://dx.doi.org/10.1103/PhysRevD.39.3443}{\textbf{D39}
  (1989) 3443}.

\bibitem{Branco:1994jx}
G.~C. Branco and J.~I. Silva-Marcos, \emph{{NonHermitian Yukawa couplings?}},
  \MYhref[journalLinks]{http://dx.doi.org/10.1016/0370-2693(94)91069-3}{Phys.
  Lett.
  }\MYhref[journalLinks]{http://dx.doi.org/10.1016/0370-2693(94)91069-3}{\textbf{B331}
  (1994) 390--394}.

\bibitem{Harayama:1996am}
K.~Harayama and N.~Okamura, \emph{{Exact parametrization of the mass matrices
  and the KM matrix}},
  \MYhref[journalLinks]{http://dx.doi.org/10.1016/0370-2693(96)01079-9}{Phys.Lett.
  }\MYhref[journalLinks]{http://dx.doi.org/10.1016/0370-2693(96)01079-9}{\textbf{B387}
  (1996) 614--622},
  \MYhref[eprintLinks]{http://arxiv.org/abs/hep-ph/9605215}{{\ttfamily
  arXiv:hep-ph/9605215 [hep-ph]}}.

\bibitem{Harayama:1996jr}
K.~Harayama, N.~Okamura, A.~Sanda and Z.-Z. Xing, \emph{{Getting at the quark
  mass matrices}},
  \MYhref[journalLinks]{http://dx.doi.org/10.1143/PTP.97.781}{Prog.Theor.Phys.
  }\MYhref[journalLinks]{http://dx.doi.org/10.1143/PTP.97.781}{\textbf{97}
  (1997) 781--790},
  \MYhref[eprintLinks]{http://arxiv.org/abs/hep-ph/9607461}{{\ttfamily
  arXiv:hep-ph/9607461 [hep-ph]}}.

\bibitem{Mohapatra:1998ka}
R.~N. Mohapatra and S.~Nussinov, \emph{{Bimaximal neutrino mixing and neutrino
  mass matrix}},
  \MYhref[journalLinks]{http://dx.doi.org/10.1103/PhysRevD.60.013002}{Phys.Rev.
  }\MYhref[journalLinks]{http://dx.doi.org/10.1103/PhysRevD.60.013002}{\textbf{D60}
  (1999) 013002},
  \MYhref[eprintLinks]{http://arxiv.org/abs/hep-ph/9809415}{{\ttfamily
  arXiv:hep-ph/9809415 [hep-ph]}}.

\bibitem{Lam:2001fb}
C.~Lam, \emph{{A $2-3$ symmetry in neutrino oscillations}},
  \MYhref[journalLinks]{http://dx.doi.org/10.1016/S0370-2693(01)00465-8}{Phys.Lett.
  }\MYhref[journalLinks]{http://dx.doi.org/10.1016/S0370-2693(01)00465-8}{\textbf{B507}
  (2001) 214--218},
  \MYhref[eprintLinks]{http://arxiv.org/abs/hep-ph/0104116}{{\ttfamily
  arXiv:hep-ph/0104116 [hep-ph]}}.

\bibitem{Kitabayashi:2002jd}
T.~Kitabayashi and M.~Yasue, \emph{{$S(2L)$ permutation symmetry for
  left-handed $\mu$ and $\tau$ families and neutrino oscillations in an
  $SU(3)_{L} \times SU(1)_{N}$ gauge model}},
  \MYhref[journalLinks]{http://dx.doi.org/10.1103/PhysRevD.67.015006}{Phys.Rev.
  }\MYhref[journalLinks]{http://dx.doi.org/10.1103/PhysRevD.67.015006}{\textbf{D67}
  (2003) 015006},
  \MYhref[eprintLinks]{http://arxiv.org/abs/hep-ph/0209294}{{\ttfamily
  arXiv:hep-ph/0209294 [hep-ph]}}.

\bibitem{Koide:2003rx}
Y.~Koide, \emph{{Universal texture of quark and lepton mass matrices with an
  extended flavor $2<->3$ symmetry}},
  \MYhref[journalLinks]{http://dx.doi.org/10.1103/PhysRevD.69.093001}{Phys.Rev.
  }\MYhref[journalLinks]{http://dx.doi.org/10.1103/PhysRevD.69.093001}{\textbf{D69}
  (2004) 093001},
  \MYhref[eprintLinks]{http://arxiv.org/abs/hep-ph/0312207}{{\ttfamily
  arXiv:hep-ph/0312207 [hep-ph]}}.

\bibitem{Haba:2006hc}
N.~Haba and W.~Rodejohann, \emph{{A Supersymmetric contribution to the neutrino
  mass matrix and breaking of mu-tau symmetry}},
  \MYhref[journalLinks]{http://dx.doi.org/10.1103/PhysRevD.74.017701}{Phys.
  Rev.
  }\MYhref[journalLinks]{http://dx.doi.org/10.1103/PhysRevD.74.017701}{\textbf{D74}
  (2006) 017701},
  \MYhref[eprintLinks]{http://arxiv.org/abs/hep-ph/0603206}{{\ttfamily
  arXiv:hep-ph/0603206 [hep-ph]}}.

\bibitem{Xing:2006xa}
Z.-z. Xing, H.~Zhang and S.~Zhou, \emph{{Nearly Tri-bimaximal Neutrino Mixing
  and CP Violation from $\mu-\tau$ Symmetry Breaking}},
  \MYhref[journalLinks]{http://dx.doi.org/10.1016/j.physletb.2006.08.045}{Phys.
  Lett.
  }\MYhref[journalLinks]{http://dx.doi.org/10.1016/j.physletb.2006.08.045}{\textbf{B641}
  (2006) 189--197},
  \MYhref[eprintLinks]{http://arxiv.org/abs/hep-ph/0607091}{{\ttfamily
  arXiv:hep-ph/0607091 [hep-ph]}}.

\bibitem{GomezIzquierdo:2007vn}
J.~C. Gomez-Izquierdo and A.~Perez-Lorenzana, \emph{{Softly broken $\mu
  \longleftrightarrow \tau$ symmetry in the minimal see-saw model}},
  \MYhref[journalLinks]{http://dx.doi.org/10.1103/PhysRevD.77.113015}{Phys.
  Rev.
  }\MYhref[journalLinks]{http://dx.doi.org/10.1103/PhysRevD.77.113015}{\textbf{D77}
  (2008) 113015},
  \MYhref[eprintLinks]{http://arxiv.org/abs/0711.0045}{{\ttfamily
  arXiv:0711.0045 [hep-ph]}}.

\bibitem{GomezIzquierdo:2009id}
J.~C. Gomez-Izquierdo and A.~Perez-Lorenzana, \emph{{A left-right symmetric
  model with $\mu\leftrightarrow\tau$ symmetry}},
  \MYhref[journalLinks]{http://dx.doi.org/10.1103/PhysRevD.82.033008}{Phys.
  Rev.
  }\MYhref[journalLinks]{http://dx.doi.org/10.1103/PhysRevD.82.033008}{\textbf{D82}
  (2010) 033008},
  \MYhref[eprintLinks]{http://arxiv.org/abs/0912.5210}{{\ttfamily
  arXiv:0912.5210 [hep-ph]}}.

\bibitem{Xing:2010ez}
Z.-z. Xing and Y.-L. Zhou, \emph{{A Generic Diagonalization of the $3 \times 3$
  Neutrino Mass Matrix and Its Implications on the $\mu-\tau$ Flavor Symmetry
  and Maximal CP Violation}},
  \MYhref[journalLinks]{http://dx.doi.org/10.1016/j.physletb.2010.09.020}{Phys.
  Lett.
  }\MYhref[journalLinks]{http://dx.doi.org/10.1016/j.physletb.2010.09.020}{\textbf{B693}
  (2010) 584--590},
  \MYhref[eprintLinks]{http://arxiv.org/abs/1008.4906}{{\ttfamily
  arXiv:1008.4906 [hep-ph]}}.

\bibitem{He:2011kn}
H.-J. He and F.-R. Yin, \emph{{Common Origin of $\mu-\tau$ and CP Breaking in
  Neutrino Seesaw, Baryon Asymmetry, and Hidden Flavor Symmetry}},
  \MYhref[journalLinks]{http://dx.doi.org/10.1103/PhysRevD.84.033009}{Phys.
  Rev.
  }\MYhref[journalLinks]{http://dx.doi.org/10.1103/PhysRevD.84.033009}{\textbf{D84}
  (2011) 033009},
  \MYhref[eprintLinks]{http://arxiv.org/abs/1104.2654}{{\ttfamily
  arXiv:1104.2654 [hep-ph]}}.

\bibitem{Grimus:2012hu}
W.~Grimus and L.~Lavoura, \emph{{mu-tau Interchange symmetry and lepton
  mixing}},
  \MYhref[journalLinks]{http://dx.doi.org/10.1002/prop.201200118}{Fortsch.
  Phys.
  }\MYhref[journalLinks]{http://dx.doi.org/10.1002/prop.201200118}{\textbf{61}
  (2013) 535--545},
  \MYhref[eprintLinks]{http://arxiv.org/abs/1207.1678}{{\ttfamily
  arXiv:1207.1678 [hep-ph]}}.

\bibitem{Garg:2013xwa}
S.~K. Garg and S.~Gupta, \emph{{Corrections for tribimaximal, bimaximal and
  democratic neutrino mixing matrices}},
  \MYhref[journalLinks]{http://dx.doi.org/10.1007/JHEP10(2013)128}{JHEP
  }\MYhref[journalLinks]{http://dx.doi.org/10.1007/JHEP10(2013)128}{\textbf{10}
  (2013) 128}, \MYhref[eprintLinks]{http://arxiv.org/abs/1308.3054}{{\ttfamily
  arXiv:1308.3054 [hep-ph]}}.

\bibitem{Gupta:2013it}
S.~Gupta, A.~S. Joshipura and K.~M. Patel, \emph{{How good is $\mu$-$\tau$
  symmetry after results on non-zero $\theta_{13}$?}},
  \MYhref[journalLinks]{http://dx.doi.org/10.1007/JHEP09(2013)035}{JHEP
  }\MYhref[journalLinks]{http://dx.doi.org/10.1007/JHEP09(2013)035}{\textbf{09}
  (2013) 035}, \MYhref[eprintLinks]{http://arxiv.org/abs/1301.7130}{{\ttfamily
  arXiv:1301.7130 [hep-ph]}}.

\bibitem{Luo:2014upa}
S.~Luo and Z.-z. Xing, \emph{{Resolving the octant of $\theta_{23}$ via
  radiative $\mu-\tau$ symmetry breaking}},
  \MYhref[journalLinks]{http://dx.doi.org/10.1103/PhysRevD.90.073005}{Phys.
  Rev.
  }\MYhref[journalLinks]{http://dx.doi.org/10.1103/PhysRevD.90.073005}{\textbf{D90}
  (2014) 7 073005},
  \MYhref[eprintLinks]{http://arxiv.org/abs/1408.5005}{{\ttfamily
  arXiv:1408.5005 [hep-ph]}}.

\bibitem{Xing:2015fdg}
Z.-z. Xing and Z.-h. Zhao, \emph{{A review of ?-? flavor symmetry in neutrino
  physics}},
  \MYhref[journalLinks]{http://dx.doi.org/10.1088/0034-4885/79/7/076201}{Rept.
  Prog. Phys.
  }\MYhref[journalLinks]{http://dx.doi.org/10.1088/0034-4885/79/7/076201}{\textbf{79}
  (2016) 7 076201},
  \MYhref[eprintLinks]{http://arxiv.org/abs/1512.04207}{{\ttfamily
  arXiv:1512.04207 [hep-ph]}}.

\bibitem{Rivera-Agudelo:2015vza}
D.~C. Rivera-Agudelo and A.~Pérez-Lorenzana, \emph{{Generating $\theta_{13}$
  from sterile neutrinos in $\mu-\tau$ symmetric models}},
  \MYhref[journalLinks]{http://dx.doi.org/10.1103/PhysRevD.92.073009}{Phys.
  Rev.
  }\MYhref[journalLinks]{http://dx.doi.org/10.1103/PhysRevD.92.073009}{\textbf{D92}
  (2015) 7 073009},
  \MYhref[eprintLinks]{http://arxiv.org/abs/1507.07030}{{\ttfamily
  arXiv:1507.07030 [hep-ph]}}.

\bibitem{Zhao:2016orh}
Z.-h. Zhao, \emph{{On the breaking of mu-tau flavor symmetry}}, in
  \emph{{Conference on New Physics at the Large Hadron Collider Singapore,
  Singapore, February 29-March 4, 2016}} (2016)  page~.,
  \MYhref[eprintLinks]{http://arxiv.org/abs/1605.04498}{{\ttfamily
  arXiv:1605.04498 [hep-ph]}},
  \urlprefix\url{http://inspirehep.net/record/1459073/files/arXiv:1605.04498.pdf}.

\bibitem{Borgohain:2017inp}
H.~Borgohain and M.~K. Das, \emph{{Neutrinoless double beta decay and lepton
  flavour violation in broken $\mu-\tau$ symmetric neutrino mass models}},
  \MYhref[journalLinks]{http://dx.doi.org/10.1007/s10773-017-3458-8}{Int. J.
  Theor. Phys.
  }\MYhref[journalLinks]{http://dx.doi.org/10.1007/s10773-017-3458-8}{\textbf{56}
  (2017) 9 2911--2934},
  \MYhref[eprintLinks]{http://arxiv.org/abs/1705.00922}{{\ttfamily
  arXiv:1705.00922 [hep-ph]}}.

\bibitem{Garg:2017mjk}
S.~K. Garg, \emph{{Consistency of perturbed Tribimaximal, Bimaximal and
  Democratic mixing with Neutrino mixing data}}  (2017),
  \MYhref[eprintLinks]{http://arxiv.org/abs/1712.02212}{{\ttfamily
  arXiv:1712.02212 [hep-ph]}}.

\bibitem{Borgohain:2018uhf}
H.~Borgohain and M.~K. Das, \emph{{Perturbations to $\mu-\tau$ symmetry, lepton
  Number Violation and baryogenesis in left-right symmetric Model}}  (2018),
  \MYhref[eprintLinks]{http://arxiv.org/abs/1803.05710}{{\ttfamily
  arXiv:1803.05710 [hep-ph]}}.

\bibitem{Samanta:2018hqm}
R.~Samanta and M.~Chakraborty, \emph{{A minimally broken residual TBM-Klein
  symmetry and baryogenesis via leptogenesis}}  (2018),
  \MYhref[eprintLinks]{http://arxiv.org/abs/1802.04751}{{\ttfamily
  arXiv:1802.04751 [hep-ph]}}.

\bibitem{Terrazas:2018pyl}
E.~R.~L. Terrazas and A.~Pérez-Lorenzana, \emph{{Dirac neutrino mixings from
  hidden $\mu-\tau$ symmetry}}  (2018),
  \MYhref[eprintLinks]{http://arxiv.org/abs/1802.02249}{{\ttfamily
  arXiv:1802.02249 [hep-ph]}}.

\bibitem{Garg:2018rfz}
S.~K. Garg, \emph{{A Systematic Analysis of Perturbations for Hexagonal Mixing
  Matrix}}  (2018),
  \MYhref[eprintLinks]{http://arxiv.org/abs/1806.06658}{{\ttfamily
  arXiv:1806.06658 [hep-ph]}}.

\bibitem{Garg:2018jsg}
S.~K. Garg, \emph{{Model independent Analysis of Dirac CP Violating Phase for
  some well known mixing scenarios}}  (2018),
  \MYhref[eprintLinks]{http://arxiv.org/abs/1806.08239}{{\ttfamily
  arXiv:1806.08239 [hep-ph]}}.

\bibitem{Ahn:2008hy}
Y.~H. Ahn, S.~K. Kang, C.~S. Kim and T.~P. Nguyen, \emph{{Bridges of Low Energy
  observables with Leptogenesis in mu-tau Reflection Symmetry}}  (2008),
  \MYhref[eprintLinks]{http://arxiv.org/abs/0811.1458}{{\ttfamily
  arXiv:0811.1458 [hep-ph]}}.

\bibitem{Chen:2015siy}
P.~Chen, G.-J. Ding, F.~Gonzalez-Canales and J.~W.~F. Valle, \emph{{Generalized
  $\mu-\tau$ reflection symmetry and leptonic CP violation}},
  \MYhref[journalLinks]{http://dx.doi.org/10.1016/j.physletb.2015.12.069}{Phys.
  Lett.
  }\MYhref[journalLinks]{http://dx.doi.org/10.1016/j.physletb.2015.12.069}{\textbf{B753}
  (2016) 644--652},
  \MYhref[eprintLinks]{http://arxiv.org/abs/1512.01551}{{\ttfamily
  arXiv:1512.01551 [hep-ph]}}.

\bibitem{Chen:2016ica}
P.~Chen, G.-J. Ding, F.~Gonzalez-Canales and J.~W.~F. Valle, \emph{{Classifying
  CP transformations according to their texture zeros: theory and
  implications}},
  \MYhref[journalLinks]{http://dx.doi.org/10.1103/PhysRevD.94.033002}{Phys.
  Rev.
  }\MYhref[journalLinks]{http://dx.doi.org/10.1103/PhysRevD.94.033002}{\textbf{D94}
  (2016) 3 033002},
  \MYhref[eprintLinks]{http://arxiv.org/abs/1604.03510}{{\ttfamily
  arXiv:1604.03510 [hep-ph]}}.

\bibitem{Nishi:2016wki}
C.~C. Nishi and B.~L. Sánchez-Vega, \emph{{Mu-tau reflection symmetry with a
  texture-zero}},
  \MYhref[journalLinks]{http://dx.doi.org/10.1007/JHEP01(2017)068}{JHEP
  }\MYhref[journalLinks]{http://dx.doi.org/10.1007/JHEP01(2017)068}{\textbf{01}
  (2017) 068}, \MYhref[eprintLinks]{http://arxiv.org/abs/1611.08282}{{\ttfamily
  arXiv:1611.08282 [hep-ph]}}.

\bibitem{Zhao:2017yvw}
Z.-h. Zhao, \emph{{Breakings of the neutrino $\mu-\tau$ reflection symmetry}},
  \MYhref[journalLinks]{http://dx.doi.org/10.1007/JHEP09(2017)023}{JHEP
  }\MYhref[journalLinks]{http://dx.doi.org/10.1007/JHEP09(2017)023}{\textbf{09}
  (2017) 023}, \MYhref[eprintLinks]{http://arxiv.org/abs/1703.04984}{{\ttfamily
  arXiv:1703.04984 [hep-ph]}}.

\bibitem{Liu:2017frs}
Z.-C. Liu, C.-X. Yue and Z.-h. Zhao, \emph{{Neutrino $\mu-\tau$ reflection
  symmetry and its breaking in the minimal seesaw}},
  \MYhref[journalLinks]{http://dx.doi.org/10.1007/JHEP10(2017)102}{JHEP
  }\MYhref[journalLinks]{http://dx.doi.org/10.1007/JHEP10(2017)102}{\textbf{10}
  (2017) 102}, \MYhref[eprintLinks]{http://arxiv.org/abs/1707.05535}{{\ttfamily
  arXiv:1707.05535 [hep-ph]}}.

\bibitem{Zhao:2018vxy}
Z.-h. Zhao, \emph{{Modifications to the neutrino mixing given by the mu-tau
  reflection symmetry}}  (2018),
  \MYhref[eprintLinks]{http://arxiv.org/abs/1803.04603}{{\ttfamily
  arXiv:1803.04603 [hep-ph]}}.

\bibitem{Nath:2018hjx}
N.~Nath, Z.-z. Xing and J.~Zhang, \emph{{$ \mu-\tau $ Reflection Symmetry
  Embedded in Minimal Seesaw}},
  \MYhref[journalLinks]{http://dx.doi.org/10.1140/epjc/s10052-018-5751-y}{Eur.
  Phys. J.
  }\MYhref[journalLinks]{http://dx.doi.org/10.1140/epjc/s10052-018-5751-y}{\textbf{C78}
  (2018) 4 289},
  \MYhref[eprintLinks]{http://arxiv.org/abs/1801.09931}{{\ttfamily
  arXiv:1801.09931 [hep-ph]}}.

\bibitem{Fritzsch:1999ee}
H.~Fritzsch and Z.-z. Xing, \emph{{Mass and flavor mixing schemes of quarks and
  leptons}},
  \MYhref[journalLinks]{http://dx.doi.org/10.1016/S0146-6410(00)00102-2}{Prog.
  Part. Nucl. Phys.
  }\MYhref[journalLinks]{http://dx.doi.org/10.1016/S0146-6410(00)00102-2}{\textbf{45}
  (2000) 1--81},
  \MYhref[eprintLinks]{http://arxiv.org/abs/hep-ph/9912358}{{\ttfamily
  arXiv:hep-ph/9912358}}.

\bibitem{Verma:2015mgd}
R.~Verma and S.~Zhou, \emph{{Quark Flavor Mixings from Hierarchical Mass
  Matrices}},
  \MYhref[journalLinks]{http://dx.doi.org/10.1140/epjc/s10052-016-4117-6}{Eur.
  Phys. J.
  }\MYhref[journalLinks]{http://dx.doi.org/10.1140/epjc/s10052-016-4117-6}{\textbf{C76}
  (2016) 5 272},
  \MYhref[eprintLinks]{http://arxiv.org/abs/1512.06638}{{\ttfamily
  arXiv:1512.06638 [hep-ph]}}.

\bibitem{Pati:1974yy}
J.~C. Pati and A.~Salam, \emph{{Lepton Number as the Fourth Color}},
  \MYhref[journalLinks]{http://dx.doi.org/10.1103/PhysRevD.10.275,
  10.1103/PhysRevD.11.703.2}{Phys. Rev.
  }\MYhref[journalLinks]{http://dx.doi.org/10.1103/PhysRevD.10.275,
  10.1103/PhysRevD.11.703.2}{\textbf{D10} (1974) 275--289}, [Erratum: Phys.
  Rev.D11,703(1975)].

\bibitem{Mohapatra:1974gc}
R.~N. Mohapatra and J.~C. Pati, \emph{{A Natural Left-Right Symmetry}},
  \MYhref[journalLinks]{http://dx.doi.org/10.1103/PhysRevD.11.2558}{Phys. Rev.
  }\MYhref[journalLinks]{http://dx.doi.org/10.1103/PhysRevD.11.2558}{\textbf{D11}
  (1975) 2558}.

\bibitem{Senjanovic:1975rk}
G.~Senjanovic and R.~N. Mohapatra, \emph{{Exact Left-Right Symmetry and
  Spontaneous Violation of Parity}},
  \MYhref[journalLinks]{http://dx.doi.org/10.1103/PhysRevD.12.1502}{Phys. Rev.
  }\MYhref[journalLinks]{http://dx.doi.org/10.1103/PhysRevD.12.1502}{\textbf{D12}
  (1975) 1502}.

\bibitem{Senjanovic:1978ev}
G.~Senjanovic, \emph{{Spontaneous Breakdown of Parity in a Class of Gauge
  Theories}},
  \MYhref[journalLinks]{http://dx.doi.org/10.1016/0550-3213(79)90604-7}{Nucl.
  Phys.
  }\MYhref[journalLinks]{http://dx.doi.org/10.1016/0550-3213(79)90604-7}{\textbf{B153}
  (1979) 334--364}.

\bibitem{Chetyrkin:2000yt}
K.~G. Chetyrkin, J.~H. Kuhn and M.~Steinhauser, \emph{{RunDec: A Mathematica
  package for running and decoupling of the strong coupling and quark masses}},
  \MYhref[journalLinks]{http://dx.doi.org/10.1016/S0010-4655(00)00155-7}{Comput.
  Phys. Commun.
  }\MYhref[journalLinks]{http://dx.doi.org/10.1016/S0010-4655(00)00155-7}{\textbf{133}
  (2000) 43--65},
  \MYhref[eprintLinks]{http://arxiv.org/abs/hep-ph/0004189}{{\ttfamily
  arXiv:hep-ph/0004189 [hep-ph]}}.

\bibitem{Lesgourgues2006307}
J.~Lesgourgues and S.~Pastor, \emph{Massive neutrinos and cosmology},
  \MYhref[journalLinks]{http://dx.doi.org/https://doi.org/10.1016/j.physrep.2006.04.001}{Physics
  Reports
  }\MYhref[journalLinks]{http://dx.doi.org/https://doi.org/10.1016/j.physrep.2006.04.001}{\textbf{429}
  (2006) 6 307--379}, ISSN 0370-1573.

\bibitem{Otten086201}
E.~W. Otten and C.~Weinheimer, \emph{Neutrino mass limit from tritium $\beta$
  decay}, Reports on Progress in Physics \textbf{71} (2008) 8 086201.

\bibitem{RevModPhys.80.481}
F.~T. Avignone, S.~R. Elliott and J.~Engel, \emph{Double beta decay, majorana
  neutrinos, and neutrino mass},
  \MYhref[journalLinks]{http://dx.doi.org/10.1103/RevModPhys.80.481}{Rev. Mod.
  Phys.
  }\MYhref[journalLinks]{http://dx.doi.org/10.1103/RevModPhys.80.481}{\textbf{80}
  (2008) 481--516}.

\bibitem{Agostini:2018tnm}
M.~Agostini et~al. (GERDA), \emph{{Improved Limit on Neutrinoless
  Double-$\beta$ Decay of $^{76}$Ge from GERDA Phase II}},
  \MYhref[journalLinks]{http://dx.doi.org/10.1103/PhysRevLett.120.132503}{Phys.
  Rev. Lett.
  }\MYhref[journalLinks]{http://dx.doi.org/10.1103/PhysRevLett.120.132503}{\textbf{120}
  (2018) 13 132503},
  \MYhref[eprintLinks]{http://arxiv.org/abs/1803.11100}{{\ttfamily
  arXiv:1803.11100 [nucl-ex]}}.

\bibitem{KamLAND-Zen:2016pfg}
A.~Gando et~al. (KamLAND-Zen), \emph{{Search for Majorana Neutrinos near the
  Inverted Mass Hierarchy Region with KamLAND-Zen}},
  \MYhref[journalLinks]{http://dx.doi.org/10.1103/PhysRevLett.117.109903,
  10.1103/PhysRevLett.117.082503}{Phys. Rev. Lett.
  }\MYhref[journalLinks]{http://dx.doi.org/10.1103/PhysRevLett.117.109903,
  10.1103/PhysRevLett.117.082503}{\textbf{117} (2016) 8 082503}, [Addendum:
  Phys. Rev. Lett.117,no.10,109903(2016)],
  \MYhref[eprintLinks]{http://arxiv.org/abs/1605.02889}{{\ttfamily
  arXiv:1605.02889 [hep-ex]}}.

\bibitem{Agostini:2013mzu}
M.~Agostini et~al. (GERDA), \emph{{Results on Neutrinoless Double-$\beta$ Decay
  of $^{76}$Ge from Phase I of the GERDA Experiment}},
  \MYhref[journalLinks]{http://dx.doi.org/10.1103/PhysRevLett.111.122503}{Phys.
  Rev. Lett.
  }\MYhref[journalLinks]{http://dx.doi.org/10.1103/PhysRevLett.111.122503}{\textbf{111}
  (2013) 12 122503},
  \MYhref[eprintLinks]{http://arxiv.org/abs/1307.4720}{{\ttfamily
  arXiv:1307.4720 [nucl-ex]}}.

\bibitem{Agostini:2017iyd}
M.~Agostini et~al., \emph{{Background free search for neutrinoless double beta
  decay with GERDA Phase II}},
  \MYhref[journalLinks]{http://dx.doi.org/10.1038/nature21717}{Nature544,47(2017)
  }\MYhref[journalLinks]{http://dx.doi.org/10.1038/nature21717}{ (2017)},
  [Nature544,47(2017)],
  \MYhref[eprintLinks]{http://arxiv.org/abs/1703.00570}{{\ttfamily
  arXiv:1703.00570 [nucl-ex]}}.

\end{thebibliography}

\end{document}